\pdfoutput=1
\RequirePackage{ifpdf}
\ifpdf 
\documentclass[pdftex]{sigma}
\else
\documentclass{sigma}
\fi

\usepackage{mathtools}
\usepackage[all]{xy}
\usepackage{tikz}
\usepackage{dsfont}
\usetikzlibrary{patterns}

\numberwithin{equation}{section}

\newtheorem{Theorem}{Theorem}[section]

\newtheorem{Lemma}[Theorem]{Lemma}
\newtheorem{Proposition}[Theorem]{Proposition}
 { \theoremstyle{definition}
\newtheorem{Definition}[Theorem]{Definition}
\newtheorem{Note}[Theorem]{Note}
\newtheorem{Example}[Theorem]{Example}
\newtheorem{Remark}[Theorem]{Remark} }

\def\p{\partial}
\def\cl{{\cal L}}

\def\<{\langle}
\def\>{\rangle}

\def\cO{\mathcal{O}}

\def\be{\begin{equation}}
\def\ee{\end{equation}}
\def\beq{\be\begin{array}{c}}
\def\eeq{\end{array}\ee}
\def\bes{\be\begin{split}}
\def\ees{\end{split} \ee}
\def\bs{\begin{split}}
\def\es{\end{split} }

\def\b{{\beta}}
\def\a{{\alpha}}
\def\g{{ \gamma}}

\def\e{{\epsilon}}

\def\v{{\varphi}}
\def\G{\Gamma}

\begin{document}

\allowdisplaybreaks

\newcommand{\arXivNumber}{2204.06896}

\renewcommand{\PaperNumber}{072}

\FirstPageHeading

\ShortArticleName{Tropical Mirror}

\ArticleName{Tropical Mirror}

\Author{Andrey LOSEV~$^{\rm abc}$ and Vyacheslav LYSOV~$^{\rm cd}$}

\AuthorNameForHeading{A.~Losev and V.~Lysov}

\Address{$^{\rm a)}$~National Research University Higher School of Economics, Laboratory of Mirror Symmetry,\\
\hphantom{$^{\rm a)}$}~NRU HSE, 6 Usacheva Str., Moscow, 119048, Russia}
\Address{$^{\rm b)}$~Wu Wen-Tsun Key Lab of Mathematics, Chinese Academy of Sciences,\\
\hphantom{$^{\rm b)}$}~USTC, No.~96, JinZhai Road Baohe District, Hefei, Anhui, 230026, P.R.~China}
\Address{$^{\rm c)}$~Shanghai Institute for Mathematics and Interdisciplinary Sciences,\\
\hphantom{$^{\rm c)}$}~Building 3, 62 Weicheng Road, Yangpu District, Shanghai, 200433, P.R.~China}
\EmailD{\href{mailto:aslosev2@gmail.com}{aslosev2@gmail.com}, \href{mailto:lysov@simis.cn}{lysov@simis.cn}}

\Address{$^{\rm d)}$~Okinawa Institute of Science and Technology,\\
\hphantom{$^{\rm d)}$}~1919-1 Tancha, Onna-son, Okinawa 904-0495, Japan}

\ArticleDates{Received July 30, 2023, in final form July 24, 2024; Published online August 04, 2024}

\Abstract{We describe the tropical curves in toric varieties and define the tropical Gromov--Witten invariants. We introduce amplitudes for the higher topological quantum mechanics (HTQM) on special trees and show that the amplitudes are equal to the tropical Gromov--Witten invariants. We show that the sum over the amplitudes in $A$-model HTQM equals the total amplitude in B-model HTQM, defined as a deformation of the $A$-model HTQM by the mirror superpotential. We derived the mirror superpotentials for the toric varieties and showed that they coincide with the superpotentials in the mirror Landau--Ginzburg theory. We construct the mirror dual states to the evaluation observables in the tropical Gromov--Witten theory.}

\Keywords{mirror symmetry; Gromov--Witten invariants; tropical geometry; topological quantum mechanics on trees}

\Classification{14J33; 14T20; 14N35; 81Q35}

\section{Introduction}
The enumerative problems in algebraic geometry over complex numbers often allow complex tropicalization. In particular, Mikhalkin \cite{Mikhalkin2004, Mikh} showed that counting the complex curves on a toric surface becomes the counting of graphs. The higher dimensional version was developed by \cite{nishinou2006toric}, and the intersection theoretic interpretation was done by \cite{gross2018intersection} and \cite{ranganathan2017skeletons}. The TQFT formulation for the tropical multiplicities for the curve counting, including descendant invariants, was provided by Mandel--Ruddat \cite{mandel2020descendant, mandel2023tropical}.

The main focus of the present paper is the tropicalization of the Gromov--Witten invariants. Gromov--Witten invariants are naturally related to different phenomena, such as counting of the holomorphic maps, topological string theory, mirror symmetry, WDVV equations, and many others. We will discuss the tropicalization of these phenomena and show that it usually leads to simplifications.

The complex curves in toric variety $X$ in the tropical limit become graphs embedded via piece-wise linear maps. Hence, the counting problem for such curves, passing through the cycles on $X$, becomes an enumerative problem. In particular, Gathmann and Markwig \cite{Hannah} used tropical geometry to prove the Kontsevich--Manin recursive relation \cite{kontsevich1994gromov} for numbers $N_d$, of degree $d$, genus zero curves in $\mathbb{CP}^2$ passing through $3d-1$ points.

The Gromov--Witten invariants over complex numbers admit type-$A$ topological string theory \cite{Witten:1988xj} representation. We represent the tropical GW invariants using the higher topological quantum mechanics (HTQM), introduced in \cite{losev2019tqft}. Our HTQM naturally emerges from topological string theory in the limit of maximally degenerate complex structure, the very long strings. However, its ability to describe all GW invariants is surprising. Moreover, Losev and Shardin~\cite{Losev_2007} showed that the HTQM, similar to the one we describe in the present paper, provides a~solution to the WDVV equations.

In physics literature \cite{vafa2003mirror}, the mirror symmetry is a relation between the $A$-model, given by the GW invariants for toric space $X$, and the $B$-model provided by the amplitudes in Landau--Ginzburg theory with exponential mirror superpotential. In math literature \cite{givental1995quantum}, the $B$-model has an alternative description, given by periods of ``classical'' oscillatory integrals. The common feature of the two descriptions is that the mirror superpotential and argument of exponent in oscillatory integrals have the same function, uniquely constructed for a toric space $X$. Frenkel and Losev \cite{Frenkel:2005ku} derived the mirror superpotential using two-dimensional conformal field theory (CFT) methods to perform the sum over instantons in the $A$-type topological string. The tropicalization of their construction is the amplitude resummation in the HTQM.

We show that the sum over divisor states in the HTQM describes a deformation of the HTQM by a holomorphic superpotential. Moreover, we perform the summation explicitly and express the mirror superpotential in terms of the compactification data for the toric space $X$. For a~given toric space $X$, our mirror superpotential matches with the mirror superpotential in LG theory, derived using the gauged linear sigma model \cite{vafa2003mirror}, with CFT derivation \cite{Frenkel:2005ku} from physics literature and superpotentials in oscillatory integrals in math literature \cite{givental1995quantum}. For $X = \mathbb{P}^2$, our superpotential matches the mirror superpotential derived by Gross \cite{gross2010mirror} using the tropical curve counting with Mikhalkin's vertex multiplicities.

Though we used the logic of the mirror construction from \cite{Frenkel:2005ku}, the simplifications of the $A$- and $B$-models due to the tropicalization allowed us not only to reproduce the known mirror superpotentials but also to construct the mirror states for the evaluation observables. The mirror states for the tropical observables on a toric variety are mixed degree differential forms. The degree zero component of the mirror state gives K. Saito's good section \cite{saito1983period} for the mirror superpotential.

The structure of the present paper is as follows. In Section~\ref{sec2}, we briefly introduce tropical geometry and describe tropical curves in projective space and the corresponding moduli. In the next section, we describe the tropical Gromov--Witten theory. Section~\ref{sec4} presents the higher topological quantum mechanics on trees, defines the amplitudes, and proves equality between the tropical GW invariants and the $A$-model HTQM amplitudes. Section~\ref{sec5} shows an equality between the amplitudes in $A$-and $B$-model HTQMs. The next section demonstrates the various steps of the tropical mirror relation for tropical Gromov--Witten invariants on $\mathbb{P}^1$. In Section~\ref{sec7}, we briefly summarize the main results.

\section{ Introduction to tropical geometry}\label{sec2}
Here, we briefly review the notion of tropical curves in projective spaces. Readers familiar with tropical curves may refer to this section for notations used in later sections. To motivate the notations introduced in later parts of this section, we will start with a few simple examples of tropicalization for smooth curves in projective space.

\subsection{Real tropical numbers}
The {\it tropical numbers} is the set $ [-\infty, +\infty) = \mathbb{R}\cup \{-\infty\}$ with arithmetic operations
\[
x+_Ty = \max(x,y),\qquad x\ast_Ty = x+y.
\]
 The operations extended to include $-\infty$ via
\[
(-\infty)+_Tx = x +_T (-\infty) = x,\qquad (-\infty)\ast_Tx = x\ast_T (-\infty) = -\infty.
\]
The set of tropical numbers is a semigroup with respect to tropical addition $+_T$. It is commutative, associative and admits a tropical zero $0_T = -\infty$, but there is no inverse, i.e.,
\[
x+_Ty = 0_T \ \Longrightarrow \ x=0_T\qquad \hbox{or}\qquad y = 0_T.
\]
More details on bottom-up construction for tropical geometry can be found in Mikhalkin's work~\cite{Mikhalkin2004}. In the present paper, we will use the physicist approach to tropical numbers as a~certain scaling limit of $\mathbb{R}^+$, equipped with standard addition $+$ and multiplication $\ast$. To specify the limit, we introduce an exponential parametrization
\[
X = {\rm e}^{\frac{x}{\e}},\qquad Y = {\rm e}^{\frac{y}{\e}},\qquad W = {\rm e}^{\frac{w}{\e}},
\]
with parameter $\e>0$ and take the limit $\e \to 0$. Indeed, the addition operation on $\mathbb{R}^+$ in exponential parametrization after the limit $\e \to 0$ becomes the tropical addition, i.e.,
\[
w=\lim_{\e \to 0} \e \ln \bigl({\rm e}^{\frac{x}{\e}}+{\rm e}^{\frac{y}{\e}}\bigr) = \max (x,y) = x+_Ty.
\]
Similarly, the usual multiplication on $\mathbb{R}^+$ through exponential parametrization becomes tropical multiplication in the limit, i.e.,
\[
w = \lim_{\e \to 0}\e \ln \bigl({\rm e}^{\frac{x}{\e}}\ast {\rm e}^{\frac{y}{\e}}\bigr) =x+y = x\ast_Ty.
\]

\subsection[Tropical curves in C\^{}2]{Tropical curves in $\boldsymbol{\mathbb{C}^2}$} \label{sec_trop_curves_c2}
For complex numbers, we generalize the tropical limit using the cylindrical parametrization
\begin{gather}\label{cyl_param_trop}
W = {\rm e}^{\frac{r}{\e} + {\rm i} \phi},\qquad r \in \mathbb{R},\quad \phi \in S^1.
\end{gather}
The degree $d$ complex curve in Euclidean space $\mathbb{C}^2$ is a zero locus of a degree $d$ polynomial in $2$ variables $P_d (X, Y)=0$. Below, we describe the tropical limit for two simple cases:
\begin{itemize}\itemsep=0pt
\item degree 1 curve
\begin{gather}\label{deg_1_curve}
P_1(X,Y)=Y-A(X-B)=0;
\end{gather}
\item degree 2 curve
\begin{gather}\label{deg_2_curve}
P_2(X,Y)=Y - A(X-B)(X-C) = 0.
\end{gather}
\end{itemize}
The cylindrical parametrization for $X, Y$-coordinates and complex coefficients is
\begin{gather*}
X={\rm e}^{\frac{x}{\e} + {\rm i} \phi_x},\qquad Y={\rm e}^{\frac{y}{\e} + {\rm i} \phi_y},\qquad
A={\rm e}^{\frac{r_A}{\e} + {\rm i} \phi_A},\qquad B={\rm e}^{\frac{r_B}{\e} + {\rm i} \phi_B},\qquad C={\rm e}^{\frac{r_C}{\e} + {\rm i} \phi_C}.
\end{gather*}
For degree 1 curve (\ref{deg_1_curve}), we can solve for the radial and angular parts
\begin{gather*}
y = r_A + \e \ln\bigl|{\rm e}^{\frac{x}{\e}+{\rm i}\phi_x}-{\rm e}^{\frac{r_B}{\e} + {\rm i} \phi_B}\bigr|, \\
\phi_y=\phi_A +\operatorname{arg} \bigl( {\rm e}^{\frac{x}{\e} + {\rm i} \phi_x}-{\rm e}^{\frac{r_B}{\e}+{\rm i}\phi_B}\bigr).
\end{gather*}
The limit $\e \to 0$ of the curve projected to $(x,y)$-plane is the union of three rays:
\begin{itemize}\itemsep=0pt
\item $x<r_B$ region:
\begin{gather*}
y(x, \phi_x)=r_A +\lim_{\e\to 0} \e \ln\bigl|{\rm e}^{\frac{x}{\e}+{\rm i}\phi_x}-{\rm e}^{\frac{r_B}{\e} + {\rm i} \phi_B}\bigr|=r_A+r_B,\\
\phi_y (x, \phi_x) = \phi_A +\lim_{\e\to 0}\operatorname{arg}\bigl({\rm e}^{\frac{x}{\e} + {\rm i} \phi_x}-{\rm e}^{\frac{r_B}{\e} + {\rm i} \phi_B}\bigr)=\phi_A+\phi_B +\pi.
\end{gather*}
\item $x=r_B$ region:
\begin{gather*}
y(x, \phi_x)=r_A+r_B +\lim_{\e\to 0} \e \ln\biggl|\sin \frac {\phi_x-\phi_B}{2}\biggr|,\\
\phi_y(x, \phi_x)=\phi_A+\operatorname{arg} \bigl( {\rm e}^{ {\rm i} \phi_x}-{\rm e}^{ {\rm i} \phi_B}\bigr).
\end{gather*}
For $\phi_x\in(\phi_B-\delta, \phi_B+\delta)$, $y(\phi_x)\in(0, -\infty)$, hence we have a vertical line on the radial projection of the curve. 
\item $x>r_B$ region:
\begin{gather*}
y(x, \phi_x)=r_A+x,\qquad \phi_y(x, \phi_x)=\phi_A+\phi_x.
\end{gather*}
\end{itemize}
We can perform a similar analysis for the degree 2 curve (\ref{deg_2_curve}). The radial shape of the curve depends on the complex parameter $\frac{B}{C}$. The case $r_C > r_B$ is almost identical to the degree one curve(see picture below). The case $r_B=r_C$ and $\phi_C \neq \phi_B$ require careful analysis for the vertical ray $x=r_B$. The radial coordinate is given by
\begin{gather}\label{parabola_radial_aligned}
y(x, \phi_x)=r_A+r_B +r_C+\lim_{\e\to 0} \e \ln \biggl|\sin \frac{\phi_x-\phi_B}{2}\biggr|+\lim_{\e\to 0} \e \ln \biggl|\sin \frac{\phi_x-\phi_C}{2}\biggr|.
\end{gather}
We have two regions $\phi_x\in(\phi_B-\delta, \phi_B+\delta)$ and $\phi_x\in(\phi_C-\delta, \phi_C+\delta)$, each describing a vertical ray $y(\phi_x) \in (0, -\infty)$. Hence, we have two vertical rays. In order to analyze the radial coordinate for other values of $\phi_C$, we rewrite the expression in (\ref{parabola_radial_aligned}) in the form
\begin{gather*}
y(x, \phi_x)=r_A+r_B +r_C+\lim_{\e\to 0} \e \ln \biggl|\cos \bigg( \frac{\phi_C-\phi_B}{2}\bigg)-\cos\bigg(\phi_x-\frac{\phi_B+ \phi_C}{2} \bigg)\biggr|.
\end{gather*}
The radial coordinate has two maximums: one at $\phi_x=\frac{\phi_B+\phi_C}{2}+\frac12 \pi$ with the radial coordinate $r_A+r_B +r_C$, the other at $\phi_x=\frac{\phi_B+\phi_C}{2}$ with a smaller value of the radial coordinate
\begin{gather*}
r_A+r_B +r_C+2 \lim_{\e\to 0} \e \ln \biggl|\sin \bigg( \frac{\phi_C-\phi_B}{4}\bigg) \biggr|.
\end{gather*}
We can consider a limit $\phi_B\to\phi_C$, such that
\begin{gather*}
\phi_C-\phi_B = {\rm e}^{-\tau/\e},\qquad \tau>0.
\end{gather*}
In such limit the radial coordinate of the mid point $\phi_x= \frac{\phi_B+ \phi_C}{2}$ has a finite value $y=r_A+r_B +r_C-2 \tau $. The plot of the radial function for such scaling is presented in the right part of the picture below:
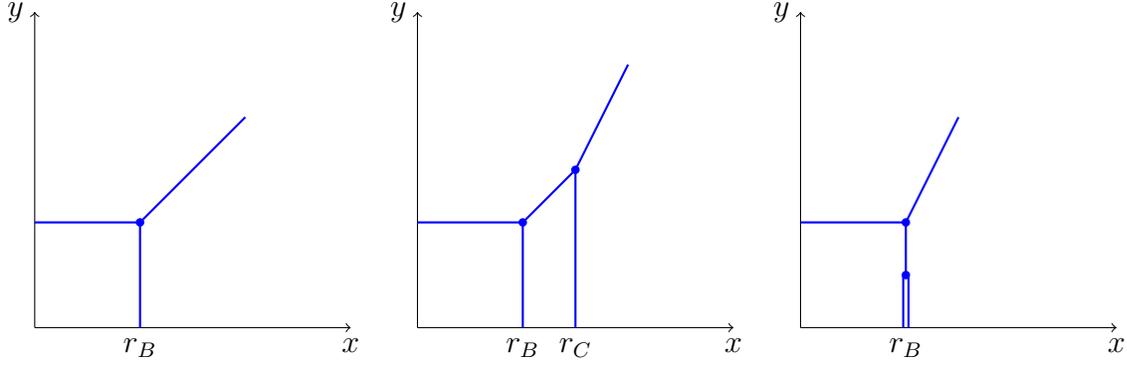
\[
\begin{tikzpicture}[scale=0.7]
\node (x) at (0,-2) [anchor=north] {$r_B$};
\node (x) at (-2,4)[anchor=east] {$y$};
\node (x) at (4,-2)[anchor=north]{$x$};
\draw [->](-2, -2) --(4,-2) ;
\draw[->] (-2, -2) -- (-2, 4);
\draw[color=blue, thick] (-2, 0) -- (0, 0)-- (0,-2);
\draw[color=blue, thick](0, 0)-- (2,2);
\filldraw[blue] (0,0) circle (2pt);
\end{tikzpicture}\quad
\begin{tikzpicture}[scale=0.7]
\node (x) at (-2,4) [anchor=east] {$y$};
\node (x) at (4,-2) [anchor=north] {$x$};
\node (x) at (0,-2) [anchor=north] {$r_B$};
\node (x) at (1,-2) [anchor=north] {$r_C$};
\draw [->](-2, -2) --(4,-2) ;
\draw[->] (-2, -2) -- (-2, 4);
\draw[color=blue, thick] (-2, 0) -- (0, 0)-- (0,-2);
\draw[color=blue, thick](0, 0)-- (1,1)--(2,3);
\draw[color=blue, thick] (1,1)--(1,-2);
\filldraw[blue] (0,0) circle (2pt);
\filldraw[blue] (1,1) circle (2pt);
\end{tikzpicture}\quad
\begin{tikzpicture}[scale=0.7]
\node (x) at (0,-2) [anchor=north] {$r_B$};
\node (x) at (-2,4)[anchor=east] {$y$};
\node (x) at (4,-2)[anchor=north]{$x$};
\draw [->](-2, -2) --(4,-2) ;
\draw[->] (-2, -2) -- (-2, 4);
\draw[color=blue, thick] (-2, 0) -- (0, 0);
\draw[color=blue, thick](0, 0)-- (1,2);
\draw[color=blue, thick] (0, 0)-- (0,-1);
\draw[color=blue, thick] (0.05, -1)-- (0.05,-2);
\draw[color=blue, thick] (-0.05, -1)-- (-0.05,-2);
\filldraw[blue] (0,0) circle (2pt);
\filldraw[blue] (0,-1) circle (2pt);
\end{tikzpicture}
\]

\subsection{Toric varieties}
\begin{Definition}
A compact toric variety $X$ of complex dimension $N$ is a holomorphic compactification at infinity of ${\mathbb{C^\ast}}^N$ to a compact space by a union of algebraic tori of lower dimensions. We represent $\mathbb{C}^{\ast N} = \mathbb{R}^N \times \mathbb{T}^N$, the radial part with coordinates $r^j$, and the angular part with coordinates $\phi^j$ for $j=1,\dots ,N$. The radial part of codimension-1 torus is a hyperplane in $\mathbb{R}^N$, described by a $N$-dimensional integer {\it primitive normal vector} $\vec{b} = \bigl(b^1,\dots,b^N\bigr) \in \mathbb{Z}^N$, i.e., such that gcd$\bigl(b^1, b^2,\dots ,b^N\bigr)=1$. For a given toric variety $X$, we denote by $B_X$ the set of primitive normal vectors for all its compactifying hyperplanes.
\end{Definition}
\begin{Remark} There are interesting toric varieties, thought singular, with non-primitive normal vectors $\vec{b}_a$, but we will not discuss them in the present paper.
\end{Remark}
\begin{Example}
The projective space $\mathbb{P}^N$ is the compactification of $\mathbb{C}^{\ast N}$ by $N+1$ hyperplanes. The corresponding primitive normal vectors are
\begin{gather*}
\vec{b}_{1} = (1,0,\dots,0),\qquad \vec{b}_2 = (0,1,0,\dots,0), \qquad \dots, \qquad
\vec{b}_N = (0,\dots,0,1),\\ \vec{b}_{N+1} = (-1,-1,\dots,-1).
\end{gather*}
\end{Example}
\begin{Example} The $\mathbb{P}^1\times \mathbb{P}^1$ is the compactification of $\mathbb{C}^{\ast 2}$ by the four lines. The corresponding primitive normal vectors are
\[
\vec{b}_{1} = (1,0),\qquad \vec{b}_2 = (0,1),\qquad \vec{b}_3 = (-1,0),\qquad \vec{b}_4 = (0,-1).
\]
\end{Example}
In the pictures below, we depict the compactifications and the normal vectors for $\mathbb{P}^1$, $\mathbb{P}^2$ and $\mathbb{P}^1 \times \mathbb{P}^1$:
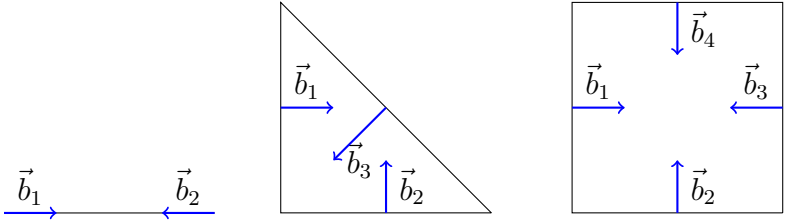
\[
\begin{tikzpicture}[scale=.7]
\draw(0,2)-- (4, 2);
\node (x) at (0.5,2.5) {$\vec{b}_1$};
\draw[color=blue, thick, ->] (0, 2) -- (1, 2);
\node (x) at (3.5,2.5) {$\vec{b}_2$};
\draw[color=blue, thick, ->] (4, 2) -- (3, 2);
\end{tikzpicture}
\qquad
\begin{tikzpicture}[scale=0.7]
\draw(4,0)-- (0, 0) -- (0, 4)--(4,0);
\node (x) at (0.5,2.5) {$\vec{b}_1$};
\draw[color=blue, thick, ->] (0, 2) -- (1, 2);
\node (x) at (2.5,0.5) {$\vec{b}_2$};
\draw[color=blue, thick, ->] (2, 0) -- (2, 1);
\node (x) at (1.5,1) {$\vec{b}_3$};
\draw[color=blue, thick, ->] (2, 2) -- (1, 1);
\end{tikzpicture}\qquad
\begin{tikzpicture}[scale=0.7]
\draw(4,0)-- (0, 0) -- (0, 4)--(4,4)--(4,0);
\node (x) at (0.5,2.5) {$\vec{b}_1$};
\draw[color=blue, thick, ->] (0, 2) -- (1, 2);
\node (x) at (2.5,0.5) {$\vec{b}_2$};
\draw[color=blue, thick, ->] (2, 0) -- (2, 1);
\node (x) at (3.5,2.5) {$\vec{b}_3$};
\draw[color=blue, thick, ->] (4, 2) -- (3, 2);
\node (x) at (2.5,3.5) {$\vec{b}_4$};
\draw[color=blue, thick, ->] (2, 4) -- (2, 3);
\end{tikzpicture}
\]
We can express the topological data of toric variety $X$ in terms of the compactifying data. In~particular, Proposition~4.5 of Fulton's book~\cite{fulton1993introduction} relates the homology $H_{2}(X)$ to dimension of~$X$ and
the cardinality $|B_X|$ of the set $B_X$ via
\begin{gather}\label{toric_betti_numbers}
\dim H_{2}(X)=|B_X|-\dim X.
\end{gather}

\subsection{Tropical curves in projective spaces}
A degree one tropical curve in $\mathbb{C}^2$ from Section \ref{sec_trop_curves_c2} can be turned into a tropical curve in compactifications of $\mathbb{C}^2$. For $\mathbb{P}^2$, we add a compactifying line \big(with $\vec{b} =(-1,-1) $\big) at infinity, and the $45^0$-ray of the tropical curve ends on this divisor. For $\mathbb{P}^1\times \mathbb{P}^1$, we add horizontal and vertical lines at infinity, as shown in the picture. The $45^0$-ray at some point splits into two rays, horizontal and vertical, ending on the corresponding lines:
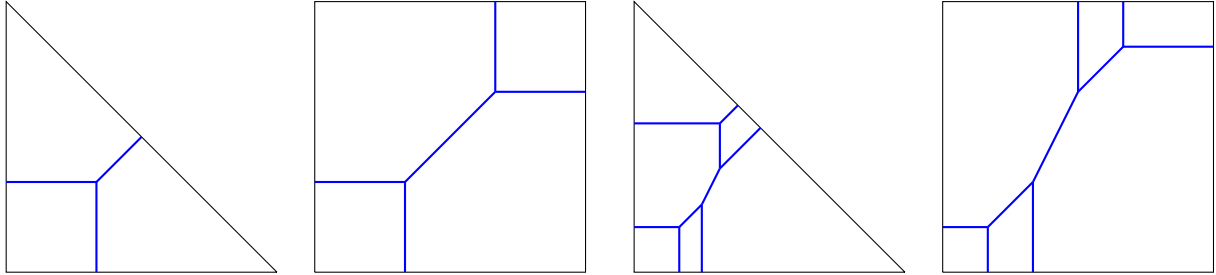
\[
\begin{tikzpicture}[scale=0.5]
\draw(4,-2)-- (-2, -2) -- (-2, 4)--(4,-2);
\draw[color=blue, thick] (-2, 0) -- (0, 0)-- (0,-2);
\draw[color=blue, thick](0, 0)-- (1,1);
\end{tikzpicture}\qquad
\begin{tikzpicture}[scale=0.5]
\draw(-2,-2)-- (4, -2) -- (4, 4)--(-2,4)--(-2,-2);
\draw[color=blue, thick] (-2, 0) -- (0, 0)-- (0,-2);
\draw[color=blue, thick](0, 0)-- (2,2);
\draw[color=blue, thick](2, 2)-- (2,4);
\draw[color=blue, thick](2, 2)-- (4,2);
\end{tikzpicture} \qquad
\begin{tikzpicture}[scale=0.5]
\draw(4,-2)-- (-2, -2) -- (-2, 4)--(4,-2);
\draw[color=blue, thick](-2, -1)--(-1, -1);
\draw[color=blue, thick](-1, -1)--(-1,-2);
\draw[color=blue, thick](-1, -1)--(-0.5,-0.5);
\draw[color=blue, thick](-0.5,-0.5)--(-.5,-2);
\draw[color=blue, thick](-0.5, -0.5)--(-0.5+0.4,-0.5+0.8);
\draw[color=blue, thick](-0.5+0.4,-0.5+0.8)--(-0.1,-0.5+0.8+1);
\draw[color=blue,thick](-0.5+0.4,-0.5+0.8)--(-0.5+0.4+0.9,-0.5+0.8+0.9);
\draw[color=blue, thick](-2,-0.5+0.8+1)--(-0.1,-0.5+0.8+1);
\draw[color=blue, thick](-0.1,-0.5+0.8+1)--(-0.1+0.4,-0.5+0.8+1+0.4);
\end{tikzpicture}\qquad
\begin{tikzpicture}[scale=0.5]
\draw(-2,-2)-- (4, -2) -- (4, 4)--(-2,4)--(-2,-2);
\draw[color=blue, thick] (-2, -1) -- (-1, -1);
\draw[color=blue, thick](-1, -1)-- (-1,-2);
\draw[color=blue, thick](-1, -1)-- (0,0);
\draw[color=blue, thick](0,0)--(0,-2);
\draw[color=blue, thick](0,0)-- (1,2)--(1,4);
\draw[color=blue, thick](1,2)--(2,3)--(2,4);
\draw[color=blue, thick](2,3) -- (4,3);
\end{tikzpicture}
\]
Let us consider two distinct degree-one tropical curves $\mathbb{P}^2$, represented by the two trees below. Two degree-one curves is a singular degree two tropical curve due to the 4-valent vertex. The resolution of tropical singularity replaces a single 4-valent vertex with a pair of 3-valent vertices connected by an additional edge. There are two possible ways of doing so, presented in the picture below. The last two pictures represent the corresponding degree-two smooth tropical curves:
\[
\begin{tikzpicture}[scale=0.5]
\draw(4,-2)-- (-2, -2) -- (-2, 4)--(4,-2);
\draw[color=blue, thick] (-2, 0) -- (0, 0)-- (0,-2);
\draw[color=blue, thick](0, 0)-- (1,1);
\draw[color=green, thick] (-2, 1) -- (-1, 1)-- (-1,-2);
\draw[color=green, thick](-1, 1)-- (0,2);
\draw (-1,0) circle (0.3);
\end{tikzpicture}
\begin{tikzpicture}[scale=0.24]
\draw[color=green, thick](0, -2)-- (0,0);
\draw[color=blue, thick] (-2, 0) -- (0, 0);
\draw[color=red, thick] (0, 0) -- (1, 1);
\draw[color=blue, thick](1, 1)-- (3,1);
\draw[color=green, thick] (1, 1) -- (1, 3);
\draw (0.5,0.5) circle (2.7);
\draw[color=red, thick] (0, 0+6) -- (-1, 1+6);
\draw[color=blue, thick] (-3, 1+6) -- (-1, 1+6);
\draw[color=blue, thick] (0, 0+6) -- (2, 0+6);
\draw[color=green, thick] (0, 0+6) -- (0, -2+6);
\draw[color=green, thick] (-1, 1+6) -- (-1, 3+6);
\draw (-0.5,0.5+6) circle (2.7);
\end{tikzpicture}\qquad
\begin{tikzpicture}[scale=0.5]
\draw(4,-2)-- (-2, -2) -- (-2, 4)--(4,-2);
\draw[color=blue, thick](-1, 1)-- (0,2);
\draw[color=blue, thick](-1, 0)-- (0,-1);
\draw[color=blue, thick] (-2, 0) -- (-1, 0);
\draw[color=blue, thick] (0, -1)-- (0,-2);
\draw[color=blue, thick] (-2, 1) -- (-1, 1);
\draw[color=blue, thick] (-1, 1)-- (-1,0);
\draw[color=blue, thick] (0, -1)-- (1,-1);
\draw[color=blue, thick] (1, -1)-- (2,0);
\draw[color=blue, thick] (1, -1)-- (1,-2);
\end{tikzpicture}\qquad
\begin{tikzpicture}[scale=0.5]
\draw(4,-2)-- (-2, -2) -- (-2, 4)--(4,-2);
\draw[color=blue, thick](-2, -1)-- (-1,-1);
\draw[color=blue, thick](-1, -1)-- (-1,-2);
\draw[color=blue, thick](-1, -1)-- (0,0);
\draw[color=blue, thick](0, 0)-- (0,1);
\draw[color=blue, thick](0, 0)-- (1,0);
\draw[color=blue, thick](1, 0)-- (1,-2);
\draw[color=blue, thick](1, 0)-- (1.5,0.5);
\draw[color=blue, thick](0, 1)-- (-2,1);
\draw[color=blue, thick](0, 1)-- (0.5,1.5);
\end{tikzpicture}
\]
Two other possible relative positions for a pair of distinct degree one tropical curves in $\mathbb{P}^2$. Below, we present the possible resolutions for the 4-valent vertices:
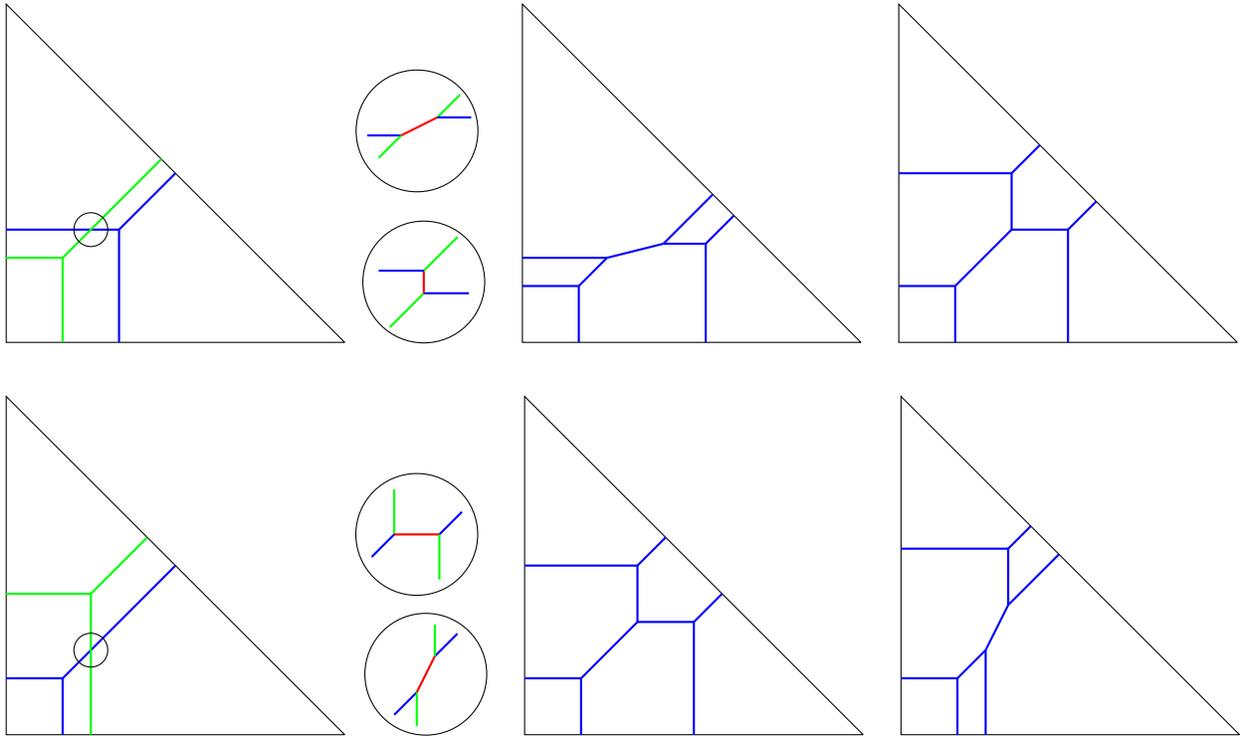
\begin{gather*}
\begin{tikzpicture}[scale=0.5]
\draw(4,-2)-- (-2, -2) -- (-2, 4)--(4,-2);
\draw[color=blue, thick] (-2, 0) -- (0, 0)-- (0,-2);
\draw[color=blue, thick](0, 0)-- (1,1);
\draw[color=green, thick] (-2, -0.5) -- (-1, -.5)--(-1,-2);
\draw[color=green, thick](-1, -.5)-- (-1+1.75,-.5+1.75);
\draw (-.5,0) circle (0.3);
\end{tikzpicture}
\begin{tikzpicture}[scale=0.24]
\draw[color=green, thick](-1.5, -1.5)-- (0,0);
\draw[color=blue, thick] (0, 0) -- (2, 0);
\draw[color=red, thick] (0, 0) -- (0, 1);
\draw[color=green, thick](0, 1)-- (1.5,2.5);
\draw[color=blue, thick] (0, 1) -- (-2, 1);
\draw (0,0.5) circle (2.7);
\draw[color=blue, thick] (-2.5, 1+6) -- (-1, 1+6);
\draw[color=green, thick] (-2, 0+6) -- (-1, 1+6);
\draw[color=red, thick] (-1, 1+6) -- (0.6, 1.8+6);
\draw[color=green,thick](0.6, 1.8+6)--(0.6+1, 1.8+1+6);
\draw[color=blue,thick](0.6, 1.8+6)--(0.6+1.5, 1.8+6);
\draw (-.3,1.2+6) circle (2.7);
\end{tikzpicture}\qquad
\begin{tikzpicture}[scale=0.5]
\draw(4,-2)-- (-2, -2) -- (-2, 4)--(4,-2);
\draw[color=blue, thick](-2, -1)-- (-1,-1);
\draw[color=blue, thick](-1, -1)-- (-1,-2);
\draw[color=blue, thick](-1, -1)-- (-0.5,-0.5);
\draw[color=blue, thick](-2, -0.5)-- (-0.5,-0.5);
\draw[color=blue, thick] (-0.5,-0.5)--(0.5,-0.25);
\draw[color=blue, thick] (0.5+0.87,-0.25+0.87)--(0.5,-0.25);
\draw[color=blue, thick] (1.25,-0.25)--(0.5,-0.25);
\draw[color=blue, thick] (1.25,-0.25)--(1.25+0.5,-0.25+0.5);
\draw[color=blue, thick] (1.25,-0.25)--(1.25,-2);
\end{tikzpicture}\qquad
\begin{tikzpicture}[scale=0.5]
\draw(4,-2)-- (-2, -2) -- (-2, 4)--(4,-2);
\draw[color=blue, thick](-2, -1)-- (-1,-1);
\draw[color=blue, thick](-1, -1)-- (-1,-2);
\draw[color=blue, thick](-1, -1)-- (0,0);
\draw[color=blue, thick](0, 0)-- (0,1);
\draw[color=blue, thick](0, 0)-- (1,0);
\draw[color=blue, thick](1, 0)-- (1,-2);
\draw[color=blue, thick](1, 0)-- (1.5,0.5);
\draw[color=blue, thick](0, 1)-- (-2,1);
\draw[color=blue, thick](0, 1)-- (0.5,1.5);
\end{tikzpicture}
\\
\begin{tikzpicture}[scale=0.5]
\draw(4,-2)-- (-2, -2) -- (-2, 4)--(4,-2);
\draw[color=blue, thick] (-2, -1) -- (-1, -1)-- (-1,-2);
\draw[color=blue, thick](-1, -1)-- (1,1);
\draw[color=green, thick] (-2, 0.5) -- (-0.5, 0.5)--(-0.5,-2);
\draw[color=green, thick](-0.5, 0.5)-- (-0.5+1, 0.5+1);
\draw (-.5,-0.5) circle (0.3);
\end{tikzpicture}
\begin{tikzpicture}[scale=0.24]
\draw[color=blue, thick](-1, -1)-- (0,0);
\draw[color=green, thick] (0, 0) -- (0, -1.5);
\draw[color=red, thick] (0, 0) -- (0.8, 1.6);
\draw[color=blue, thick] (1.8, 2.6) -- (0.8, 1.6);
\draw[color=green, thick] (0.8, 3) -- (0.8, 1.6);
\draw (0.4,0.8) circle (2.7);
\draw[color=green, thick] (-1, 5+4) -- (-1, 3+4);
\draw[color=blue, thick] (-2, 2+4) -- (-1, 3+4);
\draw[color=red, thick] (-1, 3+4) -- (1, 3+4);
\draw[color=blue, thick] (1, 3+4) -- (2, 4+4);
\draw[color=green, thick] (1, 3+4) -- (1, 1+4);
\draw (0,3+4) circle (2.7);
\end{tikzpicture}\qquad
\begin{tikzpicture}[scale=0.5]
\draw(4,-2)-- (-2, -2) -- (-2, 4)--(4,-2);
\draw[color=blue, thick](-2, -1)-- (-1,-1);
\draw[color=blue, thick](-1, -1)-- (-1,-2);
\draw[color=blue, thick](-1, -1)-- (0,0);
\draw[color=blue, thick](0, 0)-- (0,1);
\draw[color=blue, thick](0, 0)-- (1,0);
\draw[color=blue, thick](1, 0)-- (1,-2);
\draw[color=blue, thick](1, 0)-- (1.5,0.5);
\draw[color=blue, thick](0, 1)-- (-2,1);
\draw[color=blue, thick](0, 1)-- (0.5,1.5);
\end{tikzpicture}\qquad
\begin{tikzpicture}[scale=0.5]
\draw(4,-2)-- (-2, -2) -- (-2, 4)--(4,-2);
\draw[color=blue, thick] (-2, -1) -- (-1, -1);
\draw[color=blue,thick](-1, -1)--(-1,-2);
\draw[color=blue, thick](-1, -1)-- (-0.5,-0.5);
\draw[color=blue, thick](-0.5,-0.5)--(-.5,-2);
\draw[color=blue, thick](-0.5, -0.5)-- (-0.5+0.4,-0.5+0.8);
\draw[color=blue, thick](-0.5+0.4,-0.5+0.8)--(-0.1,-0.5+0.8+1);
\draw[color=blue,thick](-0.5+0.4,-0.5+0.8)--(-0.5+0.4+0.9,-0.5+0.8+0.9);
\draw[color=blue, thick](-2,-0.5+0.8+1)--(-0.1,-0.5+0.8+1);
\draw[color=blue, thick](-0.1,-0.5+0.8+1)--(-0.1+0.4,-0.5+0.8+1+0.4);
\end{tikzpicture}
\end{gather*}
In the case of three distinct degree-one tropical curves in $\mathbb{P}^2$, the resolution of singularities can give us a tropical curve of genus one or zero.

The four 4-valent singularities on the left picture below can be resolved to get a genus one tropical curve. In the right picture, we present a singular degree-three curve as an intersection of a smooth degree-two tropical curve and a degree-one tropical curve. As shown in the right part of the picture, we can resolve the singularity to get a smooth tropical curve of degree three and genus zero:
\[
\begin{tikzpicture}[scale=0.5]
\draw(4,-2)-- (-2, -2) -- (-2, 4)--(4,-2);
\draw[color=blue, thick] (-2, 0) -- (0, 0)-- (0,-2);
\draw[color=blue, thick](0, 0)-- (1,1);
\draw[color=green, thick] (-2, 1) -- (-1, 1)-- (-1,-2);
\draw[color=green, thick](-1, 1)-- (0,2);
\draw[color=red, thick] (-2, -1) -- (2, -1)-- (2,-2);
\draw[color=red, thick](2, -1)-- (2+0.5,-1+0.5);
\end{tikzpicture}\qquad
\begin{tikzpicture}[scale=0.5]
\draw(4,-2)-- (-2, -2) -- (-2, 4)--(4,-2);
\draw[color=blue, thick](-1.2, -1) -- (-1,-1);
\draw[color=blue, thick](-1, -1)-- (-1,-2);
\draw[color=blue, thick](-1, -1)-- (-0.5,-0.5);
\draw[color=blue, thick](-1.2, -0.5)-- (-0.5,-0.5);
\draw[color=blue, thick] (-0.5,-0.5)--(0.5,-0.25);
\draw[color=blue, thick] (0.5+0.87,-0.25+0.87)--(0.5,-0.25);
\draw[color=blue, thick] (1.25,-0.25)--(0.5,-0.25);
\draw[color=blue, thick] (1.25,-0.25)--(1.25+0.5,-0.25+0.5);
\draw[color=blue, thick] (1.25,-0.25)--(1.25,-2);
\draw[color=blue, thick] (-1.2, -1) -- (-1.2, -.5);
\draw[color=blue, thick] (-1.2, -0.5) -- (-1.2-.3, -.5+0.3);
\draw[color=blue, thick] (-1.5, -.2)--(-2, -0.2) ;
\draw[color=blue, thick] (-1.5, -.2)--(-1.5, 1) ;
\draw[color=blue, thick] (-2, 1)--(-1.5, 1) ;
\draw[color=blue, thick](-1.5, 1)--(-1.5+1.27, 1+1.27) ;
\draw[color=blue, thick] (-1.2, -1) -- (-1.5,-1.3);
\draw[color=blue, thick] (-2, -1.3) -- (-1.5,-1.3);
\draw[color=blue, thick] (-1.5, -2) -- (-1.5,-1.3);
\end{tikzpicture}\qquad
\begin{tikzpicture}[scale=0.5]
\draw(4,-2)-- (-2, -2) -- (-2, 4)--(4,-2);
\draw[color=blue, thick](-2, -1)-- (-1,-1);
\draw[color=blue, thick](-1, -1)-- (-1,-2);
\draw[color=blue, thick](-1, -1)-- (-0.5,-0.5);
\draw[color=blue, thick](-2, -0.5)-- (-0.5,-0.5);
\draw[color=blue, thick] (-0.5,-0.5)--(0.5,-0.25);
\draw[color=blue, thick] (0.5+0.87,-0.25+0.87)--(0.5,-0.25);
\draw[color=blue, thick] (1.25,-0.25)--(0.5,-0.25);
\draw[color=blue, thick] (1.25,-0.25)--(1.25+0.5,-0.25+0.5);
\draw[color=blue, thick] (1.25,-0.25)--(1.25,-2);
\draw[color=green, thick] (-2, 1) -- (0, 1)-- (0,-2);
\draw[color=green, thick] (0, 1)-- (0.5,1.5);
\end{tikzpicture}\qquad
\begin{tikzpicture}[scale=0.5]
\draw(4,-2)-- (-2, -2) -- (-2, 4)--(4,-2);
\draw[color=blue, thick](-2, -1)-- (-1,-1);
\draw[color=blue, thick](-1, -1)-- (-1,-2);
\draw[color=blue, thick](-1, -1)-- (-0.5,-0.5);
\draw[color=blue, thick](-2, -0.5)-- (-0.5,-0.5);
\draw[color=blue, thick] (-0.5,-0.5)--(0,-0.37);
\draw[color=blue, thick] (-2, 1) -- (0, 1)-- (0.5,1.5) ;
\draw[color=blue, thick] (0, 1)--(0,-0.37)--(0.5,-0.37) ;
\draw[color=blue, thick] (0.5,-0.37)--(0.5,-2) ;
\draw[color=blue, thick] (0.5,-0.37)--(0.5+0.5,-0.37+0.25);
\draw[color=blue, thick] (0.5+0.5,-0.37+0.25)-- (0.5+0.5+0.55,-0.37+0.25+0.55);
\draw[color=blue, thick] (0.5+0.5,-0.37+0.25)-- (0.5+0.5+0.5,-0.37+0.25);
\draw[color=blue,thick](0.5+0.5+0.5,-0.37+0.25)--(0.5+0.5+0.5,-2);
\draw[color=blue,thick](0.5+0.5+0.5,-0.37+0.25)--(0.5+0.5+0.5+0.3,-0.37+0.25+0.3);
\end{tikzpicture}
\]

\subsection{Tropical curves via graphs}\label{sec_trop_curve_graph}

This section will briefly review the definition of the tropical curves in toric varieties. For more details, see Mikhalkin \cite{mikhalkin2009tropical}.

A graph $\Gamma$ is a collection of vertices $v \in V(\Gamma)$ and edges $e \in E(\Gamma)$. A directed graph is a graph with directed edges, i.e., each edge is incoming to one vertex and outgoing to another. For each vertex $v$, we can introduce a set $E_{{\rm in}}(v)$ of incoming edges and a set $E_{{\rm out}}(v)$ of outgoing edges connected to this vertex. Each vertex $v$ has a {\it valence}, the number of edges connecting to this vertex. We will use $n_k(\Gamma)$ for the number of $k$-valent vertices in a graph~$\Gamma$. It is convenient to decompose all edges into internal and external. An {\it external edge} $e \in E(\Gamma)$ is an edge connected to a 1-valent vertex. Similarly, an internal edge connects two vertices with valencies bigger than~1. We will denote $I(\Gamma)$, the set of internal edges of the graph $\Gamma$. The Euler characteristic (genus) $g(\G) = |E(\Gamma)|-|V(\Gamma)|+1$ of a connected graph equals the number of loops of the graph~$\G$. A~tree is a graph without loops.

The topological data for a connected tropical curve of genus zero in general position is a~tree~$\Gamma$ with only 3- and 1-valent vertices. The homological class of the curve $\beta$ determines the number of 1-valent vertices.

The discrete data of a connected tropical curve in general position in toric variety $X$ is an equivalence class of {\it consistent} directed trees $\vec{\G}$, consistently decorated with an integer vector $\vec{m}_e \in \mathbb{Z}^N$ for each edge $e\in E(\G)$. The consistency conditions are the following:
\begin{itemize}\itemsep=0pt
\item Each external edge $e$ is outgoing for 1-valent vertex and is decorated with a vector equal to one of the primitive normals for the compactification polytope, i.e., $\vec{m}_e \in B_{X}$.
\item An integer vector on each internal edge $e$ is non-vanishing $\vec{m}_e \neq \vec{0}$.
\item For each 3-valent vertex $v$ the integer vectors on edges connected to it obey the {\it vertex balance condition}:
\begin{gather}\label{vert_bal_con}
\sum _{e \in E_{{\rm in}}(v) } \vec{m}_e = \sum _{e \in E_{{\rm out}}(v) } \vec{m}_e.
\end{gather}
\end{itemize}
Two directed trees are equivalent because they differ by a simultaneous flip of orientation and a change $\vec{m}_e \to - \vec{m}_e$ for some internal edges.
\begin{Remark} Our definition for the integer vector $\vec{m}_e$ slightly differs from the tropical geometry literature. For example, instead of a single vector $\vec{m}$ for each edge author \cite{gross2011tropical} uses a pair $(w,\vec{u})$, a primitive vector $\vec{u}$ in the direction of $\vec{m}$ and integer weight $w$, i.e., $\vec{m}=w\cdot \vec{u}$. However, the balancing condition in these notations exactly matches our balancing condition (\ref{vert_bal_con}). Moreover, our additional requirement for the integer vectors on external edges allows us to use the balancing condition to uniquely restore the primitive vectors and weights for all internal edges.\looseness=1
\end{Remark}
\begin{Remark}
An equivalence relation for decorated directed graphs is well known in the scattering theory in physics. The direction corresponds to particles being either incoming or outgoing, while an integer vector corresponds to the particle's momentum. In scattering theory, the incoming particle with incoming momentum $\vec{p}$ is equivalent to the outgoing particle with momentum~$-\vec{p}$.
\end{Remark}
The geometric data of the tropical curve with discrete data $\vec{\Gamma}$ is the equivalence class of directed trees decorated with integer vectors and equipped with the following data:
\begin{itemize}\itemsep=0pt
\item For each internal edge $e\in E(\G)$ we assign a positive number $\tau_e\in\overline{\mathbb{R}^{+}\setminus\{0\}}$, compactified by infinity, referred to as the {\it length} of the edge $e$ and an angle $\varphi_e \in S^1$, referred as the {\it twist} of the edge $e$.
\item We assign a distinguished vertex, $v_c$, referred to as the {\it sink}. The sink is assigned with a~{\it sink location}, a point in toric variety $X$ with the radial location $\vec{r}_c \in \mathbb{R}^N$ and the angular location $\vec{\phi}_c \in \mathbb{T}^N$.
\item {\it Equivalence relation}: We change the sink vertex from $v_c$ to $v_c'$ with simultaneous change in sink location
\begin{gather}\label{root_relocation}
\vec{r}^{\,\prime}_c=\vec{r}_c+\sum_{e \in \gamma (v_c, v_c')} \pm \vec{m}_e \tau_e, \qquad \vec{\phi}^{\,\prime}_c=\vec{\phi}_c+\sum_{e \in \gamma (v_c, v_c')} \pm \vec{m}_e \varphi_e,
\end{gather}
where $\gamma^\G$ is the (unique) path from $v_c$ to $v_c'$ on the tree. The term in the sum over edges is weighted $+1$ when the direction of the edge $e$ is aligned with the path $\gamma^\G (v_c, v_c')$ and $-1$ otherwise.
\end{itemize}

\subsection{Moduli space of tropical curves}

In this section, we will briefly review the moduli space tropical curves. For more details, see Mikhalkin \cite{mikhalkin2009tropical}. For our discussion of moduli space, we assume that the tropical curve is connected, has genus 0, and non-trivial homological class $\beta$, so the corresponding graph is a tree.

The moduli space $\mathcal{M}_{0}(X, \beta)$ of the tropical curves of homological class $\beta \in H_{2}(X)$, also denoted as a degree of curve and genus zero in toric space $X$ is a compactification of the union of components, labeled by the discrete types of tropical curves
\[
\mathcal{M}_{0}(X,\beta)=\overline{ \bigcup_{\vec{\Gamma}} \mathcal{M}_{0}(X,\beta; \vec{\G})}.
\]
The tropical curves with zero-length edges compactify the moduli space. There are two types of such tropical curves: the ones with the 4-and higher-valent vertices and the ones with the non-primitive vectors on the external edges.

Each component of the moduli space $\mathcal{M}_{0}\bigl(X,\beta; \vec{\G}\bigr)$ is a (non-compact) toric variety with the radial part
\[
\mathcal{M}^R_{0}\bigl(X,\beta; \vec{\Gamma}\bigr)= \mathbb{R}^{\dim X} \times \bigl(\mathbb{R}^+\bigr)^{I(\Gamma)}.
\]
The \smash{$\bigl(\mathbb{R}^+\bigr)^{I(\Gamma)}$}-factor describes the possible values of internal edge lengths, while the $\mathbb{R}^{\dim X}$ describes the radial location of the sink. The angular part of the moduli space is
\[
\mathcal{M}^\phi_{0}\bigl(X,\beta; \vec{\G}\bigr) = \bigl(S^1\bigr)^{\dim X+ I(\Gamma)}.
\]
The \smash{$\bigl(S^1\bigr)^{I(\Gamma)}$}-factor describes the possible values of internal edge twists, while the \smash{$\bigl(S^1\bigr)^{\dim X}$} describes the angular location of the sink.
\begin{Example} The moduli space of degree one tropical curves in $\mathbb{P}^2$ has a single component, while the moduli space of bi-degree (1,1) tropical curves in $\mathbb{P}^1\times \mathbb{P}^1$ has two components corresponding to the tropical curves below. The two trees have different integer vectors assigned to the integral edge, $(1,1)$ and $(1,-1)$ correspondingly:
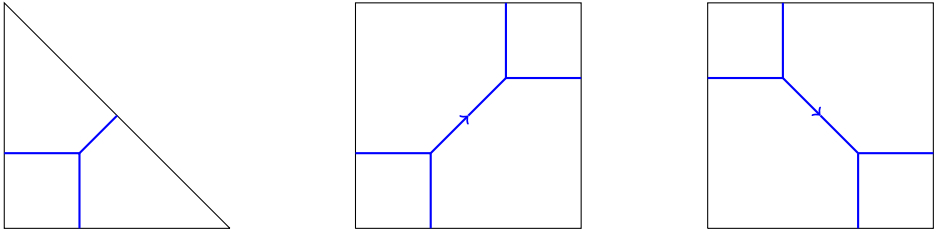
\[
\begin{tikzpicture}[scale=0.5]
\draw(4,-2)-- (-2, -2) -- (-2, 4)--(4,-2);
\draw[color=blue, thick] (-2, 0) -- (0, 0)-- (0,-2);
\draw[color=blue, thick](0, 0)-- (1,1);
\end{tikzpicture}\qquad
\begin{tikzpicture}[scale=0.5]
\draw(-2,-2)-- (4, -2) -- (4, 4)--(-2,4)--(-2,-2);
\draw[color=blue, thick] (-2, 0) -- (0, 0)-- (0,-2);
\draw[color=blue, thick, ->](0, 0)-- (1,1);
\draw[color=blue, thick](1, 1)-- (2,2);
\draw[color=blue, thick](2, 2)-- (2,4);
\draw[color=blue, thick](2, 2)-- (4,2);
\end{tikzpicture}\qquad
\begin{tikzpicture}[scale=0.5]
\draw(-2,-2)-- (4, -2) -- (4, 4)--(-2,4)--(-2,-2);
\draw[color=blue, thick] (-2, 2) -- (0, 2)-- (0,4);
\draw[color=blue, thick, ->](0, 2)-- (1,1);
\draw[color=blue, thick](1, 1)-- (2,0);
\draw[color=blue, thick](2, 0)-- (2,-2);
\draw[color=blue, thick](2, 0)-- (4,0);
\end{tikzpicture}
\]
\end{Example}
\begin{Example}
The moduli space $\mathcal{M}_{0}(\mathbb{P}^2,\beta)$ of degree 2 tropical curves in $\mathbb{P}^2$ is a union of 7 components, corresponding to the trees below. Below the pictures, we provided the list of integer vectors for each internal edge:
\begin{gather*}
\begin{tikzpicture}[scale=0.5]
\draw(4,-2)-- (-2, -2) -- (-2, 4)--(4,-2);
\draw[color=blue, thick](-1, 1)-- (0,2);
\draw[color=blue, thick](-1, 0)-- (0,-1);
\draw[color=blue, thick] (-2, 0) -- (-1, 0);
\draw[color=blue, thick] (0, -1)-- (0,-2);
\draw[color=blue, thick] (-2, 1) -- (-1, 1);
\draw[color=blue, thick] (-1, 1)-- (-1,0);
\draw[color=blue, thick] (0, -1)-- (1,-1);
\draw[color=blue, thick] (1, -1)-- (2,0);
\draw[color=blue, thick] (1, -1)-- (1,-2);
\node at (1,-3) {$(1,0), (1,-1), (0,1)$};
\end{tikzpicture}\quad
\begin{tikzpicture}[scale=0.5]
\draw(4,-2)-- (-2, -2) -- (-2, 4)--(4,-2);
\draw[color=blue, thick](-2, -1)-- (-1,-1);
\draw[color=blue, thick](-1, -1)-- (-1,-2);
\draw[color=blue, thick](-1, -1)-- (0,0);
\draw[color=blue, thick](0, 0)-- (0,1);
\draw[color=blue, thick](0, 0)-- (1,0);
\draw[color=blue, thick](1, 0)-- (1,-2);
\draw[color=blue, thick](1, 0)-- (1.5,0.5);
\draw[color=blue, thick](0, 1)-- (-2,1);
\draw[color=blue, thick](0, 1)-- (0.5,1.5);
\node at (1,-3) {$(1,1), (1,0), (0,1)$};
\end{tikzpicture}\quad
\begin{tikzpicture}[scale=0.5]
\draw(4,-2)-- (-2, -2) -- (-2, 4)--(4,-2);
\draw[color=blue, thick](-2, -1)-- (-1,-1);
\draw[color=blue, thick](-1, -1)-- (-1,-2);
\draw[color=blue, thick](-1, -1)-- (-0.5,-0.5);
\draw[color=blue, thick](-2, -0.5)-- (-0.5,-0.5);
\draw[color=blue, thick] (-0.5,-0.5)--(0.5,-0.25);
\draw[color=blue, thick] (0.5+0.87,-0.25+0.87)--(0.5,-0.25);
\draw[color=blue, thick] (1.25,-0.25)--(0.5,-0.25);
\draw[color=blue, thick] (1.25,-0.25)--(1.25+0.5,-0.25+0.5);
\draw[color=blue, thick] (1.25,-0.25)--(1.25,-2);
\node at (1,-3) {$(1,1), (1,2), (1,0)$};
\end{tikzpicture}\quad
\begin{tikzpicture}[scale=0.5]
\draw(4,-2)-- (-2, -2) -- (-2, 4)--(4,-2);
\draw[color=blue, thick] (-2, -1) -- (-1, -1);
\draw[color=blue, thick](-1, -1)-- (-1,-2);
\draw[color=blue, thick](-1, -1)-- (-0.5,-0.5);
\draw[color=blue, thick](-0.5,-0.5)--(-.5,-2);
\draw[color=blue, thick](-0.5, -0.5)-- (-0.5+0.4,-0.5+0.8);
\draw[color=blue, thick](-0.5+0.4,-0.5+0.8)--(-0.1,-0.5+0.8+1);
\draw[color=blue,thick](-0.5+0.4,-0.5+0.8)--(-0.5+0.4+0.9,-0.5+0.8+0.9);
\draw[color=blue, thick](-2,-0.5+0.8+1)--(-0.1,-0.5+0.8+1);
\draw[color=blue, thick](-0.1,-0.5+0.8+1)--(-0.1+0.4,-0.5+0.8+1+0.4);
\node at (1,-3) {$(1,1), (1,2), (0,1)$};
\end{tikzpicture}
\\
\begin{tikzpicture}[scale=0.5]
\draw(4,-2)-- (-2, -2) -- (-2, 4)--(4,-2);
\draw[color=blue, thick] (-2, -1) -- (-1, -1);
\draw[color=blue,thick](-1, -1)--(-1,-2);
\draw[color=blue, thick](-1, -1)-- (-0.5,-0.5);
\draw[color=blue, thick](-0.5,-0.5)--(-.5,-2);
\draw[color=blue, thick](-0.5, -0.5)-- (-0.5+0.4,-0.5+0.8);
\draw[color=blue, thick](-0.5+0.4,-0.5+0.8)--(-2,-0.5+0.8);
\draw[color=blue, thick] (-0.5+0.4,-0.5+0.8)-- (-0.5+0.4+0.4,-0.5+0.8+0.4);
\draw[color=blue,thick](-0.5+0.4+0.4-0.1,-0.5+0.8+0.4)--(-0.5+0.4+0.9,-0.5+0.8+0.9+0.05);
\draw[color=blue,thick](-0.5+0.4+0.4,-0.5+0.8+0.4-0.1)--(-0.5+0.4+0.9+0.1,-0.5+0.8+0.85);
\node at (1,-3) {$(1,1), (1,2), (2,2)$};
\end{tikzpicture}\quad
\begin{tikzpicture}[scale=0.5]
\draw(4,-2)-- (-2, -2) -- (-2, 4)--(4,-2);
\draw[color=blue, thick](-1, 1)-- (0,2);
\draw[color=blue, thick](-1, 0)-- (0,-1);
\draw[color=blue, thick] (-1, 1)-- (-1,0);
\draw[color=blue, thick] (-2, 1) -- (-1, 1);
\draw[color=blue, thick] (-2, 0) -- (-1, 0);
\draw[color=blue, thick] (0, -1)-- (0+1.5,-1+1.5);
\draw[color=blue, thick] (0, -1)-- (0,-1.4);
\draw[color=blue, thick] (0.05, -2)-- (0.05,-1.4);
\draw[color=blue, thick] (-0.05, -2)-- (-0.05,-1.4);
\node at (1,-3) {$(1,0), (1,-1), (0,2)$};
\end{tikzpicture}\quad
\begin{tikzpicture}[scale=0.5]
\draw(4,-2)-- (-2, -2) -- (-2, 4)--(4,-2);
\draw[color=blue, thick] (0, -1)-- (0,-2);
\draw[color=blue, thick] (1, -1)-- (1,-2);
\draw[color=blue, thick](-1, 0)-- (0,-1);
\draw[color=blue, thick] (0, -1)-- (1,-1);
\draw[color=blue, thick] (1, -1)-- (2,0);
\draw[color=blue, thick] (-1.4, 0) -- (-1, 0);
\draw[color=blue, thick] (-1.4, 0.05) -- (-2, 0.05);
\draw[color=blue, thick] (-1.4, -0.05) -- (-2, -0.05);
\draw[color=blue, thick](-1, 0)-- (-1+1.5,0+1.5);
\node at (1,-3) {$(2,0), (1,-1), (0,1)$};
\end{tikzpicture}
\end{gather*}
Below, we show the two smooth tropical curves of degree two and a limiting curve with zero (dotted) internal edge length. The limiting curve is a smooth tropical curve with a non-primitive integer vector on the external edge. By construction, such a tropical curve provides a compactification of the moduli space $\mathcal{M}_{0}\bigl(\mathbb{P}^2,\beta\bigr)$:
\[
\begin{tikzpicture}[scale=0.5]
\draw(4,-2)-- (-2, -2) -- (-2, 4)--(4,-2);
\draw[color=blue, thick] (-2, -1) -- (-1, -1);
\draw[color=blue, thick](-1, -1)--(-1,-2);
\draw[color=blue, thick](-1, -1)-- (-0.5,-0.5);
\draw[color=blue, thick](-0.5,-0.5)--(-.5,-2);
\draw[color=blue, thick](-0.5, -0.5)-- (-0.5+0.4,-0.5+0.8);
\draw[color=blue, dotted, thick](-0.5+0.4,-0.5+0.8)--(-0.1,-0.5+0.8+1);
\draw[color=blue,thick] (-0.5+0.4,-0.5+0.8)-- (-0.5+0.4+0.9,-0.5+0.8+0.9);
\draw[color=blue, thick](-2,-0.5+0.8+1)--(-0.1,-0.5+0.8+1);
\draw[color=blue, thick](-0.1,-0.5+0.8+1)--(-0.1+0.4,-0.5+0.8+1+0.4);
\end{tikzpicture}\qquad
\begin{tikzpicture}[scale=0.5]
\draw(4,-2)-- (-2, -2) -- (-2, 4)--(4,-2);
\draw[color=blue, thick] (-2, -1) -- (-1, -1);
\draw[color=blue, thick](-1, -1)-- (-1,-2);
\draw[color=blue, thick](-1, -1)-- (-0.5,-0.5);
\draw[color=blue, thick](-0.5,-0.5)--(-.5,-2);
\draw[color=blue, thick](-0.5, -0.5)-- (-0.5+0.4,-0.5+0.8);
\draw[color=blue, thick](-0.5+0.4,-0.5+0.8)--(-2,-0.5+0.8);
\draw[color=blue, ultra thick] (-0.5+0.4,-0.5+0.8)-- (-0.5+0.4+0.9,-0.5+0.8+0.9);
\end{tikzpicture}\qquad
\begin{tikzpicture}[scale=0.5]
\draw(4,-2)-- (-2, -2) -- (-2, 4)--(4,-2);
\draw[color=blue, thick] (-2, -1) -- (-1, -1);
\draw[color=blue, thick](-1, -1)-- (-1,-2);
\draw[color=blue, thick](-1, -1)-- (-0.5,-0.5);
\draw[color=blue, thick](-0.5,-0.5)--(-.5,-2);
\draw[color=blue, thick](-0.5, -0.5)-- (-0.5+0.4,-0.5+0.8);
\draw[color=blue, thick](-0.5+0.4,-0.5+0.8)--(-2,-0.5+0.8);
\draw[color=blue, dotted, thick] (-0.5+0.4,-0.5+0.8)-- (-0.5+0.4+0.4,-0.5+0.8+0.4);
\draw[color=blue,thick] (-0.5+0.4+0.4-0.1,-0.5+0.8+0.4)-- (-0.5+0.4+0.9,-0.5+0.8+0.9+0.05);
\draw[color=blue,thick] (-0.5+0.4+0.4,-0.5+0.8+0.4-0.1)-- (-0.5+0.4+0.9+0.1,-0.5+0.8+0.85);
\end{tikzpicture}
\]
\end{Example}

\subsection{Moduli space with marked points} \label{trop_moduli_mark_section}

This section will briefly review the moduli space of tropical curves with marked points in toric varieties. For more details, see Mikhalkin \cite{mikhalkin2009tropical}.

We construct the moduli space $\mathcal{M}_{0,n}(X,\beta)$ for tropical curves of degree $\beta$ and genus zero with $n$ marked points recursively using the following fibration:
\[
\xymatrix@1{
\vec{\G}\ar[r]&\mathcal{M}_{0,n}\bigl(X,\beta; \vec{\G}\bigr) \ar[d] \\&\mathcal{M}_{0,n-1}\bigl(X,\beta; \vec{\G}\bigr).
}
\]
We already constructed the moduli space without marked points $\mathcal{M}_{0,0}\bigl(X,\beta; \vec{\G}\bigr)=\mathcal{M}_{0}\bigl(X,\beta; \vec{\G}\bigr)$. The moduli space with a single marked point is the union
\[
\mathcal{M}_{0,1}\bigl(X,\beta;\vec{\Gamma}\bigr)=\bigcup_{e\in E(\G)} \mathcal{M}_{0}\bigl(X,\beta; D_e^1\vec{\G}\bigr).
\]
The union is taken over decorations $D_e^1\vec{\Gamma}$ of a tree $\Gamma$ by a marked point at edge $e\in E(\G)$.
\begin{Example}
A degree one tropical curve in $\mathbb{P}^2$ can be decorated in three ways, so the corresponding moduli space is a union over the three decorations $D_e^1 \vec{\Gamma}$, depicted below:
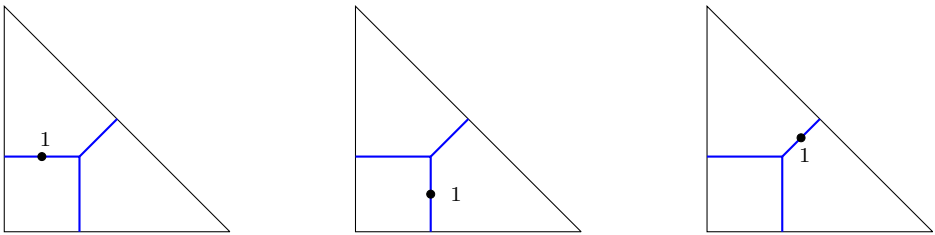
\[
\begin{tikzpicture}[scale=0.5]
\draw(4,-2)-- (-2, -2) -- (-2, 4)--(4,-2);
\draw[color=blue, thick] (-2, 0) -- (0, 0)-- (0,-2);
\draw[color=blue, thick](0, 0)-- (1,1);
\filldraw[black] (-1,0) circle (3pt) node[anchor=south]{$\;_1$};
\end{tikzpicture}\qquad
\begin{tikzpicture}[scale=0.5]
\draw(4,-2)-- (-2, -2) -- (-2, 4)--(4,-2);
\draw[color=blue, thick] (-2, 0) -- (0, 0)-- (0,-2);
\draw[color=blue, thick](0, 0)-- (1,1);
\filldraw[black] (0,-1) circle (3pt)node[anchor=west]{$\;_1$};
\end{tikzpicture}\qquad
\begin{tikzpicture}[scale=0.5]
\draw(4,-2)-- (-2, -2) -- (-2, 4)--(4,-2);
\draw[color=blue, thick] (-2, 0) -- (0, 0)-- (0,-2);
\draw[color=blue, thick](0, 0)-- (1,1);
\filldraw[black] (0.5,0.5) circle (3pt)node[anchor=north]{$\;_1$};
\end{tikzpicture}
\]
\end{Example}
The moduli space of a decorated tropical curve $D_e^1\vec{\G}$ has two additional moduli: the radial $t \in (0, \tau_e)$ and angular $\varphi \in (0, \phi_e)$, locations of the marked point on edge $e$ with the radial $\tau_e$ and the angular moduli $\phi_e$. Hence, the radial part of the moduli space is
\[
\mathcal{M}^R_{0}\bigl(X,\beta; D_e^1\vec{\G}\bigr)=\mathcal{M}^R_{0}\bigl(X,\beta; \vec{\G}\bigr)\times(0, \tau_e)=\mathbb{R}^{\dim X} \times \bigl(\mathbb{R}^+\bigr)^{I(\Gamma)} \times (0, \tau_e).
\]
The moduli space is compactified by $t = 0$ and $t=\tau_e$ for each internal edge $e$ and just $t=0$ for external edges.

It is more convenient to give an alternative description of the moduli space $ \mathcal{M}_{0,1}\bigl(X,\beta; \vec{\G}\bigr)$ using the following proposition.
\begin{Proposition}\label{prop_trop_curve_marked}
A decoration $D^1_e\vec{\G}$ of the tropical curve $\vec{\G}$ is a tropical curve $\vec{\Gamma}_e$, constructed from $\vec{\G}$ by adding an additional $2$-valent vertex on the edge $e \in \G$.
\end{Proposition}
\begin{proof}
For a directed tree $\vec{\G}$, the new tree $\vec{\Gamma}_e$ is also a directed tree if we choose directions, as shown in the picture below. The new directed tree $\vec{\G}_e$ obeys the balancing condition (\ref{vert_bal_con}) at every vertex, including the 2-valent one, if we assign the same integer vector on both incoming and outgoing edges to the 2-valent vertex:
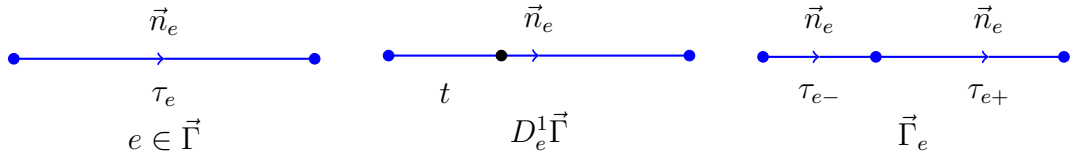
\[
\begin{tikzpicture}
\draw[color=blue, thick, ->] (0, 0) -- (2, 0);
\draw[color=blue, thick](2, 0)-- (4,0);
\filldraw[blue] (0,0) circle (2 pt);
\filldraw[blue] (4,0) circle (2 pt);
\node at (2,+0.5) {$ \vec{n}_e$};
\node at (2,-0.5) {$ \tau_e$};
\node at (2,-1) {$e \in \vec{\G}$};
\end{tikzpicture}\qquad
\begin{tikzpicture}
\draw[color=blue, thick, ->] (0, 0) -- (2, 0);
\draw[color=blue, thick](2, 0)-- (4,0);
\filldraw[black] (1.5,0) circle (2 pt);
\filldraw[blue] (0,0) circle (2 pt);
\filldraw[blue] (4,0) circle (2 pt);
\node at (2,+0.5) {$ \vec{n}_e$};
\node at (0.75,-0.5) {$ t$};
\node at (2,-1) {$D_e^1\vec{\G}$};
\end{tikzpicture}\qquad
\begin{tikzpicture}
\draw[color=blue, thick, ->] (0, 0) -- (0.75, 0);
\draw[color=blue, thick] (0.75, 0) -- (1.5, 0);
\draw[color=blue, thick, ->](1.5, 0)-- (3,0);
\draw[color=blue, thick ](4, 0)-- (3,0);
\filldraw[blue] (1.5,0) circle (2 pt);
\filldraw[blue] (0,0) circle (2 pt);
\filldraw[blue] (4,0) circle (2 pt);
\node at (0.75,+0.5) {$ \vec{n}_e$};
\node at (3,+0.5) {$ \vec{n}_e$};
\node at (0.75,-0.5) {$\tau_{e-} $};
\node at (3,-0.5) {$\tau_{e+} $};
\node at (2,-1) {$\vec{\G}_e$};
\end{tikzpicture}
\]
We combine the radial location of the marked point $t \in (0, \tau_e)$ and the length $\tau_e$ of the edge~$e$ into lengths $\tau_{e\pm}$ of the two edges connected to the 2-valent vertex. In particular,
$\tau_{e-}=t$, $\tau_{e+}=\tau_e -t$.
The same procedure can be applied to the angular moduli. We have successfully decorated the directed tree $\G_e$ with the integer vectors and (radial and angular) moduli. Hence, $\vec{\G}_e$ is a tropical curve.
\end{proof}

Using Proposition \ref{prop_trop_curve_marked}, we identify the moduli space for a decorated tropical curve $D^1_e \vec{\G}$ with the moduli space of a tropical curve $\vec{\G}_e$ with an additional 2-valent vertex
\begin{gather}\label{mod_space_decorated_trop_curve}
\mathcal{M}^R_{0}\bigl(X,\beta; D_e^1\vec{\G}\bigr)=\mathcal{M}^R_{0}\bigl(X,\beta; \vec{\G}_e\bigr)=\mathbb{R}^{\dim X} \times \bigl(\mathbb{R}^+\bigr)^{I(\Gamma_e)}.
\end{gather}
The relation (\ref{mod_space_decorated_trop_curve}) implies that the moduli spaces for the tropical curves $\vec{\G}_e$ are the same for all choices of edges $e$. Hence, it is convenient to introduce the notation $\vec{\G}_1$ for a tropical curve with a single 2-valent vertex. In this notation, the moduli space
\[
\mathcal{M}_{0,1}(X,\beta)=\bigcup_ {\vec{\G}} \mathcal{M}_{0,1}\bigl(X,\beta; \vec{\Gamma}\bigr)= \bigcup_ {\vec{\G}_1}\mathcal{M}_{0}\bigl(X,\beta; \vec{\Gamma}_1\bigr).
\]
We can iterate the argument for a marked point decoration to describe the moduli space of a~tropical curve with an arbitrary number of marked points in the form
\begin{gather}\label{trop_moduli_marked}
\mathcal{M}_{0,n}(X,\beta)= \bigcup_{\vec{\Gamma}_n} \mathcal{M}_{0}\bigl(X,\beta; \vec{\G}_n\bigr).
\end{gather}
The individual components of the moduli space (\ref{trop_moduli_marked}) are
\begin{gather}\label{trop_moduli_marked_components}
\mathcal{M}_{0}\bigl(X,\beta; \vec{\G}_n\bigr) = \bigl(\mathbb{R}^{+} \times S^1\bigr)^{I(\G_n)} \times \mathbb{R}^{\dim X} \times \mathbb{T}^{\dim X}.
\end{gather}
The dimension of moduli space (\ref{trop_moduli_marked}) is
\[
\dim_{\mathbb{C}} \mathcal{M}_{0,n}(X,\beta)=\dim_{\mathbb{C}} \mathcal{M}_{0}\bigl(X,\beta; \vec{\Gamma}_n\bigr)=\dim X + I(\Gamma_n).
\]
For a graph $\G$, the number of internal edges $I(\G)$ is smaller than the number of total edges $E(\G)$ by the number of leaves $n_1(\G)$, i.e., $I(\G)=E(\G) - n_1(\G)$. Each edge of a graph connects two vertices. Hence, there is a relation
\[
3n_3(\Gamma)+2n_2(\Gamma)+n_1(\Gamma)=2E(\Gamma).
\]
All graphs are connected and have zero genus, so
\[
0=g(\G)=E(\Gamma)-V(\Gamma) +1 =E(\G)-n_2(\Gamma)-n_3(\Gamma) -n_1(\Gamma) +1.
\]
Hence, we express the dimension of the moduli space via
\begin{gather}\label{trop_mod_dim}
\dim_{\mathbb{C}} \mathcal{M}_{0,n}(X,\beta)=\dim X+E(\G_n) - n_1(\G_n)=\dim X + n_2(\G_n) +n_1(\G_n)-3.
\end{gather}
The number of 2-valent vertices $n_2(\G_n) $ is the number of marked points $n$, while the number of leaves is the tropical intersection number between the tropical curve of degree $\beta$ and the compactifying hyperplanes
\begin{gather}\label{number_leaves_degree_relation}
n_1(\G_n) = \sum_{b\in B_X} \beta \cdot H_{b}.
\end{gather}
The universal formula for the (virtual) dimension of the moduli space of complex curves is
\[
\operatorname{vdim}_{\mathbb{C}} \mathcal{M}_{g,n} (X, \beta) = \int_{\beta} c_1 (T_X)+(\dim X-3)(1-g) +n.
\]
In the case of genus zero, degree $\beta$ curve in toric space $X$, the universal formula simplifies into
\begin{gather}\label{eq_complex_moduli_space_dim}
\dim_{\mathbb{C}} \mathcal{M}_{0,n} (X, \beta ) = \sum_{b\in B_X} \beta \cdot H_{b}+\dim X -3 +n.
\end{gather}
The moduli space dimension (\ref{eq_complex_moduli_space_dim}) for complex curves matches with the tropical moduli space dimension (\ref{trop_mod_dim}) and (\ref{number_leaves_degree_relation}).

\section{Tropical Gromov--Witten theory}

\subsection{Enumerative problems}\label{GW_theory_complex}
This section will briefly review the two types of invariants: the lowest component and the top component Gromov--Witten invariants.
\begin{Definition} The {\it lowest component Gromov--Witten invariant}, denoted as $n^X_{\beta} (C_1,\dots, C_n)$ is the number of holomorphic maps $\phi\colon \mathbb{CP}^1 \to X$ of degree $\beta$, such that the image of the set of (fixed) points $z_1,\dots,z_n$ belongs to cycles $C_1,\dots, C_n$ on $X$, i.e., $\phi (z_1) \in C_1,\dots,\phi(z_n) \in C_n$.
\end{Definition}
One can show that the invariant $n^X_{\beta} (C_1,\dots, C_n)$ does not depend on the choice of $n$ points $z_1,\dots,z_n$ as long as they are distinct. The key feature of the GW invariants is the existence of the integral representation.
\begin{Proposition}[{Konstevich--Manin~\cite{kontsevich1994gromov}}]The lowest component of the Gromov--Witten invariant has a representation
\[
n^X_{\beta} (C_1,\dots,C_n)=\int_{\overline{\mathcal{M}_0(X,\beta)}} \bigwedge_{\alpha=1}^n {\rm ev}_{\a}^\ast\g_\a,
\]
where $\overline{\mathcal{M}_0(X,\beta)}$ is the moduli space of stable holomorphic maps $\mathbb{CP}^1 \to X$ of degree $\beta$, equipped with the evaluation maps
\[
{\rm ev}_{\a}\colon \ \overline{\mathcal{M}_0(X,\beta)} \to X\colon (\phi\colon \mathbb{CP}^1 \to X, z_1,\dots,z_n)\mapsto \phi(z_\a).
\]
The $\g_\a$ are special, sufficiently smooth representatives of the Poincar\'e dual classes to the cycles~$C_\a$.
\end{Proposition}
\begin{Example} The lowest component Gromov--Witten invariant for degree 0 curves is the number of intersection points for collection of cycles $C_1,\dots, C_n$ on $X$, i.e.,
\[
n^X_{0} (C_1,\dots,C_n) = \int_{X}\bigwedge_{\alpha=1}^n\g_\a = \# (C_1\cap C_2\cap\cdots\cap C_n).
\]
\end{Example}
\begin{Definition}The {\it top component Gromov--Witten invariant} $N^X_{\beta} (C_1,\dots,C_n)$ is the number of curves of degree $\beta$, genus zero in complex space $X$, passing through the cycles $ C_1,\dots,C_n$.
\end{Definition}
\begin{Proposition}[{Konstevich--Manin~\cite{kontsevich1994gromov}}]\label{prop_KM_moduli_space_integral}
 The top component Gromov--Witten invariant has representation
\begin{gather}\label{top_compon_GW}
N^X_{\beta} (C_1,\dots,C_n) = \int_{\overline{\mathcal{M}_{0,n}(X,\beta)}} \bigwedge_{\alpha=1}^n {\rm ev}_\a^\ast \g_\a,
\end{gather}
where $\overline{\mathcal{M}_{0,n}(X,\beta)}$ is the moduli space of stable holomorphic maps $\mathbb{CP}^1 \to X$ of degree $\beta$ with $n$ marked points, equipped with the evaluation map
\[
{\rm ev}_\a\colon \ \overline{\mathcal{M}_{0,n}(X,\beta)}\to X\colon (\phi\colon \mathbb{CP}^1 \to X, z_1,\dots,z_n ) \mapsto \phi (z_\a).
\]
The $\g_\a$ are special $($sufficiently smooth$)$ representatives of the Poincar\'e dual classes to the cycles~$C_\a$.
\end{Proposition}

\subsection{Tropical observables}
The observables in GW invariants are sufficiently smooth representatives of Poincar\'e dual classes to the complex cycles $C_\a$. In cylindrical parametrization (\ref{cyl_param_trop}), a $(k,k)$ form in complex coordinates $W$, $\bar{W}$ becomes a $(k,k)$ form in radial/angular coordinates. Furthermore, a smooth differential form on $X$ becomes a singular form on $\mathbb{R}^N$ in tropical limit.
\begin{Example}
The Fubini--Studi form $\omega_{\rm FS}$ on $\mathbb{P}^1$ is a smooth representative for the Poincar\'e dual class to the point cycle. The cylindrical parametrization for $\omega_{\rm FS}$ is given by
\begin{gather}\label{eq_fs_form_complex_coord}
\omega_{\rm FS}= \frac{{\rm i}}{4\pi}\frac{{\rm d}W\wedge \overline{{\rm d}W}}{(1+|W|^2)^2}= \frac{1}{\e^2} \frac{{\rm e}^{\frac{2r}{ \e}}{\rm d}r \wedge {\rm d}\phi}{2\pi \bigl(1+ {\rm e}^{\frac{2r}{\e}}\bigr)^2}.
\end{gather}
The tropical limit $\e \to 0$ turns a smooth form (\ref{eq_fs_form_complex_coord}) into a singular one, supported at a single point $r=0$, i.e.,
\begin{gather}\label{p1_tro_observ}
\lim_{\e \to 0} \omega_{\rm FS}= \frac{1}{2\pi} \delta (r) {\rm d}r{\rm d}\phi.
\end{gather}
\end{Example}
We can use the compact $U(1)^N$-action on toric space of complex dimension $N$ to turn an arbitrary form $\omega$ into $U(1)^N$-invariant form $\bar{\omega}$ via averaging over the group action
\[
\bar{\omega} = \frac{1}{\operatorname{Vol} [U(1)^N] } \int_{U(1)^N} {\rm d}\mu(g) g\cdot \omega.
\]
In particular, for the differential form $\g_P$, the Poincar\'e dual to the point, the averaging is
\begin{gather}\label{point_class_average}
\g_P = \prod_{i=1}^N\delta\bigl(r^i-r^i_p\bigr) \delta\bigl(\phi^i-\phi^i_p\bigr) {\rm d}r^i {\rm d}\phi^i \quad \Longrightarrow\quad \bar{\g}_P = \frac{1}{(2\pi)^N} \prod_{i=1}^N \delta\bigl(r^i-r^i_p\bigr) {\rm d}r^i {\rm d}\phi^i.
\end{gather}
More generally, the tropical limit of a complex cycle $C$ on $X$ is the torus fibration over the polyhedral complex, a collection of polyhedra glued in a particular way. The tropical limit of a smooth representative in the Poincar\'e dual
class of $C$ is a singular form supported on the corresponding polyhedral complex.
\begin{Remark}
A differential form on a toric variety $X$ has an alternative description as a function on the superspace $T[1]X$. In the present paper, we will use two notations interchangeably, i.e.,
\[
\omega=\omega\bigl(\vec{r}, {\rm d}\vec{r}, {\rm d}\vec{\phi}\,\bigr).
\]
\end{Remark}

\subsection{Tropical evaluation map}\label{trop_eval_section}
The tropical evaluation map
\[
{\rm ev}_\a \colon \ \mathcal{M}_{0,n}\bigl(X,\beta;\vec{\Gamma}\bigr)\to X
\]
on the sink vertex returns is the location of the sink, i.e.,
\[
{\rm ev}_R \colon \ \mathcal{M}_{0,n}\bigl(X,\beta;\vec{\Gamma}\bigr)\to X\colon \bigl(\vec{r}_c,\vec{\phi}_c, \tau_1,\varphi_1,\dots,\tau_{I(\Gamma)},\varphi_{I(\Gamma)}\bigr) \mapsto \bigl(\vec{r}_c, \vec{\phi}_c\bigr).
\]
We can use the sink relocation formulae (\ref{root_relocation}) to define the tropical evaluation map for any vertex $\a$ via the following formula
\begin{gather}
{\rm ev}_\a \colon \ \mathcal{M}_{0,n}\bigl(X,\beta;\vec{\Gamma}\bigr) \to X\colon \bigl(\vec{r}_c,\vec{\phi}_c, \tau_1,\varphi_1,\dots,\tau_{I(\Gamma)},\varphi_{I(\Gamma)}\bigr) \mapsto \bigl(\vec{r}_\a, \vec{\phi}_\a\bigr),\nonumber \\
\vec{r}_\a = \vec{r}_c + \sum_{e\in \gamma_\G (R, v)}\pm \vec{m}_e \tau_e, \qquad \vec{\phi}_\a = \vec{\phi}_c + \sum_{e\in \gamma_\G (R, v)} \pm \vec{m}_e \v_e,\label{ev_map_trop}
\end{gather}
where $\gamma_\Gamma (\a, R)$ is the oriented path from sink $R$ to vertex $\a$. For the $U(1)^N$-invariant form, the pullback map is
\[
{\rm ev}^\ast_\a \g\bigl(\vec{r}, {\rm d}\vec{r}, {\rm d}\vec{\phi}\bigr)=\g\bigl(\vec{r}_{\a}, {\rm d}\vec{r}_{\a}, {\rm d}\vec{\phi}_{\a}\bigr) .
\]
\begin{Example} Let us consider a degree one tropical curve in $\mathbb{P}^1$ with 3 marked points. The discrete data of the tropical curve $\vec{\G}_3$ with three 2-valent vertices is ordering 3 points on a line. Let us fix the ordering and directions as shown in the picture below:
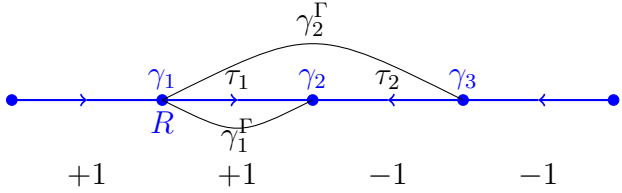
\[
\begin{tikzpicture}[scale=1]
\draw[blue, thick, ->] (0, 0) -- (1, 0);
\draw[ blue, thick](1, 0)-- (2,0);
\draw[blue,thick, ->](2, 0)-- (3,0);
\draw[blue, thick](3, 0)-- (5,0);
\draw [blue, thick, ->](6, 0)-- (5,0);
\draw[blue, thick](7, 0)-- (6,0);
\draw[ blue, thick, ->](8, 0)-- (7,0);
\filldraw[blue] (0,0) circle (2pt);
\filldraw[blue] (2,0) circle (2pt) node[anchor=south]{$\g_1$} node[anchor=north]{$R$};
\filldraw[blue] (4,0) circle (2pt) node[anchor=south]{$\g_2$};
\filldraw[blue] (6,0) circle (2pt) node[anchor=south]{$\g_3$};
\filldraw[blue] (8,0) circle (2pt);
\node at (3,0) [anchor=south]{$\tau_1$};
\node at (5,0) [anchor=south]{$\tau_2$};
\node at (1,-1) {$+1$};
\node at (3,-1) {$+1$};
\node at (5,-1) {$-1$};
\node at (7,-1) {$-1$};
\draw(2,0)..controls (4,1)..(6,0);
\node at (4, 0.7) [anchor=south] {$\gamma_2^\G$};
\draw(2,0)..controls (3,-0.5)..(4,0);
\node at (3, -0.1) [anchor=north] {$\gamma_1^\G$};
\end{tikzpicture}
\]
The tropical evaluation map pullbacks for three $U(1)$-invariant observables are given by
\begin{gather}
{\rm ev}^\ast_1\g_1 = \g_1(r_c, {\rm d}r_c, {\rm d}\phi_c),\nonumber\\
{\rm ev}^\ast_2 \g_2 = \g_2(r_c+\tau_1, {\rm d}r_c+{\rm d}\tau_1, {\rm d}\phi_c+{\rm d}\varphi_1),\nonumber\\
{\rm ev}^\ast_3 \g_3 = \g_3(r_c+\tau_1-\tau_2, {\rm d}r_c+{\rm d}\tau_1-d\tau_2, {\rm d}\phi_c+{\rm d}\varphi_1 -{\rm d}\varphi_2).\label{ev_p1_2pt}
\end{gather}
The signs in the expressions (\ref{ev_p1_2pt}) reflect the fact that an edge between points 1 and 2 is aligned with the path $\gamma^\G_1 = \gamma_\G(R,2)$, while only the first edge is aligned with the path $\gamma^\G_2 = \gamma_\G(R,3)$.
\end{Example}
\begin{Example} Let us consider a degree-one tropical curve with two marked points in $\mathbb{P}^2$. Let us choose two discrete types of tropical curves with marked point locations presented in the two pictures below:
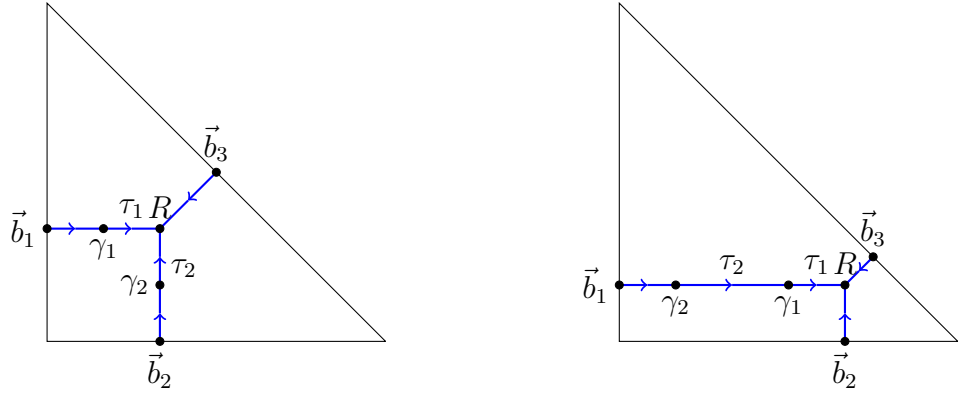
\[
\begin{tikzpicture}[scale=0.7]
\draw(4,-2)-- (-2, -2) -- (-2, 4)--(4,-2);
\draw[color=blue, thick, ->] (-2, 0) -- (-1.5, 0);
\draw[color=blue, thick] (-1.5, 0) -- (-1, 0);
\draw[color=blue, thick, ->] (-1, 0) -- (-.5, 0);
\draw[color=blue, thick] (-0.5, 0) -- (0, 0);
\draw[color=blue, thick, ->] (0, -2) -- (0,-1.5);
\draw[color=blue, thick] (0, -1) -- (0,-1.5);
\draw[color=blue, thick, ->] (0, -1) -- (0, -0.5);
\draw[color=blue, thick] (0, -0.5) -- (0,0);
\draw[color=blue, thick, ->](1,1)--(0.5, 0.5);
\draw[color=blue, thick](0.5, 0.5)-- (0,0);
\filldraw[black] (0,0) circle (2pt) node[anchor=south]{$R$};
\filldraw[black] (-1,0) circle (2pt) node[anchor=north]{$\g_1$};
\filldraw[black] (0,-1) circle (2pt) node[anchor=east]{$\g_2$};
\filldraw[black] (1,1) circle (2pt) node[anchor=south]{$\vec{b}_3$};
\filldraw[black] (-2,0) circle (2pt) node[anchor=east]{$\vec{b}_1$};
\filldraw[black] (0,-2) circle (2pt) node[anchor=north]{$\vec{b}_2$};
\node at (-0.5,0) [anchor=south]{$\tau_1$};
\node at (0,-0.7) [anchor=west]{$\tau_2$};
\end{tikzpicture}\qquad
\begin{tikzpicture}[scale=0.7]
\draw(4,-2)-- (-2, -2) -- (-2, 4)--(4,-2);
\draw[color=blue, thick, ->] (-2, -1) -- (-1.5, -1);
\draw[color=blue, thick](-1.5, -1) -- (-1, -1);
\draw[color=blue, thick, ->] (-1, -1) -- (0, -1);
\draw[color=blue, thick](0, -1) -- (1, -1);
\draw[color=blue, thick,->] (1, -1) -- (1.5, -1);
\draw[color=blue, thick](1.5, -1) -- (2, -1);
\draw[color=blue, thick,->](2, -2)-- (2,-1.5);
\draw[color=blue, thick](2, -1.5)-- (2,-1);
\draw[color=blue, thick](2, -1)-- (2+0.25,-1+0.25);
\draw[color=blue, thick,->](2+0.5,-1+0.5)-- (2+0.25,-1+0.25);
\filldraw[black] (2,-1) circle (2pt) node[anchor=south]{$R$};
\filldraw[black] (-1,-1) circle (2pt) node[anchor=north]{$\g_2$};
\filldraw[black] (1,-1) circle (2pt) node[anchor=north]{$\g_1$};
\filldraw[black] (2+0.5,-1+0.5)circle (2pt) node[anchor=south]{$\vec{b}_3$};
\filldraw[black] (-2,-1) circle (2pt) node[anchor=east]{$\vec{b}_1$};
\filldraw[black] (2,-2) circle (2pt) node[anchor=north]{$\vec{b}_2$};
\node at (1.5,-1) [anchor=south]{$\tau_1$};
\node at (0,-1) [anchor=south]{$\tau_2$};
\end{tikzpicture}
\]
The evaluation maps for the tropical curve on the left picture are
\begin{gather*}
{\rm ev}^\ast_{1}\g_1 =\g_1\bigl(\vec{r}_c-\vec{b}_1\tau_1, {\rm d}\vec{r}_c-\vec{b}_1{\rm d}\tau_1, {\rm d}\vec{\phi}_c-\vec{b}_1{\rm d}\varphi_1\bigr),\\
{\rm ev}^\ast_{2}\g_2 =\g_2\bigl(\vec{r}_c-\vec{b}_2\tau_2, {\rm d}\vec{r}_c-\vec{b}_2{\rm d}\tau_2, {\rm d}\vec{\phi}_c-\vec{b}_2{\rm d}\varphi_2\bigr).
\end{gather*}
The evaluation maps for the tropical curve in the right picture are
\begin{gather*}
{\rm ev}^\ast_{1}\g_1 =\g_1\bigl(\vec{r}_c-\vec{b}_1\tau_1, {\rm d}\vec{r}_c-\vec{b}_1{\rm d}\tau_1, {\rm d}\vec{\phi}_c-\vec{b}_1{\rm d}\varphi_1\bigr),\\
{\rm ev}^\ast_{2}\g_2 =\g_2\bigl(\vec{r}_c-\vec{b}_1\tau_1 -\vec{b}_1\tau_2, {\rm d}\vec{r}_c-\vec{b}_1{\rm d} \tau_1-\vec{b}_1{\rm d}\tau_2, {\rm d}\vec{\phi}_c-\vec{b}_1{\rm d}\varphi_1 -\vec{b}_1{\rm d}\varphi_2\bigr).
\end{gather*}
\end{Example}

\subsection{Tropical moduli space integral}\label{moduli_space_integr_section}
In Section \ref{GW_theory_complex}, we described the solution to an enumerative problem in terms of the moduli space integral (\ref{top_compon_GW}). In Sections \ref{trop_moduli_mark_section} and \ref{trop_eval_section}, we defined the tropical moduli space and the tropical evaluation map, so the tropical GW invariant $TN^{X}_{\beta} (C_1,\dots, C_n)$ can be defined in the following way.
\begin{Definition} The number of tropical curves of degree $\beta$ and genus zero on toric space $X$ passing through tropical cycles $C_1,\dots,C_n$ is given by
\begin{gather}\label{form_def_trop_GW_invariants}
TN^{X}_{\beta} (C_1,\dots,C_n)=\sum_{\vec{\G}_n } \int_{\mathcal{M}_n \bigl(X, \beta; \vec{\G}_n\bigr)}\bigwedge_{\alpha=1}^n {\rm ev}^\ast_\a \g_\a,
\end{gather}
where ${\rm ev}$ is the tropical evaluation map (\ref{ev_map_trop}), $\mathcal{M}_n \bigl(X, \beta; \vec{\G}_n\bigr)$ is the tropical moduli space and $\g_1,\dots,\g_n$ are representatives of Poincar\'e dual classes to the cycles $C_1,\dots,C_n$.
\end{Definition}
Although our Definition~\ref{form_def_trop_GW_invariants} was motivated by Proposition~\ref{prop_KM_moduli_space_integral}, we did not include the compactification of the tropical moduli space. We conjecture that there is no contribution from the codimension one and higher components due to good integrand behavior at zero edge length. Our conjecture, at least for genus zero invariants, is later justified by matching the known results from the compactified moduli space integrals in a complex geometry setup. We suspect that we might need to carefully include the higher codimension integrals on the tropical moduli space for the genus one and higher.

It is convenient to introduce another notation $\<\g_1,\dots,\g_n\>^X_\beta$ for the tropical GW invariant~(\ref{form_def_trop_GW_invariants}) to indicate a choice of representatives $\g_1,\dots,\g_n$.
\begin{Example}
The number of tropical degrees one curves in $\mathbb{P}^1$, passing through two distinct point cycles, is given by
\begin{gather}\label{p1_trop_GW_full}
TN^{\mathbb{P}^1}_{1} (p_1, p_2)= \sum_{\vec{\G}_2 } \int_{(S^1)^2} \int_{\mathbb{R} \times \mathbb{R}^{+}}{\rm ev}^\ast_1 \g_1\wedge {\rm ev}^\ast_2 \g_2.
\end{gather}
The two observables $\g_1$ and $\g_2$ are $U(1)^2$-averaged representatives (\ref{point_class_average}) of the Poincar\'e dual class to a point cycle. The sum in (\ref{p1_trop_GW_full}) is taken over two trees $\vec{\G}_2^{(12)}$ and $\vec{\G}_2^{(21)}$ depicted below:
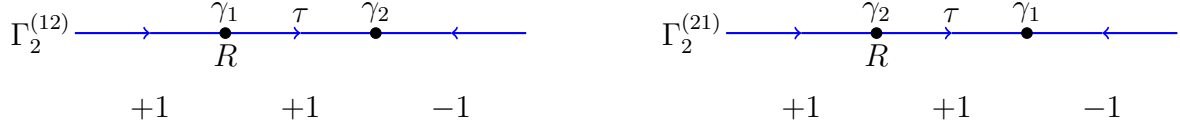
\[
\begin{tikzpicture}[scale=1]
\node at (-1,0) [anchor=west]{$\G_2^{(12)}$};
\draw[blue, thick, ->] (0, 0) -- (1, 0);
\draw[ blue, thick](1, 0)-- (2,0);
\draw[blue,thick, ->](2, 0)-- (3,0);
\draw[blue, thick](3, 0)-- (5,0);
\draw [blue, thick, ->](6, 0)-- (5,0);
\filldraw[black] (2,0) circle (2pt) node[anchor=south]{$\g_1$} node[anchor=north]{$R$};
\filldraw[black] (4,0) circle (2pt) node[anchor=south]{$\g_2$};
\node at (3,0) [anchor=south]{$\tau$};
\node at (1,-1) {$+1$};
\node at (3,-1) {$+1$};
\node at (5,-1) {$-1$};
\end{tikzpicture}\qquad
\begin{tikzpicture}[scale=1]
\node at (-1,0) [anchor=west]{$\G_2^{(21)}$};
\draw[blue, thick, ->] (0, 0) -- (1, 0);
\draw[blue, thick](1, 0)-- (2,0);
\draw[blue,thick,->](2, 0)--(3,0);
\draw[blue,thick](3, 0)--(5,0);
\draw[blue,thick,->](6, 0)--(5,0);
\filldraw[black] (2,0) circle (2pt) node[anchor=south]{$\g_2$} node[anchor=north]{$R$};
\filldraw[black] (4,0) circle (2pt) node[anchor=south]{$\g_1$};
\node at (3,0) [anchor=south]{$\tau$};
\node at (1,-1) {$+1$};
\node at (3,-1) {$+1$};
\node at (5,-1) {$-1$};
\end{tikzpicture}
\]
Using the evaluation map (\ref{ev_p1_2pt}), we can evaluate the contributions for each tropical curve. The contribution for the curve $\G_2^{(12)}$ is
\begin{gather}
TN^{\mathbb{P}^1}_{1} \bigl(p_1, p_2; \G^{(12)}_2\bigr)=\int_{(S^1)^2} \int_{\mathbb{R} \times \mathbb{R}^{+} } \g_1( r_c,\phi_c)\wedge\g_2(r_c+\tau, \phi_c+\varphi)\nonumber\\
\qquad{}=\int_{(S^1)^2} \int_{\mathbb{R} \times \mathbb{R}^{+} }\frac{1}{2\pi} \delta ( r_c-r_1) {\rm d}r_c {\rm d}\phi_c\wedge \frac{1}{2\pi} \delta ( r_c+\tau -r_2) ({\rm d}r_c +{\rm d}\tau) ({\rm d}\phi_c+{\rm d}\v) \nonumber\\
\qquad{}=\frac{1}{(2\pi)^2 } \int_{(S^1)^2} {\rm d}\phi_c {\rm d}\v\int_{\mathbb{R}}{\rm d}r_c \int^\infty_0 {\rm d}\tau\, \delta ( r_c-r_1) \delta ( r_c+\tau -r_2) \nonumber\\
\qquad{}=\int^\infty_0 {\rm d}\tau\, \delta(r_1+\tau-r_2)=\Theta (r_2-r_1),\label{p1_trop_GW}
\end{gather}
where we introduced the step function
\[
\Theta (a)=\int^\infty_0 {\rm d}x \delta (x-a)=
\begin{cases}
1, & a>0, \\
0, & a<0.
\end{cases}
\]
The contribution from the tropical curve $\G_2^{(21)}$ is
\[
TN^{\mathbb{P}^1}_{1} \bigl(p_1, p_2; \G^{(21)}_2\bigr)=\int_{(S^1)^2} \int_{\mathbb{R} \times \mathbb{R}^{+}}\g_2( r_c,\phi_c)\wedge\g_1(r_c+\tau, \phi_c+\varphi)=\Theta (r_1-r_2).
\]
The tropical GW evaluates into
\[
TN^{\mathbb{P}^1}_{1} (p_1, p_2)=\Theta (r_2-r_1)+\Theta (r_1-r_2)=1 \qquad\hbox{if}\quad r_1\neq r_2.
\]
The individual contributions from the left and right graphs describe the contributions from two distinct tropical curves labeled by the ordering of marked points. For $r_2>r_1$, the second point is to the left of the first point on the real line. Hence, only tropical curves that preserve this property do contribute. There is only one such curve, represented by the left graph.
\end{Example}
\begin{Example}
The number of degree 1 tropical curves in $\mathbb{P}^2$ passing through two distinct point cycles is
\begin{gather}\label{trop_GW_p2_2_points}
TN^{\mathbb{P}^2}_{1} (p_1, p_2)=
\sum_{\vec{\G}_2 } \int_{(S^1)^4} \int_{\mathbb{R}^2 \times (\mathbb{R}^{+})^2}{\rm ev}^\ast _1 \g_1 \wedge {\rm ev}_2^\ast \g_2.
\end{gather}
The two observables $\g_1$ and $\g_2$ are $U(1)^2$-averaged representatives (\ref{point_class_average}) of the Poincar\'e dual class of the point. The sum is taken over 12 tropical curves $\vec{\G}_2$. There are two types of tropical curves $\vec{\G}_2$: six with two marked points on the same edge and six with marked points on two different edges. Below, we will show that the contributions from the tropical curves with two marked points on the same edge vanish, so the only nontrivial contributions come from the six tropical curves depicted below:
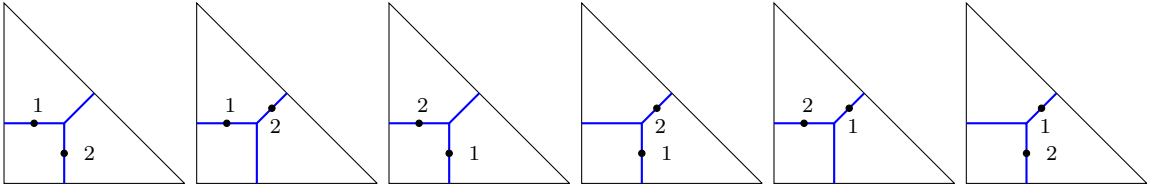
\begin{gather*}
\begin{tikzpicture}[scale=0.4]
\draw(4,-2)-- (-2, -2) -- (-2, 4)--(4,-2);
\draw[color=blue, thick] (-2, 0) -- (0, 0)-- (0,-2);
\draw[color=blue, thick](0, 0)-- (1,1);
\filldraw[black] (-1,0) circle (3pt) node[anchor=south]{$\;_1$};
\filldraw[black] (0,-1) circle (3pt)node[anchor=west]{$\;_2$};
\end{tikzpicture}\
\begin{tikzpicture}[scale=0.4]
\draw(4,-2)-- (-2, -2) -- (-2, 4)--(4,-2);
\draw[color=blue, thick] (-2, 0) -- (0, 0)-- (0,-2);
\draw[color=blue, thick](0, 0)-- (1,1);
\filldraw[black] (-1,0) circle (3pt) node[anchor=south]{$\;_1$};
\filldraw[black] (0.5,0.5) circle (3pt)node[anchor=north]{$\;_2$};
\end{tikzpicture}\,
\begin{tikzpicture}[scale=0.4]
\draw(4,-2)-- (-2, -2) -- (-2, 4)--(4,-2);
\draw[color=blue, thick] (-2, 0) -- (0, 0)-- (0,-2);
\draw[color=blue, thick](0, 0)-- (1,1);
\filldraw[black] (0,-1) circle (3pt)node[anchor=west]{$\;_1$};
\filldraw[black] (-1,0) circle (3pt) node[anchor=south]{$\;_2$};
\end{tikzpicture}\
\begin{tikzpicture}[scale=0.4]
\draw(4,-2)-- (-2, -2) -- (-2, 4)--(4,-2);
\draw[color=blue, thick] (-2, 0) -- (0, 0)-- (0,-2);
\draw[color=blue, thick](0, 0)-- (1,1);
\filldraw[black] (0,-1) circle (3pt)node[anchor=west]{$\;_1$};
\filldraw[black] (0.5,0.5) circle (3pt)node[anchor=north]{$\;_2$};
\end{tikzpicture}\,
\begin{tikzpicture}[scale=0.4]
\draw(4,-2)-- (-2, -2) -- (-2, 4)--(4,-2);
\draw[color=blue, thick] (-2, 0) -- (0, 0)-- (0,-2);
\draw[color=blue, thick](0, 0)-- (1,1);
\filldraw[black] (0.5,0.5) circle (3pt)node[anchor=north]{$\;_1$};
\filldraw[black] (-1,0) circle (3pt) node[anchor=south]{$\;_2$};
\end{tikzpicture}\,
\begin{tikzpicture}[scale=0.4]
\draw(4,-2)-- (-2, -2) -- (-2, 4)--(4,-2);
\draw[color=blue, thick] (-2, 0) -- (0, 0)-- (0,-2);
\draw[color=blue, thick](0, 0)-- (1,1);
\filldraw[black] (0.5,0.5) circle (3pt)node[anchor=north]{$\;_1$};
\filldraw[black] (0,-1) circle (3pt) node[anchor=west]{$\;_2$};
\end{tikzpicture}
\end{gather*}
The tropical evaluation map from Section \ref{trop_eval_section} for two observables on the same edge (with integer vector $\vec{b}$) is
\begin{gather}\label{ev_p2_same}
{\rm ev}^\ast _1 \g_1 \wedge {\rm ev}_2^\ast \g_2=\g_1\bigl(\vec{r}_c-\vec{b}\tau_1,\vec{\phi}_c-\vec{b}\varphi_1\bigr)\wedge\g_2\bigl(\vec{r}_c-\vec{b}(\tau_1 +\tau_2), \vec{\phi}_c-\vec{b}(\varphi_1 +\varphi_2)\bigr).
\end{gather}
The radial part of the form (\ref{ev_p2_same}) vanishes
\begin{gather*}
\prod_{i=1}^2\delta\bigl(r^i_c- b^i \tau_1 -r^i_1\bigr)\bigl({\rm d}r^i_c-b^i {\rm d}\tau_1\bigr)\wedge \prod_{i=1}^2\delta\bigl(r^i_c-b^i \tau_1-b^i \tau_2-r_2^i\bigr) \bigl({\rm d}r^i_c-b^i {\rm d}\tau_1-b^i {\rm d}\tau_2\bigr)\\
\qquad{}\propto\bigl( {\rm d}r^1_c-b^1 {\rm d}\tau_1 \bigr) \bigl( {\rm d}r^2_c-b^2 {\rm d}\tau_1\bigr)\bigl({\rm d}r^1_c - b^1 {\rm d}\tau_1 -b^1{\rm d}\tau_2\bigr)\bigl({\rm d}r^2_c - b^2 {\rm d}\tau_1 -b^2 {\rm d}\tau_2\bigr)\\
\qquad{} = {\rm d}r^1_c\bigl(-b^2 {\rm d}\tau_1\bigr) \bigl(- b^1 {\rm d}\tau_2 \bigr) {\rm d}r^2_c - b^1 {\rm d}\tau_1 {\rm d}r^2_c {\rm d}r^1_c \bigl(-b^2 {\rm d}\tau_2\bigr)=0.
\end{gather*}
In case two observables on different edges with integer vectors $\vec{b}_1$ and $\vec{b}_2$,
\begin{gather}\label{ev_p2_diff}
{\rm ev}^\ast _1 \g_1 \wedge {\rm ev}_2^\ast \g_2= \omega_1\bigl(\vec{r}_c-\vec{b}_1\tau_1,\vec{\phi}_c-\vec{b}_1\varphi_1\bigr)\wedge\omega_2\bigl(\vec{r}_c-\vec{b}_2\tau_2 , \vec{\phi}_c-\vec{b}_2 \varphi_2\bigr).
\end{gather}
The radial part of the form (\ref{ev_p2_diff}) is
\begin{gather*}
\prod_{i=1}^2\delta \bigl(r^i_c-b_1^i\tau_1 -r^i_1\bigr)\bigl({\rm d}r^i_c-b_1^i {\rm d}\tau_1 \bigr) \wedge \prod_{i=1}^2 \delta \bigl(r^i_c-b_2^i\tau_2-r_2^i\bigr)\bigl({\rm d}r^i_c - b_2^i {\rm d}\tau_2\bigr)\\
\qquad{} = - \Biggl[\prod_{i=1}^2\delta \bigl(r^i_c- b_1^i \tau_1 -r^i_1\bigr)\delta \bigl(r^i_c-b_2^i \tau_2-r_2^i\bigr)\Biggr]\bigl(\vec{b}_1\times \vec{b}_2\bigr){\rm d}r^1_c{\rm d}r^2_c {\rm d}\tau_1 {\rm d}\tau_2,
\end{gather*}
where we introduced the 2-dimensional vector product
\begin{gather}\label{2d_vect_prod}
\vec{b}_1\wedge \vec{b}_2 = b_1^1b_2^2 - b_1^2b_2^1.
\end{gather}
The angular part of the form (\ref{ev_p2_diff}) is
\begin{gather}\label{eq_prod_angular_forms_2_points_p2_2}
\frac{1}{(2\pi)^4}\prod_{i=1}^2 \bigl( {\rm d}\phi^i_c-b_1^i {\rm d}\v_1 \bigr) \wedge \prod_{i=1}^2\bigl({\rm d}\phi^i_c - b_2^i {\rm d}\v_2\bigr)=- \frac{1}{(2\pi)^4}\bigl(\vec{b}_1\wedge \vec{b}_2\bigr){\rm d}\phi^1_c{\rm d}\phi^2_c{\rm d}\v_1 {\rm d}\v_2.
\end{gather}
\begin{Remark}
 The equation (\ref{eq_prod_angular_forms_2_points_p2_2}) shows that the product of the pulled-back observables~(\ref{ev_p2_diff}) is proportional to the vector product (\ref{2d_vect_prod}) on the integer vectors $\vec{b}_1$ and $\vec{b}_2$, assigned to the edges with the point observables. Furthermore, the moduli space integral (\ref{trop_GW_p2_2_points}) for a fixed discrete type tropical curve will be proportional to the square of the vector product. In Section~2 of Mikhalkin--Rau \cite{mikhalkin2009tropical}, authors introduced the weight factor for a cubic vertex $w_1w_2|\vec{u}_1\wedge \vec{u}_2|$ to use it in the tropical curve counting. In our notations, $\vec{b}_1 = w_1 \vec{u}_1$ and $\vec{b}_2 = w_2\vec{u}_2$, so the weight factor for the cubical vertex is identical to our vector product (\ref{2d_vect_prod}).
\end{Remark}
The square of the vector product (\ref{2d_vect_prod}) for any pair of distinct unit normal vectors $\vec{b}_a\in B_{\mathbb{P}^2} =\{(1,0), (0,1), (-1,-1)\}$ is given by \smash{$\bigl(\vec{b}_a \wedge \vec{b}_c\bigr)^2 =1$}. The moduli space integral for the tropical curve $\vec{\G}_2$ with marked points on edges with integer vectors $\vec{b}_1 = (1,0)$ and $\vec{b}_2 = (0,1)$
\begin{align}
TN^{\mathbb{P}^2}_1 \bigl(p_1,p_2; \vec{\Gamma}_2\bigr)&=\int_{(S^1)^4} \int_{\mathbb{R}^2 \times (\mathbb{R}^{+})^2} {\rm ev}_1^\ast \g_1 \wedge {\rm ev}_2^\ast\g_2 \nonumber\\
&{}=\frac{1}{\bigl(2\pi\bigr)^4}\int_{(S^1)^4} {\rm d}\phi^1_c{\rm d}\phi^2_c {\rm d}\v_1 {\rm d}\v_2\int_{\mathbb{R}^2 }{\rm d}r^1_c{\rm d}r^2_c\int_{(\mathbb{R}^{+})^2 }{\rm d}\tau_1 {\rm d}\tau_2\nonumber\\
&\quad{}\times \delta \bigl(r^1_c-\tau_1 -r^1_1\bigr)\delta \bigl(r^2_c -r^2_1\bigr)\delta \bigl(r^1_c-r_2^1\bigr)\delta \bigl(r^2_c-\tau_2-r_2^2\bigr)\nonumber\\
&{} = \int^\infty_0{\rm d}\tau_1 {\rm d}\tau_2\, \delta \bigl(-\tau_1 -r^1\bigr)\delta \bigl(-\tau_2+r^2\bigr) = \Theta \bigl(-r^1\bigr) \Theta \bigl(r^2\bigr),\label{cp_2_trop_GW_G_2}
\end{align}
where we introduced the relative position $\vec{r} = \vec{r}_1- \vec{r}_2$. Similarly to the $\mathbb{P}^1$ case, an individual tree describes a tropical curve that passes through the two points with relative locations such that the relative distance vector $\vec{r} $ belongs to the second quadrant in $\mathbb{R}^2$. The remaining five tropical curves contribute to other values of the relative position $\vec{r}$. The picture below presents the relative marked point location and the corresponding tropical curves with a nonzero contribution:
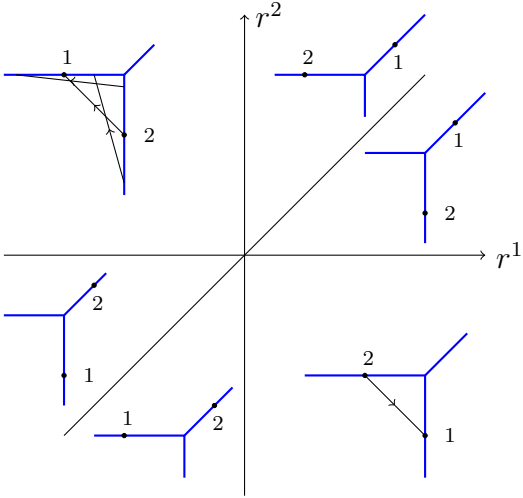
\[
\begin{tikzpicture}[scale=0.8]
\draw[->](0,-4)-- (0, 4) ;
\draw[->](-4, 0)--(4, 0);
\draw (-3,-3) -- (3,3);
\node at (0,4) [anchor=west]{$r^2$};
\node at (4,0) [anchor=west]{$r^1$};
\draw[color=blue, thick] (-2-2, 0+3) -- (0-2, 0+3)-- (0-2,-2+3);
\draw[color=blue, thick](0-2, 0+3)-- (0.5-2,0.5+3);
\filldraw[black] (-1-2,0+3) circle (1pt) node[anchor=south]{$\;_1$};
\filldraw[black] (-2,-1+3) circle (1pt)node[anchor=west]{$\;_2$};
\draw[->] (0-2,-1+3)-- (-0.5-2, -0.5+3);
\draw (-0.5-2,-0.5+3)-- (-1-2, 0+3);
\draw[->](0-2,-1.8+3)-- (-0.25-2, -0.9+3);
\draw (-0.25-2,-0.9+3)-- (-0.5-2, 0+3);
\draw[->](0-2,-0.2+3)-- (-0.9-2, -0.1+3);
\draw (-0.9-2,-0.1+3)-- (-1.8-2, 0+3);
\draw[color=blue, thick] (-2+3, 0-2) -- (0+3, 0-2)-- (0+3,-1.7-2);
\draw[color=blue, thick](0+3, 0-2)-- (0.7+3,0.7-2);
\filldraw[black] (0+3,-1-2) circle (1pt)node[anchor=west]{$\;_1$};
\filldraw[black] (-1+3,0-2) circle (1pt) node[anchor=south]{$\;_2$};
\draw[->] (-1+3,0-2)-- (-0.5+3, -0.5-2);
\draw (-.5+3,-0.5-2)-- (0+3, -1-2);
\draw[color=blue, thick](-1.5+2, 0+3) -- (0+2, 0+3)-- (0+2,-0.7+3);
\draw[color=blue, thick](0+2, 0+3)-- (1+2,1+3);
\filldraw[black] (0.5+2,0.5+3) circle (1pt)node[anchor=north]{$\;_1$};
\filldraw[black] (-1+2,0+3) circle (1pt) node[anchor=south]{$\;_2$};
\draw[color=blue, thick](-1.5-1, 0-3) -- (0-1, 0-3)-- (0-1,-0.7-3);
\draw[color=blue, thick](0-1, 0-3)-- (0.8-1,0.8-3);
\filldraw[black] (0.5-1,0.5-3) circle (1pt)node[anchor=north]{$\;_2$};
\filldraw[black] (-1-1,0-3) circle (1pt) node[anchor=south]{$\;_1$};
\draw[color=blue, thick] (-1+3, 0+1.7) -- (0+3, 0+1.7)-- (0+3,-1.5+1.7);
\draw[color=blue, thick](0+3, 0+1.7)-- (1+3,1+1.7);
\filldraw[black] (0.5+3,0.5+1.7) circle (1pt)node[anchor=north]{$\;_1$};
\filldraw[black] (0+3,-1+1.7) circle (1pt) node[anchor=west]{$\;_2$};
\draw[color=blue, thick](-1-3, 0-1) -- (0-3, 0-1)-- (0-3,-1.5-1);
\draw[color=blue, thick](0-3, 0-1)-- (0.7-3,0.7-1);
\filldraw[black] (0.5-3,0.5-1) circle (1pt)node[anchor=north]{$\;_2$};
\filldraw[black] (0-3,-1-1) circle (1pt) node[anchor=west]{$\;_1$};
\end{tikzpicture}
\]
The sum (\ref{trop_GW_p2_2_points}) over tropical curves $\vec{\G}_2$ is
\begin{align*}
TN^{\mathbb{P}^2}_{1} (p_1, p_2)&=\sum_{\vec{\G}_2} TN^{\mathbb{P}^2}_1 \bigl(p_1,p_2; \vec{\Gamma}_2\bigr)=\Theta \bigl(-r^1\bigr) \Theta \bigl(r^2\bigr) +\Theta\bigl(r^1\bigr) \Theta \bigl(-r^2\bigr)\\
&\quad{}+\Theta\bigl(r^2\bigr) \bigl(\Theta\bigl(r^1\bigr) - \Theta\bigl(r^1-r^2\bigr)\bigr)+\Theta\bigl(r^1\bigr)\bigl(\Theta\bigl(r^2\bigr) - \Theta \bigl(r^2-r^1\bigr)\bigr)\\
&\quad{}+\Theta\bigl(-r^1\bigr) \bigl(\Theta \bigl(-r^2\bigr) -\Theta\bigl(r^1-r^2\bigr)\bigr)+\Theta \bigl(-r^2\bigr) \bigl(\Theta \bigl(-r^1\bigr) -\Theta\bigl(r^2-r^1\bigr)\bigr) \\
&=1, \qquad \hbox{unless}\quad r^1 =0\quad\hbox{or}\quad r^2=0\quad \hbox{or}\quad r^1= r^2.
\end{align*}
The tropical GW invariant equals one everywhere on $\mathbb{R}^2$-plane of relative positions, except for the exceptional cases $r^1=0$, $r^2=0$, and $r^1=r^2$. There is no tropical curve passing through two points in these cases. This phenomenon was observed and discussed by Mikhalkin and Rau in~\mbox{\cite[Proposition 2.5.1]{mikhalkin2009tropical}}. The moduli space integral does not change if we replace the single point-supported representatives $\g_1$ and $\g_2$ with smooth representatives in the same cohomology class. Hence, resolving the exceptional cases issue seems plausible by using the smeared representatives $\g_1$ and $\g_2$ for tropical point cycles.
\end{Example}

\section{Higher topological quantum mechanics}\label{sec4}

Higher topological quantum mechanics (HTQM) was introduced in \cite{losev2019tqft} to generalize 1-dimen\-sion\-al TQFT by including geometric data for 1-cobordisms. This section formalizes the HTQM to describe the tropical GW invariants. In particular, we introduce an additional parity-odd symmetry. Such symmetry is the remnant from the tropicalization of a topological string theory description of the GW invariants.
\begin{Definition}
An HTQM $(V, Q, G_\pm, g)$ is a collection of the following data:
\begin{enumerate}\itemsep=0pt
\item Supervector space $V =V_0\oplus V_1$ with even, $V_0$, and odd, $V_1$ subspaces. We use notation $|v| \in \mathbb{Z}_2$ for the parity of vector $v \in V$.
\item Differential $Q\colon V \to V$, an odd operator, $|Qv| = |v|+1$, that squares to zero.
\item Quasi-homotopy $G_{+}\colon V \to V$, an odd operator, $|G_+ v| = |v|+1$, that squares to zero, i.e., $G_+^2=0$.
\item Parity-odd symmetry $G_-\colon V \to V$, an odd operator: $|G_-v| = |v|+1$, such that it squares to zero, i.e., $G_-^2 = 0$, graded commutes with differential $Q$, i.e., $\{Q, G_-\} = 0$, and graded commutes with quasi-homotopy $G_+$, i.e., $\{G_+, G_-\} = 0$.
\item Scalar product $g\colon V\times V \to \mathbb{R}$ with following properties:
\begin{itemize}\itemsep=0pt
\item parity-symmetry
\[
g(v,w) = (-1)^{|v||w|} g(w,v);
\]
\item $Q$-preservation
\[
g (Qv, w)+(-1)^{|v|} g (v,Qw) = 0;
\]
\item $G_\pm$-preservation
\[
g (G_\pm v, w)-(-1)^{|v|} g (v,G_\pm w) = 0.
\]
\end{itemize}
\end{enumerate}
If the space $V$ is infinite-dimensional, we impose certain consistency conditions on HTQM data $(V, Q, G_\pm, g)$. We define the Hamiltonian operator $H = \{ Q, G_+\}\colon V \to V$. The consistency conditions are formulated in terms of Hamiltonian:
\begin{itemize}\itemsep=0pt
\item The Hamiltonian $H$ is such that the evolution operator ${\rm e}^{-tH}$ is well defined for $t\geq0$ in the following sense:
\begin{itemize}\itemsep=0pt
\item it is a solution to the ODE
\begin{gather}\label{np_exp}
(\p_t +H) {\rm e}^{-tH} =0,\qquad {\rm e}^{-0\cdot H} =1,\qquad t \in \mathbb{R}^+\cup\{0\};
\end{gather}
\item forms a 1-parameter semi-group with multiplication
\begin{gather}\label{semi-grp}
{\rm e}^{-t_1H} {\rm e}^{-t_2 H} = {\rm e}^{-(t_1+t_2)H},\qquad \forall \ t_1, t_2 \in \mathbb{R}^+\cup\{0\}.
\end{gather}
\end{itemize}
\item We require that the $t \to \infty$ limit of the evolution operator exists and is equal to the projector on $ \ker H$, i.e.,
\[
\lim_{t\to+\infty} {\rm e}^{-tH} = \Pi_0.
\]
\item The projector $\Pi_0$ obeys
\begin{gather}\label{def_Proj_G_+}
\Pi_0 G_\pm =G_\pm\Pi_0=0.
\end{gather}
\end{itemize}
\end{Definition}
\begin{Remark}
The notation $G_\pm$ is adopted from homotopy in string theory, where $G_\pm = G_{0,L}\pm G_{0,R}$. The $G_{0, L/R}$ is the $Q$-partner of the energy-momentum tensor for the left and right modes.
\end{Remark}
\begin{Definition}The {\it propagator} $K\colon V \to V$ for HTQM $(V, Q, G_{\pm}, g)$ is an operator
\begin{gather}\label{def_HTQM_propagator}
K= \lim_{T\to \infty}\int^T_0{\rm d}t\, {\rm e}^{-t H} G_+ =\int^\infty_0{\rm d}t\, {\rm e}^{-t H} G_+.
\end{gather}
\end{Definition}
Note that the integral may diverge when the exponent vanishes for states from $\ker H$. The~$G_+$ in the expression (\ref{def_HTQM_propagator}) and the HTQM property (\ref{def_Proj_G_+}) evaluates $G_+ v =0$ on all $v \in \ker H$, hence $Kv=0$ for such states. The propagator $K$ is a homotopy, i.e.,
\[
\{ Q,K\} =\int^\infty_0{\rm d}t\, {\rm e}^{-t H} \{Q,G_+\} =- \int_{0}^\infty {\rm d} \bigl( {\rm e}^{-tH} \bigr)={\rm e}^{-tH}\big|_{0}-{\rm e}^{-tH}\big|_{\infty}= 1-\Pi_0.
\]
\begin{Example}The semi-group property (\ref{semi-grp}) may only hold at some subspace of $\mathbb{R}^+$. An~example of such a phenomenon is the one-parameter family of diffeomorphisms for
a vector field $u=x^2 \p_x$. Here vector space $V = \Omega^1(\mathbb{R})$ with the Hamiltonian $H = \cl_u= \{ \iota_u, {\rm d}\}$.
\end{Example}

\subsection{Amplitudes in HTQM}\label{sec_ampl_htqm}
\begin{Definition}
An {\it observable} $\cO$ in HTQM $(V,Q, G_\pm, g)$ is a linear operator $\cO\colon V \to V$.
\end{Definition}
\begin{Definition}
An {\it evolution operator} $U(t, {\rm d}t) \in V\otimes V^\ast \otimes \Omega^{\ast}\bigl(\mathbb{R}^+\bigr)$ is a solution to
\[
{\rm d}U(t, {\rm d}t) + [Q,U(t, {\rm d}t)] =0,\qquad U(0,{\rm d}t) = 1 + G_+ {\rm d}t.
\]
\end{Definition}
We use the exponent (\ref{np_exp}) to express the evolution operator
\[
U(t, {\rm d}t) = {\rm e}^{-t H +G_+ {\rm d}t}={\rm e}^{-tH}(1+ G_+ {\rm d}t).
\]
\begin{Note}
We will suppress the argument ${\rm d}t$ in evolution operator $U(t, {\rm d}t)$ to make our expressions less crowded.
\end{Note}
\begin{Definition}
For the HTQM $(V, Q, G_\pm, g)$, a collection of $n$ observables $\cO_\a$ and a pair of states $v$, $w$ the {\it pre-amplitude on a line} $\mathcal{PA}_{I_n}(\cO_1,\dots,\cO_n; v, w)$ is a differential form on the moduli space $\mathcal{M} (I_n)\simeq (\mathbb{R}^{+})^{n-1}$ of the $n$ ordered marked points on a line given by
\[
\mathcal{PA}_{I_n} (\cO_1,\dots,\cO_n; v, w)=g(v,\cO_1 2\pi U(\tau_2)G_- \cO_2\cdots 2\pi U(\tau_{n})G_- \cO_n w).
\]
\end{Definition}
By definition, a pre-amplitude $\mathcal{PA}_{I_n}$ is a mixed degree form on the moduli space $\mathcal{M} (I_n)$. The lowest component is a function on $\mathcal{M} (I_n)$. It is well known in Euclidean quantum mechanics as a transition amplitude (see Feynman--Hibbs--Styer~\cite{feynman2010quantum} for more details) with observables inserted at marked points. An example below gives a more detailed description.
\begin{Example}Let us consider two marked points on a line, so the moduli space $\mathcal{M}(I_2) = \mathbb{R}^+$. The coordinate on the moduli space $\tau > 0$ describes the relative position of the two ordered points. For observables $\cO_1$ and $\cO_2$ and a pair of states $v$, $w$, the pre-amplitude is
\begin{gather}\label{2_pt_pre_ampl}
\mathcal{PA}_{I_2} (\cO_1, \cO_2; v, w)=g(v,\cO_1 2\pi U(\tau) G_- \cO_2w) \in \Omega^\ast (\mathbb{R}^{+}).
\end{gather}
The degree zero component of the pre-amplitude is a function of the moduli space
\begin{gather}\label{ex_deg_0_preampl} 
\mathcal{PA}^{{\rm deg}=0}_{I_2} (\cO_1, \cO_2; v, w)=g\bigl(v,\cO_1 2\pi {\rm e}^{-\tau H} G_-\cO_2 w\bigr).
\end{gather}
We can use the (Euclidean) Heisenberg picture to replace the observables $\cO_1$ and $\cO_2$ with their time-dependent versions
\[
\cO_\a (\tau) = {\rm e}^{-\tau H} \cO_\a {\rm e}^{\tau H}
\]
to rewrite the pre-amplitude (\ref{ex_deg_0_preampl}) in the form
\begin{gather}\label{2_pt_pre_ampl_heisb_pic}
\mathcal{PA}^{{\rm deg}=0}_{I_2} (\cO_1, \cO_2; v, w)= g\bigl(v,\cO_1 (0) 2\pi\cO_2 (\tau) G_- {\rm e}^{-\tau H} w\bigr).
\end{gather}
An expression (\ref{2_pt_pre_ampl_heisb_pic}) is the matrix element between states $v$ and $w$ in Euclidean quantum mechanical transition amplitude on an interval $[0, \tau]$ with observable $\cO_1$ at a time $0$ and observable $2\pi \cO_2 G_-$ at a time $\tau$.
\end{Example}
The term ``higher'' in the higher topological quantum mechanics means we will use the full pre-amplitude instead of the degree zero component.
\begin{Definition}
{\it Generalized amplitude} $\mathcal{A}_{I_n}^\mathcal{C}$ on a line $I_n$ with $n$ marked points is an integral of the pre-amplitude $\mathcal{PA}_{I_n}$ over a chain $\mathcal{C}$ in the moduli space $\mathcal{M} (I_n)$, i.e.,
\[
\mathcal{A}_{I_n}^\mathcal{C} = \int_{\mathcal{C}} \mathcal{PA}_{I_n}.
\]
\end{Definition}
In the present paper, we will be interested in the top-degree generalized amplitudes, i.e., the integral over the whole moduli space
\[
\mathcal{A}_{I_n}(\cO_1,\dots,\cO_n; v, w) = \int_{\mathcal{M} (I_n)} \mathcal{PA}_{I_n}(\cO_1,\dots,\cO_n; v, w).
\]
We use an integral representation for the homotopy (\ref{def_HTQM_propagator}) to rewrite the amplitude in the form
\[
\mathcal{A}_{I_n}(\cO_1,\dots,\cO_n; v, w)=(2\pi)^{n-1} g (v, \cO_1K G_- \cO_2 \cdots K G_- \cO_n w).
\]
The generalized amplitudes are also present in standard quantum mechanics as the terms in perturbation theory.
\begin{Example}\label{top_ampl_htq_deformation}
Suppose an operator $\Phi\colon V \to V$ in HTQM $(V, Q, G_\pm, g)$, is nilpotent, parity-odd and commutes with $Q$ and $G_-$, i.e., $\{Q, \Phi\} =\{G_-, \Phi\} = 0$. For a parity-even number $\lambda$ we can construct a family $(V, Q - \lambda \Phi , G_\pm, g)$ of HTQM's deformations. The deformed evolution operator is a solution to
\begin{gather}\label{def_ev_equation}
{\rm d}U^\lambda(t) + \bigl[ Q - \lambda \Phi,U^\lambda(t)\bigr] =0, \qquad U(0) = 1 + G_+ {\rm d}t.
\end{gather}
The solution to the Cauchy problem (\ref{def_ev_equation}) is written using the (formal) series in $\lambda$, i.e.,
\[
U^\lambda (t) = U^{(0)}(t) + \lambda U^{(1)} (t) +\cdots+\lambda^k U^{(k)} (t) +\cdots.
\]
The $U^{(0)}(t)$ is the evolution operator in the original HTQM $(V, Q, G_\pm, g)$, while the higher terms are determined recursively solving
\[
{\rm d}U^{(k)}(t) + \bigl[Q,U^{(k)}(t)\bigr]=\bigl[\Phi,U^{(k-1)}(t)\bigr].
\]
The recursive solution is
\[
U^{(k)}(t)= \int_{t_1+t_2=t}U(t_1)\Phi U^{(k-1)}(t_2).
\]
The homotopy for deformed HTQM$_\lambda$ can be expressed as series in $\lambda$
\begin{align}
K_\lambda& = \int_{0}^\infty U_\lambda (t) \nonumber\\
&= \int_{0}^\infty U (t_1)+ \lambda \int_{0}^\infty U(t_1)\int_0^\infty \Phi U (t_2)+ \lambda^2 \int_{0}^\infty U(t_1)\int_0^\infty \Phi U (t_2)\int_0^\infty \Phi U (t_3) + \cdots \nonumber\\
&= K + \lambda K \Phi K +\lambda^2 K \Phi K \Phi K + \cdots=K\sum_{k=0}^\infty (\lambda \Phi K)^k.\label{HTQM_deformation_prop}
\end{align}
The (top-degree) generalized amplitude $\widetilde{ \mathcal{A}_{I_2} }$ in deformed HTQM $(V, Q - \lambda \Phi , G_\pm, g)$ on a line $I_2$ with two marked points is
\begin{gather}\label{2_pt_ampl_def_htqm}
\widetilde{ \mathcal{A}_{I_2}} (\cO_1, \cO_2; v,w)=g (v, \cO_1 2\pi K_\lambda G_- \cO_2 w).
\end{gather}
For a special choice of operator $\Phi=2\pi [ G_-, \cO]$, the perturbative expansion (\ref{HTQM_deformation_prop}) for the amplitude~(\ref{2_pt_ampl_def_htqm}) becomes
\begin{gather}\label{pert_exp_ampl}
\widetilde{ \mathcal{A}_{I_2}} (\cO_1, \cO_2; v,w ) = \sum_{k=0}^\infty \mathcal{A}_{I_{k+2}} (\cO_1, \underbrace { \lambda \cO,\dots, \lambda \cO}_k, \cO_2; v,w ).
\end{gather}
Equality (\ref{pert_exp_ampl}) expresses an amplitude in deformed HTQM $(V, Q - 2\pi \lambda[G_-,\cO], G_\pm, g)$ as a~sum over amplitudes in the original HTQM $(V, Q, G_\pm, g)$.
\end{Example}
\begin{Remark}
The deformation of HTQM by the multiple operators $\Phi_A$ was discussed \cite{Lysov:2002fw}. In particular, it was shown that the corresponding amplitudes solve the (anti) commutativity equation.
\end{Remark}

\subsection{HTQM on special trees}\label{sec_htqm_spec_tees}

In Section \ref{sec_trop_curve_graph}, we described generic tropical curves of genus zero with marked points as connected oriented trees $\vec{\Gamma}$ with sink and 1-, 2- and 3-valent vertices only, equipped with integer vectors and moduli. For a given tropical curve $\vec{\G}$, we can construct a special tree $\G$, defined below, by forgetting the integer vectors and moduli on the edges.
\begin{Definition} A {\it special tree} $\G$ is a directed tree with distinct sink vertex and only 1-, 2- and 3-valent vertices. The directions are as follows: the edges are outgoing to the leaves, 2-valent vertices have one incoming and one outgoing edge, and 3-valent vertices (except sink) have two incoming and one outgoing edge. The sink vertex has all incoming edges.
\end{Definition}
The line with marked points $I_n$ in Section \ref{sec_ampl_htqm} is a generalization of a special tree with no 3-valent vertices and a sink at a 2-valent vertex.
\begin{Definition} \label{def_htqm_spec_tree}
An HTQM $(V, Q, G_\pm, g, \mu_2)$ on a special tree $\Gamma$ is the decoration of a special tree with the following data:
\begin{enumerate}\itemsep=0pt
\item[(1)] each edge is decorated with the HTQM data $(V, Q, G_{\pm})$;
\item[(2)] 1-valent vertices (leaves) are decorated with states, i.e., $v_a \in V$, $a=1,\dots,n_1=|V_1(\Gamma)|$;
\item[(3)] 2-valent vertices are decorated with observables $\cO_\a \in V\otimes V^\ast$, $\a = 1,\dots,n_2 = |V_2(\Gamma)|$;
\item[(4)] 3-valent vertices are decorated with the multiplication $\mu_2\colon V\otimes V\to V $;
\item[(5)] sink 3-valent vertex is decorated with the multiplication
$\mu_3^0=g\circ \mu_2\colon \ V^{\otimes 3}\to \mathbb{R}$.
\end{enumerate}
The multiplication $\mu_2$ obeys the following consistency properties:
\begin{itemize}\itemsep=0pt
\item $\mu_2$ is graded commutative
\[
\mu_2 (v,w) = (-1)^{|v||w|} \mu_2 (w,v);
\]
\item $\mu_2$ is associative
\[
\mu_2 (\mu_2 (v,w),u) = \mu_2 (v, \mu_2(w,u));
\]
\item $(\mu_2, Q)$ obeys the Leibniz rule
\[
Q \mu_2(v,w) = \mu_2 (Qv, w)+ (-1)^{|v|} \mu_2 (v, Qw);
\]
\item $(G_-, \mu_2)$ obeys the 7-term relation
\begin{align}
&G_- \mu_2(\mu_2(v,w),u)=\mu_2(G_-\mu_2(v,w),u) + ( -1)^{|w|(|v|-1)} \mu_2 (w, G_- \mu_2(v,u)) \nonumber\\
&\qquad{}+ ( -1)^{|v|} \mu_2 (v, G_- \mu_2(w,u)) - \mu_2(G_- v, \mu_2(w,u)) \nonumber\\
&\qquad{}- (-1)^{|v|} \mu_2(v, \mu_2 (G_-w, u))- (-1)^{|u|+|v|} \mu_2(v, \mu_2(w, G_-u)). \label{def_7_term_relation}
\end{align}
\end{itemize}
\end{Definition}
For a special tree $\Gamma$, we can define an element of the tensor algebra equal to the tensor product of the elements assigned to each vertex and edge of $\G$. Each vertex $v$ is assigned with an element $V^{\ast |E_{{\rm in}}(v)|} \otimes V^{ |E_{{\rm out}}(v)|}$, i.e., has a factor $V^\ast$ for each incoming edge and a factor $V$ for each outgoing edge. Each edge is assigned with an element $V\otimes V^\ast$, factor $V$ for the vertex it is ingoing and factor $V^\ast$ for the vertex to wh it is outgoing. A special tree (directed graph, more generally) defines a map from an element of tensor algebra we described earlier to the real numbers via contraction in tensor algebra, determined by the directed edges connecting vertices. We will denote such contraction for special tree $\G$ in the following way:
\begin{gather*}
\<\;\;\>_\Gamma \colon (V\otimes V^\ast )^{\otimes E} \otimes V^{\otimes n_1} \otimes (V^\ast \otimes V)^{\otimes n_2}\otimes (V^\ast \otimes V^\ast \otimes V )^{\otimes (n_3-1)} \otimes V^\ast \otimes V^\ast \otimes V^\ast\to \mathbb{R}.
\end{gather*}
For a special tree $\G$, we define the moduli space $\mathcal{M}(\G)$, given by the lengths of internal edges,~i.e.,
\[
\mathcal{M}(\G)=(\mathbb{R}^{+})^{I(\Gamma)}.
\]
\begin{Definition}
The {\it pre-amplitude} $\mathcal{PA}_\Gamma (\cO_\a; v_a)$ on a special tree $\Gamma$ decorated with states $v_a$, operators $\cO_\a$ in HTQM $(V, Q, G_\pm, g, \mu_2)$ is a contraction in tensor algebra, defined by a special tree $\G$. The leaves of a special tree are decorated with the states $v_1,\dots,v_{n_1}\in V$, internal edges are decorated with the evolution operator $U \in V\otimes V^\ast$, external edges are decorated with the identity operator $\mathds{1} \in V\otimes V^\ast$, 2-valent vertices are decorated with observables $\cO_\a \in V\otimes V^\ast$, 3-valent vertices are decorated with multiplication $\mu_2 \in V^\ast\otimes V^\ast \otimes V$ and a sink vertex decorated with $\mu_3^0\in V^\ast \otimes V^\ast \otimes V^\ast$. We will denote such contraction as
\begin{gather}\label{htqm_pre_ampl}
\mathcal{PA}_\Gamma (\cO_\a; v_a) = \Biggl\<\mathds{1}^{\otimes n_1(\G)}\otimes U^{\otimes I(\G)}\bigotimes\limits_{a=1}^{n_1(\G)} v_a\bigotimes\limits_{\a=1}^{n_2(\G)} \cO_\a \otimes \mu_2^{\otimes (n_3(\G)-1)}\otimes \mu_3^0\Biggr\>_\Gamma.
\end{gather}
\end{Definition}
\begin{Example}The pre-amplitude (\ref{2_pt_pre_ampl}) is the pre-amplitude (\ref{htqm_pre_ampl}) on a special tree $\G$ with $n_1(\G) =2$ leaves, equipped with states $v$ and $w$ and $n_2(\G)=2$ 2-valent vertices, equipped with observables $\cO_1$ and $\cO_2$.
\end{Example}
\begin{Definition} An amplitude $\mathcal{A}_\G^{\mathcal{C}}$ on a special tree $\G$ is an integral of the pre-amplitude $\mathcal{PA}_\G$ over a cycle $\mathcal{C} \in \mathcal{M}(\G)$, i.e.,
\[
\mathcal{A}^{\mathcal{C}}_\Gamma = \int_{\mathcal{C}} \mathcal{PA}_\Gamma.
\]
\end{Definition}
In our analysis of tropical GW invariants, we will use the top degree amplitudes, i.e., integrals over the whole moduli space
\begin{gather}\label{def_ampl_htqm_tree}
\mathcal{A}_\Gamma = \int_{\mathcal{M}(\Gamma)} \mathcal{PA}_\Gamma.
\end{gather}
We can perform the moduli space integral and express the amplitude (\ref{def_ampl_htqm_tree}) using homotopy~(\ref{def_HTQM_propagator}) in the form
\begin{gather}\label{A_0_graph_homotopy}
\mathcal{A}_\Gamma (\cO_\a; v_a;Q)=\Biggl\<\mathds{1}^{\otimes n_1(\G)} \otimes (2\pi KG_-)^{\otimes I(\G)}\bigotimes\limits_{a=1}^{n_1(\G)}v_a\bigotimes\limits_{\a=1}^{n_2(\G)} \cO_\a \otimes\mu_2^{\otimes (n_3(\G)-1)}\otimes \mu_3^0 \Biggr\>_\Gamma.
\end{gather}
Definition \ref{def_htqm_spec_tree} implies that the amplitude $\mathcal{A}_\Gamma (\cO_\a; v_a)$ is independent of the location of the sink vertex \big(where $\mu_2$ is replaced by the $\mu_3^0$\big), hence for convenience we
will use the shorter notation for the same amplitude
\begin{gather}\label{A_0_graph_homotopy_short}
\mathcal{A}_\Gamma (\cO_\a; v_a)=\Biggl\<(2\pi KG_-)^{\otimes I(\G)} \bigotimes\limits_{a=1}^{n_1(\G)}v_a\bigotimes\limits_{\a=1}^{n_2(\G)} \cO_\a \otimes \mu_2^{\otimes n_3(\G)}\Biggr\>_\Gamma.
\end{gather}
The expression (\ref{A_0_graph_homotopy_short}) describes an amplitude on a special tree $\G$ with leaves decorated with states $v_a$, internal edges decorated with propagators $2\pi KG_-$, 2-valent vertices
decorated with operators $\cO_\a$, 3-valent vertices decorated with the multiplication $\mu_2$ and a scalar product $g$ at arbitrary vertex on a special tree $\G$.
\begin{Example}
Let us consider the three special trees depicted below:
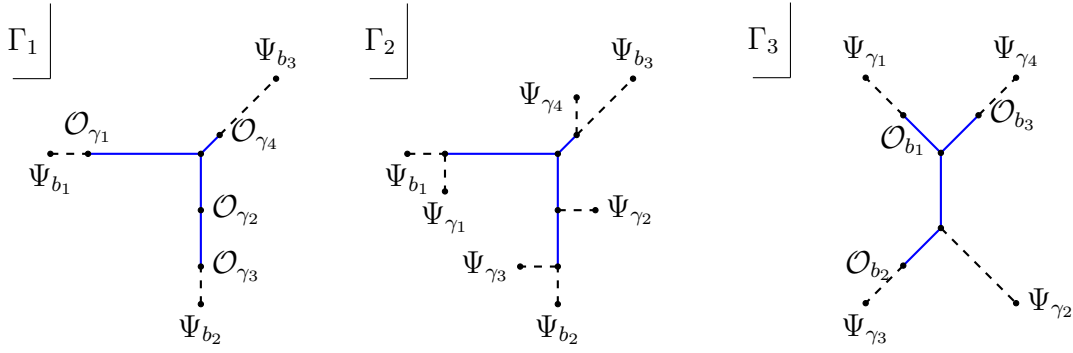
\[
\begin{tikzpicture}[scale=0.5]
\filldraw[black] (-4,3)node[anchor=east]{$\G_1$};
\draw(-5,2)-- (-4, 2) ;
\draw(-4,2)-- (-4, 4) ;
\draw[black, dashed, thick](-4,0)-- (-3, 0) ;
\draw[black, dashed, thick](0, -4) -- (0, -3);
\draw[blue, thick](0, -3) -- (0, 0);
\draw[blue, thick](-3, 0) -- (0, 0);
\draw[black, thick, dashed](0.5, 0.5) -- (2, 2);
\draw[blue, thick](0, 0) -- (0.5, 0.5);
\filldraw[black] (0,0) circle (2pt);
\filldraw[black] (-4,0) circle (2pt) node[anchor=north]{$\Psi_{b_1}$};
\filldraw[black] (0,-1.5) circle (2pt) node[anchor=west]{$\cO_{\g_2}$};
\filldraw[black] (0,-3) circle (2pt) node[anchor=west]{$\cO_{\g_3}$};
\filldraw[black] (2,2) circle (2pt) node[anchor=south]{$\Psi_{b_3}$};
\filldraw[black] (0.5,0.5) circle (2pt) node[anchor=west]{$\cO_{\g_4}$};
\filldraw[black] (0,-4) circle (2pt) node[anchor=north]{$\Psi_{b_2}$};
\filldraw[black] (-3,0) circle (2pt) node[anchor=south]{$\cO_{\g_1}$};
\end{tikzpicture}\;\;\;\;
\begin{tikzpicture}[scale=0.5]
\filldraw[black] (-4,3)node[anchor=east]{$\G_2$};
\draw(-5,2)-- (-4, 2) ;
\draw(-4,2)-- (-4, 4) ;
\draw[black, dashed, thick](-4,0)-- (-3, 0) ;
\draw[black, dashed, thick](0, -4) -- (0, -3);
\draw[blue, thick](0, -3) -- (0, 0);
\draw[blue, thick](-3, 0) -- (0, 0);
\draw[black, thick, dashed](0.5, 0.5) -- (2, 2);
\draw[blue, thick](0, 0) -- (0.5, 0.5);
\filldraw[black] (0,0) circle (2pt);
\filldraw[black] (-4,0) circle (2pt) node[anchor=north]{$\Psi_{b_1}$};
\filldraw[black] (0,-1.5) circle (2pt);
\filldraw[black] (1,-1.5) circle (2pt) node[anchor=west]{$\Psi_{\g_2}$};
\draw[color=black, thick, dashed] (0,-1.5)--(1,-1.5);
\filldraw[black] (0,-3) circle (2pt);
\filldraw[black] (-1,-3) circle (2pt) node[anchor=east]{$\Psi_{\g_3}$};
\draw[color=black, thick, dashed] (0,-3)--(-1,-3);
\filldraw[black] (2,2) circle (2pt) node[anchor=south]{$\Psi_{b_3}$};
\filldraw[black] (0.5,0.5) circle (2pt);
\filldraw[black] (0.5,1.5) circle (2pt) node[anchor=east]{$\Psi_{\g_4}$};
\draw[color=black, thick, dashed] (0.5,0.5)--(0.5,1.5);
\filldraw[black] (0,-4) circle (2pt) node[anchor=north]{$\Psi_{b_2}$};
\filldraw[black] (-3,0) circle (2pt);
\filldraw[black] (-3,-1) circle (2pt) node[anchor=north]{$\Psi_{\g_1}$};
\draw[color=black, thick, dashed] (-3,0)--(-3,-1);
\end{tikzpicture}\;\;\;\;\;\;\;
\begin{tikzpicture}[scale=0.5]
\filldraw[black](-4,3)node[anchor=east]{$\G_3$};
\draw (-5,2)-- (-4, 2) ;
\draw (-4,2)-- (-4, 4) ;
\draw[blue, thick](0, -2) -- (-1, -3);
\draw[black, thick, dashed](-1, -3) -- (-2, -4);
\draw[black, thick, dashed](0, -2) -- (2, -4);
\draw[blue, thick](0, -2) -- (0, 0);
\draw[black, dashed, thick](1, 1) -- (2, 2);
\draw[blue, thick](1, 1) -- (0, 0);
\draw[black, dashed, thick](-1, 1) -- (-2, 2);
\draw[blue, thick](0, 0) -- (-1, 1);
\filldraw[black] (0,0) circle (2pt);
\filldraw[black] (0,-2) circle (2pt);
\filldraw[black] (-2,-4) circle (2pt) node[anchor=north]{$\Psi_{\g_3}$};
\filldraw[black] (-1,-3) circle (2pt) node[anchor=east]{$\cO_{b_2}$};
\filldraw[black] (-2,2) circle (2pt) node[anchor=south]{$\Psi_{\g_1}$};
\filldraw[black] (-1,1) circle (2pt) node[anchor=north]{$\cO_{b_1}$};
\filldraw[black] (2,-4) circle (2pt) node[anchor=west]{$\Psi_{\g_2}$};
\filldraw[black] (2,2) circle (2pt) node[anchor=south]{$\Psi_{\g_4}$};
\filldraw[black] (1,1) circle (2pt) node[anchor=west]{$\cO_{b_3}$};
\end{tikzpicture}
\]
The amplitude on the first tree
\begin{gather}\label{ex_ampl_g1}
\mathcal{A}_{\G_1}=(2\pi)^4\mu_3^0 \bigl(K G_-\cO_{\g_1} \Psi_{b_1}, KG_- \cO_{\g_4} \Psi_{b_3}, K G_-\cO_{\g_2}KG_- \cO_{\g_3} \Psi_{b_2}\bigr).
\end{gather}
The amplitude of the second tree
\begin{gather}
\mathcal{A}_{\G_2}=(2\pi)^4\mu_3^0 \nonumber\\ \hphantom{\mathcal{A}_{\G_2}=}{}
\times \bigl(KG_-\mu_2(\Psi_{\g_1}, \Psi_{b_1}), KG_-\mu_2(\Psi_{\g_4}, \Psi_{b_3}), KG_-\mu_2\bigl(\Psi_{\g_2}, KG_-\mu_2\bigl(\Psi_{\g_3} ,\Psi_{b_2} \bigr)\bigr)\bigr).\label{ex_ampl_g2}
\end{gather}
The amplitude of the third tree
\begin{gather}\label{ex_ampl_g3}
\mathcal{A}_{\G_3}=(2\pi)^4\mu_3^0 \bigl(KG_- \cO_{b_1} \Psi_{\g_1}, KG_-\cO_{b_3} \Psi_{\g_4},KG_-\mu_2\bigl(\Psi_{\g_2}, KG_-\cO_{b_2} \Psi_{\g_3}\bigr)\bigr).
\end{gather}
\end{Example}
\begin{Remark} The formal construction of HTQM on special trees has several useful applications. In particular, Losev and Shadrin \cite{Losev_2007} showed that the amplitudes in HTQM on special trees with additional assumptions obey the WDVV equations.
\end{Remark}

\subsection[A-model HTQM]{$\boldsymbol{A}$-model HTQM}\label{TGQ_QM_section}
Tropical Gromov--Witten theory on a compact toric variety $X$ of complex dimension $N$ defines the particular HTQM on special trees. Following the string theory analogy, we will denote HTQM as the $A$-model. Below, we express the HTQM data $(V, Q, G_\pm, g, \mu_2)$ in terms of geometric data of $X$.
\begin{Definition}The $A$-model is the collection of the following data $(V, Q, G_\pm, g, \mu_2)$, where the vector space $V$ is the space of pairs
\[
|\omega, \vec{m}\>,\qquad \omega \in\Omega_{{\rm trop}}^\ast(X) ,\qquad \vec{m} \in \mathbb{Z}^N.
\]
The $\Omega_{{\rm trop}}^\ast(X)$ is a space of $U(1)^N$-invariant smooth differential forms on a toric variety $X$ of dimension $N$, i.e., smooth forms on $\mathbb{C}^{\ast N}$, that are extendable to the compactification. An~exis\-tence of continuation, in particular, implies that for all primitive normal vectors $\vec{b}\in B_X$, the following holds:
\begin{gather}\label{cont_form_compact}
\lim_{t\to +\infty} \omega \bigl(\vec{r}+\vec{b}t\bigr)= \text{finite}, \quad \ \lim_{t\to +\infty} \iota_{b^j \p_{\phi^j}} \omega \bigl(\vec{r}+\vec{b}t\bigr) = 0,\quad \
\lim_{t\to +\infty} \iota_{b^j \p_{r^j}} \omega \bigl(\vec{r}+\vec{b}t\bigr)= 0.
\end{gather}
The parity of a state $|\omega, \vec{m}\>$ is the parity $|\omega|$ of the form, defined as the degree of form $\deg(\omega)$ \mbox{$\bmod \, 2$}.
The differential $Q$ acts as the de Rham operator in radial direction on $\mathbb{C}^{\ast N} = {\mathbb{R}^N \times\mathbb{T}^N}$,~i.e.,
\begin{gather}\label{A_model_Q}
Q|\omega, \vec{m}\>=|{\rm d}\omega, \vec{m}\>,\qquad {\rm d}=\sum {\rm d}r^i \frac{\p}{\p r^i}.
\end{gather}
The homotopy $G_{+}$ and parity-odd symmetry $G_-$ are defined via interior derivatives
\begin{gather}\label{A_model_G_+}
G_+ |\omega, \vec{m}\>=\bigl| \iota^R_{\vec{m}}\omega, \vec{m}\bigr\>=\bigl| \iota_{m^i \p_{r^i}}\omega, \vec{m}\bigr\>, 
\end{gather}
and
\begin{gather}\label{A_model_G_-}
G_- |\omega, \vec{m}\>=\bigl| \iota^\Phi_{\vec{m}}\omega, \vec{m}\bigr\>=\bigl| \iota_{m^i \p_{\phi^i}}\omega, \vec{m}\bigr\>.
\end{gather}
The scalar product is the integration of forms
\begin{gather}\label{A_model_g}
g\colon \ V \times V \to \mathbb{R} \colon (|\omega_1,\vec{m}_1\>, |\omega_2,\vec{m}_2\>) \mapsto \delta_{\vec{m}_1+\vec{m}_2, \vec{0}}\int_{X} \omega_1\wedge \omega_2.
\end{gather}
The multiplication $\mu_2\colon V\otimes V \to V$ is the wedge product on differential forms
\begin{gather}\label{A_model_mu}
\mu_2 ( |\omega_1,\vec{m}_1\> ,|\omega_2,\vec{m}_2\>)=|\omega_1\wedge\omega_2, \vec{m}_1+\vec{m}_2\>.
\end{gather}\end{Definition}
\begin{Proposition}
The $A$-model $(V, Q, G_{\pm}, g, \mu_2)$ is the HTQM on special trees.
\end{Proposition}
\begin{proof}The operators $Q$, $G_\pm$ realized as operations (\ref{A_model_Q})--(\ref{A_model_G_-}) on differential forms immediately imply that all three are differentials, i.e., parity-odd operators, that square to zero. The~pair of differentials $G_\pm$ are interior derivatives (\ref{A_model_G_+}) and (\ref{A_model_G_-}), hence are anti-commuting. The~graded commutator of $G_-$ and $Q$
\[
\{Q, G_-\}|\omega, \vec{m}\> =\bigl|\cl^\Phi_{\vec{m}} \omega, \vec{m}\bigr\> = 0,
\]
since the form $\omega$ is invariant under the $U(1)^N$-rotations. The $A$-model scalar product (\ref{A_model_g}) is graded-symmetric
\[
g( |\omega_1,\vec{m}_1\>, | \omega_2,\vec{m}_2\>) =\delta_{\vec{m}_1+\vec{m}_2, \vec{0}}\int_{X} \omega_1\wedge \omega_2=(-1)^{|\omega_1| |\omega_2|} g( | \omega_2,\vec{m}_2\>,|\omega_1,\vec{m}_1\>),
\]
$Q$-preserving follows from compactness of $X$, i.e.,
\[
 g (Q |\omega_1,\vec{m}_1\>, | \omega_2,\vec{m}_2\>)+(-1)^{|\omega_1| } g ( |\omega_1,\vec{m}_1\>, Q| \omega_2,\vec{m}_2\>) = \delta_{\vec{m}_1+\vec{m}_2, \vec{0}}\int_{X} d(\omega_1\wedge \omega_2)=0.
\]
The $G_{\pm}$-preservation
\begin{gather*}
g( G_\pm|\omega_1,\vec{m}_1\> ,|\omega_2,\vec{m}_2\>)- (-1)^{|\omega_1|} g( |\omega_1,\vec{m}_1\> ,G_\pm|\omega_2,\vec{m}_2\>) \\
\qquad{} =\delta_{\vec{m}_1+\vec{m}_2, \vec{0}}\int_{X} \iota^{R,\Phi}_{\vec{m}_1}\omega_1\wedge \omega_2 - (-1)^{|\omega_1|}\delta_{\vec{m}_1+\vec{m}_2, \vec{0}}\int_{X}\omega_1\wedge \iota^{R,\Phi}_{\vec{m}_2}\omega_2\\
\qquad{} = \delta_{\vec{m}_1+\vec{m}_2, \vec{0}}\int_{X} \iota^{R,\Phi}_{\vec{m}_1}(\omega_1\wedge \omega_2) = 0.
\end{gather*}
The pair $(Q, \mu_2)$ for the $A$-model (\ref{A_model_Q}) and (\ref{A_model_mu}) is essentially an exterior derivative and a~wedge product, hence obeys the graded commutativity, associativity, and Leibniz rule. The interior derivative and wedge product properties imply the 7-term relation (\ref{def_7_term_relation}). The space $V$ is infinite-dimensional. Hence, we need to check the properties of the Hamiltonian, which acts as the Lie derivative in radial direction, i.e.,
\[
H|\omega, \vec{m}\> = \{ Q, G_+\} |\omega, \vec{m}\>=\bigl| \bigl\{ {\rm d}, \iota^R_{\vec{m}}\bigr\} \omega, \vec{m}\bigr\>=\bigl| \cl^R_{\vec{m}}\omega, \vec{m}\bigr\>.
\]
The non-perturbative exponent ${\rm e}^{-tH}$ defined as solution (\ref{np_exp}) is a 1-parameter family of diffeomorphisms $\Phi^t_{\vec{m}}\colon r^i\mapsto r^i-m^i t$ acting via pull-back on forms
\[
{\rm e}^{-tH}|\omega,\vec{m}\>=\bigl|\bigl(\Phi^t_{\vec{m}}\bigr)^\ast \omega, \vec{m}\bigr\>.
\]
The composition property (\ref{semi-grp}) naturally holds for diffeomorphisms. Since the vector field $\vec{m} = m^j \p_{r^j}$ is a constant vector field, the corresponding flow does not develop any singularities, hence the composition (\ref{semi-grp}) is valid for all values of $t\in \mathbb{R}^+$.

The $\ker H$ is a direct sum of vector spaces for each $\vec{m}$, i.e.,
\[
\ker H=\bigoplus_{\vec{m} \in \mathbb{Z}^N} \bigl\{|\omega, \vec{m}\>| \ \omega \in \Omega^\ast_{{\rm trop}}(X),\, \cl^R_{\vec{m}} \omega= 0\bigr\}.
\]
Let us fix $\vec{m}$ and describe the invariant forms. An $U(1)^N$-invariant form of a particular bi-degree $\omega \in \Omega^{p,q}_{{\rm trop}}(X)$ can be written in components
\[
\omega=\omega_{i_1\dots i_pj_1\dots j_q}(\vec{r}){\rm d}r^{i_1}\wedge\cdots\wedge {\rm d}r^{i_p}\wedge {\rm d}\phi^{j_1}\wedge \cdots \wedge {\rm d}\phi^{j_q}.
\]
The radial Lie derivative along the {\bf constant} vector field simplifies to the Lie derivative of each component (as a function of $\vec{r}$), i.e.,
\[
\cl^R_{\vec{m}} \omega = {\rm d}r^{i_1}\wedge\cdots \wedge {\rm d}r^{i_p}\wedge {\rm d}\phi^{j_1}\wedge\cdots \wedge {\rm d}\phi^{j_q} m^j \p_{r^j} \omega_{i_1\dots i_pj_1 \dots j_q}(\vec{r} ).
\]
The invariant forms have components independent of $\vec{r}$ in the direction of $\vec{m}$. An extension to compactification (for compact toric variety $X$) requires further restriction for the components of $\omega$, given by
\begin{gather}\label{ker_H_extra}
\iota^R_{\vec{m}} \omega=\iota^\Phi_{\vec{m}} \omega=0.
\end{gather}
When $\vec{m}$ is proportional to one of the vectors $\vec{b} \in B_X$ the extra conditions (\ref{ker_H_extra}) immediately follows from our definition of the tropical forms $\Omega^\ast_{{\rm trop}}(X)$ on $X$, since the $t\to \infty$ limit of the form in (\ref{cont_form_compact}) equal to the form itself. For more general vectors, a more careful analysis is required. Definitions (\ref{A_model_G_-}) and (\ref{A_model_G_+}) for $G_{\pm}$ action on states and extra conditions (\ref{ker_H_extra}) immediately imply $G_{\pm}\Pi_0 = 0$.
\end{proof}

\subsection[A-model states, operators, and amplitudes]{$\boldsymbol{A}$-model states, operators, and amplitudes}

The $A$-model HTQM on special trees is a tropicalization of the type-A topological string. The worldsheet description of the topological string involves the 2D conformal field theory (CFT). One of the key features of the CFT in two dimensions is the state-operator correspondence. The state-operator correspondence in the topological string description of the GW theory becomes the state-operator map in the HTQM description of the tropical GW theory.

Given a state $v \in V$ in HTQM $(V, Q, G_{\pm}, g, \mu_2)$, we can construct an operator
\begin{gather}\label{state_oper_map}
\cO_v = \mu_2(v, \cdot)\colon\ V \to V.
\end{gather}
However, not all operators $\cO\colon V \to V$ can be turned into states, but the ones relevant to the tropical GW theory have such property!
\begin{Definition} For a compactifying hyperplane with the primitive normal vector $\vec{b} \in\mathbb{Z}^N$, there is a {\it divisor state} $\Psi_{b} = \bigl|1, \vec{b}\bigr\>$. For a GW observable $\g\in \Omega_{{\rm trop}}^\ast (X)$, there is an {\it evaluation state} $\Psi_{\g} = \bigl|\g, \vec{0}\bigr\>$.
\end{Definition}
The corresponding operators are defined below.
\begin{Definition} An {\it evaluation observable} is an operator
\[
\cO_\g|\omega, \vec{m}\>=|\g\wedge \omega, \vec{m}\>=\mu_2(\Psi_\g, |\omega,\vec{m}\>),
\]
while a {\it divisor observable} is an operator
\[
\cO_{b} |\omega, \vec{m}\>=\bigl|\omega, \vec{m}+\vec{b}\bigr\>= \mu_2(\Psi_b, |\omega, \vec{m}\>).
\]
\end{Definition}
The divisor states describe the compactification of $\mathbb{C}^{\ast N}$ to the toric variety $X$ in tropical GW theory. Divisor states belong to $\ker H$, moreover,
$
G_\pm\Psi_b = Q\Psi_b=0$.
The evaluation states describe the observables in tropical GW theory. Evaluation states also belong to $\ker H$, and
$
G_\pm \Psi_\g=0$,
while
$
Q\Psi_\g = \Psi_{{\rm d}\g}$.
In Section \ref{sec_htqm_spec_tees}, we defined amplitudes in HTQM on special trees. Below, we provide several examples of the amplitudes in the $A$-model HTQM on special trees related to the tropical GW invariants on~$\mathbb{P}^1$ and~$\mathbb{P}^2$.
\begin{Example} The tropical GW invariant~(\ref{p1_trop_GW_full}) for two points on $\mathbb{P}^1$ can be constructed using the HTQM amplitudes on the special trees below:
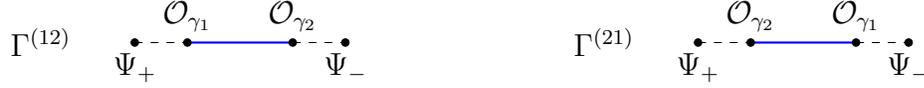
\[
\begin{tikzpicture}[scale=0.7]
\filldraw[black] (-1,0)node[anchor=east]{$\G^{(12)}$} ;
\draw [blue, thick] (1, 0)--(3,0);
\draw[black,dashed](0,0)-- (1, 0);
\draw[black,dashed](3,0)-- (4, 0);
\filldraw[black] (0,0) circle (2pt) node[anchor=north]{$\Psi_+$} ;
\filldraw[black] (1,0) circle (2pt) node[anchor=south]{$\cO_{\g_1}$} ;
\filldraw[black] (3,0) circle (2pt) node[anchor=south]{$\cO_{\g_2}$} ;
\filldraw[black] (4,0) circle (2pt) node[anchor=north]{$\Psi_-$} ;
\end{tikzpicture}\qquad
\begin{tikzpicture}[scale=0.7]
\filldraw[black] (-1,0)node[anchor=east]{$\G^{(21)}$} ;
\draw[blue, thick](1, 0)--(3,0);
\draw[black,dashed](0,0)-- (1, 0);
\draw[black,dashed](3,0)-- (4, 0);
\filldraw[black] (0,0) circle (2pt) node[anchor=north]{$\Psi_+$} ;
\filldraw[black] (1,0) circle (2pt) node[anchor=south]{$\cO_{\g_2}$} ;
\filldraw[black] (3,0) circle (2pt) node[anchor=south]{$\cO_{\g_1}$} ;
\filldraw[black] (4,0) circle (2pt) node[anchor=north]{$\Psi_-$} ;
\end{tikzpicture}
\]
The divisor states $\Psi_\pm$ describe the pair of compactification divisors with primitive normals $b_\pm = \pm 1$. The pair of evaluation observables $\cO_{\g_1}$ and $\cO_{\g_2}$ describe the tropicalized point-cycles~(\ref{p1_tro_observ}) located at $r_1$ and $r_2$.
The amplitude (\ref{A_0_graph_homotopy}) on a special tree $\Gamma^{(12)}$ is
\begin{align}
\mathcal{A}_{\Gamma^{(12)}}(\cO_{\g_1}, \cO_{\g_2}; \Psi_{+}, \Psi_{-})&{}=g(\Psi_+, \cO_{ \g_1} 2\pi K G_- \cO_{ \g_2}\Psi_-)\nonumber\\
&{}=g(|1,1\>, \cO_{ \g_1}2\pi KG_- \cO_{ \g_2} |1, -1\>)=g(|1,1\>, |\gamma_1\Theta (r_2-r) , -1\> ) \nonumber\\
&{}=\int_{{S}^1} {\rm d}\phi\int_{\mathbb{R}}{\rm d}r\, \frac{1}{2\pi}\delta(r-r_1) \Theta (r_2-r)=\Theta (r_2-r_1),\label{ampl_graph_p1_qm}
\end{align}
where we used
\begin{align*}
2\pi KG_-|\gamma_k,-1\>&=2\pi\int^\infty_0{\rm d}t\, {\rm e}^{-tH} G_+ G_- \biggl| \frac{1}{2\pi}\delta(r-r_k){\rm d}\phi {\rm d}r,-1\biggr\> \\
&=\int^\infty_0{\rm d}t\, |\delta(r-r_k + t), -1\>=|\Theta (r_k-r), -1\>.
\end{align*}
The amplitude on a special tree (\ref{ampl_graph_p1_qm}) is identical to the tropical GW invariant contribution~(\ref{p1_trop_GW}) for the same tropical curve.
\end{Example}
\begin{Example} Let us describe the amplitudes for the graphs, corresponding to tropical curves of degree 1 in $\mathbb{P}^2$ with two observables, which we used to evaluate the tropical GW invariants in~Section~\ref{moduli_space_integr_section}:
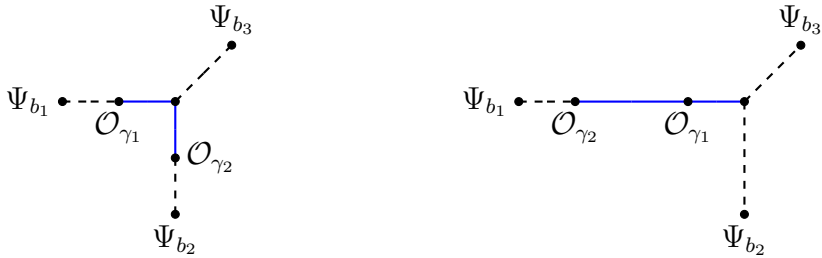
\[
\begin{tikzpicture}[scale=0.75]
\draw[color=black, thick, dashed] (-2, 0) -- (-1.5, 0);
\draw[color=black, thick, dashed] (-1.5, 0) -- (-1, 0);
\draw[color=blue, thick] (-1, 0) -- (-.5, 0);
\draw[color=blue, thick] (-0.5, 0) -- (0, 0);
\draw[color=black, thick, dashed] (0, -2) -- (0,-1.5);
\draw[color=black, thick, dashed] (0, -1) -- (0,-1.5);
\draw[color=blue, thick] (0, -1) -- (0, -0.5);
\draw[color=blue, thick] (0, -0.5) -- (0,0);
\draw[color=black, thick, dashed](1,1)--(0.5, 0.5);
\draw[color=black, thick, dashed](0.5, 0.5)-- (0,0);
\filldraw[black] (0,0) circle (2pt);
\filldraw[black] (-1,0) circle (2pt) node[anchor=north]{$\cO_{\g_1}$};
\filldraw[black] (0,-1) circle (2pt) node[anchor=west]{$\cO_{\g_2}$};
\filldraw[black] (1,1) circle (2pt) node[anchor=south]{$\Psi_{b_3}$};
\filldraw[black] (-2,0) circle (2pt) node[anchor=east]{$\Psi_{b_1}$};
\filldraw[black] (0,-2) circle (2pt) node[anchor=north]{$\Psi_{b_2}$};
\end{tikzpicture}\qquad\qquad\qquad
\begin{tikzpicture}[scale=0.75]
\draw[color=black, thick, dashed] (-2, -1) -- (-1.5, -1);
\draw[color=black, thick, dashed] (-1.5, -1) -- (-1, -1);
\draw[color=blue, thick] (-1, -1) -- (0, -1);
\draw[color=blue, thick] (0, -1) -- (1, -1);
\draw[color=blue, thick] (1, -1) -- (1.5, -1);
\draw[color=blue, thick] (1.5, -1) -- (2, -1);
\draw[color=black, thick, dashed](2, -1)-- (2,-3);
\draw[color=black, thick, dashed](2, -1)-- (2+1,-1+1);
\filldraw[black] (2,-1) circle (2pt);
\filldraw[black] (-1,-1) circle (2pt) node[anchor=north]{$\cO_{\g_2}$};
\filldraw[black] (1,-1) circle (2pt) node[anchor=north]{$\cO_{\g_1}$};
\filldraw[black] (2+1,-1+1)circle (2pt) node[anchor=south]{$\Psi_{b_3}$};
\filldraw[black] (-2,-1) circle (2pt) node[anchor=east]{$\Psi_{b_1}$};
\filldraw[black] (2,-3) circle (2pt) node[anchor=north]{$\Psi_{b_2}$};
\end{tikzpicture}
\]
The amplitude for the left graph is
\begin{gather}\label{aml_tree_htqm_p2_2pt}
\mathcal{A}_\Gamma(\cO_{\g_\a};\Psi_{b_a})=g\bigl(\Psi_{b_3},\mu_2\bigl( 2\pi KG_-\cO_{\g_1}\Psi_{b_1}, 2\pi KG_-\cO_{\g_2}\Psi_{b_2}\bigr)\bigr)= \Theta \bigl(-r^1\bigr) \Theta \bigl(r^2\bigr).
\end{gather}
We used the following relations:
\begin{gather*}
2\pi KG_-\cO_{\g_1}\Psi_{b_1} = 2\pi \int^\infty_0 {\rm d}t\,\frac{1}{(2\pi)^2}\bigl|\delta \bigl(r^1-r^1_1-t\bigr) \delta \bigl(r^2 - r^2_1\bigr) {\rm d}r^2 {\rm d}\phi^2, (1,0)\bigr\> \\ \hphantom{2\pi KG_-\cO_{\g_1}\Psi_{b_1}}{}
 = \frac{1}{2\pi}|\Theta \bigl(r^1-r^1_1\bigr) \delta \bigl(r^2 - r^2_1\bigr) {\rm d}r^2 {\rm d}\phi^2, (1,0)\>,\\
2\pi KG_-\cO_{\g_2}\Psi_{b_2} = 2\pi \int^\infty_0 {\rm d}t\, \frac{1}{(2\pi)^2}\bigl|\delta \bigl(r^1-r^1_2\bigr) \delta \bigl(r^2 - r^2_2-t\bigr) {\rm d}r^1 {\rm d}\phi^1, (0,1)\bigr\> \\ \hphantom{2\pi KG_-\cO_{\g_2}\Psi_{b_2}}{}
 = \frac{1}{2\pi} \bigl|\delta \bigl(r^1-r^1_2\bigr) \Theta \bigl(r^2 - r^2_2\bigr) {\rm d}r^1 {\rm d}\phi^1, (0,1)\bigr\>.
\end{gather*}
Again, the amplitude (\ref{aml_tree_htqm_p2_2pt}) on a special tree is identical to the tropical GW invariant contribution (\ref{cp_2_trop_GW_G_2}) for the corresponding tropical curve. These examples are a consequence of a more general relation between the $A$-model HTQM amplitudes and tropical GW invariants, which we formulate and prove in the next section.
\end{Example}

\subsection{Tropical GW invariants via HTQM amplitudes}
The theorem gives the HTQM representation for tropical GW invariants.
\begin{Theorem} \label{theo_htqm_rep_trop_gw}
An individual contribution for tropical Gromov--Witten invariant of degree $\beta\neq 0$ on toric variety $X$ and for $n\geq 3$ from the tropical curve of a given discrete type $\vec{\G}$ equal to the amplitude in $A$-model HTQM on special tree $\G$. In particular,
\begin{gather}\label{TGW_via_QM}
\int_{\mathcal{M}_{0,n}(X, \beta;\vec{\G})} \bigwedge_{\a=1}^n {\rm ev}^\ast_\a \g_\a = \mathcal{A}_\Gamma \bigl( \cO_{\g_1},\dots,\cO_{\g_n}; \underbrace {\Psi_{b_1},\dots,\Psi_{b_1}}_{d_1},\dots, \underbrace{\Psi_{b_{B}
},\dots,\Psi_{b_{B}}}_{d_{B}}\bigr),
\end{gather}
where the number $d_{a}$ of divisor states $\Psi_{b_a}$ is given by the tropical intersection number of the degree class $\beta$ and the corresponding hyperplane, i.e.,
\[
d_{a} = \beta \cdot H_{b_a}.
\]
The index $a$ runs from $1$ to cardinality $B = |B_X|$ of the set of compactifying divisors $B_X$.
\end{Theorem}
\begin{proof}We will prove the equality (\ref{TGW_via_QM}) by explicit evaluation of the HTQM amplitude. By definition (\ref{A_0_graph_homotopy_short}), amplitude on the right-hand side is
\begin{gather}\label{a_model_htqm_trop_gw}
\mathcal{A}_\Gamma (\cO_{\g_a}; \Psi_{b_a})=\Biggl\<(2\pi K G_-)^{\otimes I(\G)} \bigotimes\limits_{a=1}^{n_1(\G)}\Psi_{b_a}\bigotimes\limits_{\a=1}^{n_2(\G)} \cO_{\g_\a} \otimes \mu_2^{\otimes n_3(\G)} \Biggr\>_\Gamma.
\end{gather}
Using homotopy representation (\ref{def_HTQM_propagator}) and an additional angular variable $\varphi$, we rewrite an integral representation
\[
2\pi K G_- = 2\pi \int^\infty_0 {\rm d}\tau {\rm e}^{-\tau H} G_+G_-=\int_{S^1 \times \mathbb{R}^+} {\rm e}^{-\tau H - {\rm d}\tau G_+-{\rm d}\varphi G_-} .
\]
We can introduce an operator $\mathcal{U}(\tau, \varphi)$ on HTQM states such that
\[
\mathcal{U} (\tau, \varphi)|\omega, \vec{m}\>=\exp (-\tau H-{\rm d}\tau G_+-{\rm d}\varphi G_- )|\omega, \vec{m}\>.
\]
The amplitude (\ref{a_model_htqm_trop_gw}) has an integral representation
\begin{gather}\label{integral_rep_ampl}
\mathcal{A}_\Gamma (\cO_\a; \Psi_{b_a})=\int_{(\mathbb{R}^+\times S^1)^{I(\G)}} \Biggl\<\mathcal{U}^{\otimes I(\G)} \bigotimes\limits_{a=1}^{n_1(\G)}\Psi_{b_a}\bigotimes\limits_{\a=1}^{n_2(\G)} \cO_{\g_\a} \otimes \mu_2^{\otimes n_3(\G)} \Biggr\>_\Gamma.
\end{gather}
Note that there is a difference between the integral representation for the amplitude in Section~\ref{sec_htqm_spec_tees} and in the present proof. The operator $\mathcal{U} (\tau, \varphi)$ acts as {\it super-pullback map} on differential form part of the $A$-model HTQM states, i.e.,
\[
\mathcal{U}(\tau, \varphi)|\omega, \vec{m}\>=\bigl| \exp \bigl(-\tau \cl^R_{\vec{m}}+\iota^R_{\vec{m} } {\rm d}\tau+\iota^\Phi_{\vec{m}} {\rm d}\v \bigr)\omega,\vec{m}\bigr\>.
\]
Earlier, we evaluated the exponentiation of the Lie derivative along the constant vector field in terms of a transformation $ F_{\vec{m}}^\tau\colon \bigl(\vec{r}, \vec{\phi}\,\bigr)\mapsto \bigl(\vec{r} - \vec{m}\tau, \vec{\phi}\,\bigr)$
\[
{\rm e}^{-\tau \cl^R_{\vec{m}}}\omega \bigl(\vec{r}, {\rm d}\vec{r}, {\rm d}\vec{\phi}\,\bigr)= \bigl(F^\tau_{\vec{m}}\bigr)^\ast\omega\bigl(\vec{r}, {\rm d}\vec{r}, {\rm d}\vec{\phi}\,\bigr)= \omega\bigl(\vec{r} -\vec{m} \tau , {\rm d}\vec{r}, {\rm d}\vec{\phi}\,\bigr).
\]
The additional $G_+d\tau + G_- d\v$ terms in the $\mathcal{U} (\tau, \varphi)$ operator modify the pullback $(F^\tau_{\vec{m}})^\ast $ into
\begin{gather}\label{super_pullback}
\exp \bigl(-\tau \cl^R_{\vec{m} } + \iota^R_{\vec{m} } {\rm d}\tau + \iota^\Phi_{\vec{m}} {\rm d}\v \bigr) \omega \bigl(\vec{r}, {\rm d}\vec{r}, {\rm d}\vec{\phi}\,\bigr)=\omega \bigl(\vec{r}-\vec{m} \tau, {\rm d}\vec{r}- \vec{m} {\rm d}\tau, {\rm d}\vec{\phi} -\vec{m} {\rm d}\v\bigr).
\end{gather}
We provide a geometric description for (\ref{super_pullback}) by introducing a version of the tropical evaluation map (\ref{ev_map_trop}) on a single edge
\[
{\rm ev}_{m}\colon \ \bigl(\vec{r}, \vec{\phi}, \vec{m}, \tau, \varphi\bigr) \mapsto \bigl(\vec{r} - \vec{m}\tau, \vec{\phi} - \vec{m}\v\bigr).
\]
The action of $\mathcal{U} (\tau, \varphi)$ operator evaluates into
\begin{gather}\label{oper_u_eval_map}
\mathcal{U} (\tau, \varphi) |\omega, \vec{m}\> = \bigl| {\rm ev}_m^\ast \omega, \vec{m}\bigr\>.
\end{gather}
The action of the $2\pi G_-K$ on an internal edge $e \in I(\G)$ is replaced by
\[
2\pi K G_-|\omega, \vec{m}_e\>=\int_{S^1 \times \mathbb{R}^+} \bigl| {\rm ev}_{m_e}^\ast \omega, \vec{m}_e\bigr\>.
\]
We use (\ref{oper_u_eval_map}) to rearrange the HTQM amplitude according to the following rules:
\begin{itemize}\itemsep=0pt
\item 3-valent vertex rearrangement
\begin{gather}\label{rearr_3_valent}
\mathcal{U} \mu_2 (|\omega_1, \vec{m}_1\> ,|\omega_2, \vec{m}_2\> )=\mu_2 \bigl(\bigl|{\rm ev}^\ast_{m_1+m_2}\omega_1, \vec{m}_1\bigr\> , \bigl|{\rm ev}^\ast_{m_1+m_2}\omega_2, \vec{m}_2\bigr\>\bigr);
\end{gather}
\item 2-valent vertex rearrangement
\begin{gather*}
\mathcal{U} \cO_{\g_\a} |\omega, \vec{m}\>=\mathcal{U}|\g_\a\wedge \omega, \vec{m}\>=\cO_{{\rm ev}_m^\ast \g_\a}|{\rm ev}_m^\ast \omega, \vec{m}\> ;
\end{gather*}
\item 1-valent vertices
\begin{gather}\label{rearr_1_valent}
\mathcal{U} \Psi_b=\mathcal{U} \bigl|1, \vec{b}\bigr\>=\bigl|{\rm ev}_b^\ast 1, \vec{b}\bigr\>=\bigl|1, \vec{b}\bigr\>=\Psi_b.
\end{gather}
\end{itemize}
Hence, we gradually replace the $\mathcal{U}$ operators by the evaluation map pullbacks, starting from the sink vertex and proceeding in the direction of leaves, recursively applying the (\ref{rearr_3_valent})--(\ref{rearr_1_valent}) to rearrange the form (\ref{integral_rep_ampl}) into
\begin{gather}\label{pre_eval}
\Biggl\<\mathcal{U}^{\otimes I(\G)} \bigotimes\limits_{a=1}^{n_1(\G)}\Psi_{b_a}\bigotimes\limits_{\a=1}^{n_2(\G)} \cO_{\g_\a}\otimes \mu_2^{\otimes n_3(\G)}\Biggr>_\Gamma\!=\Biggl\<\mathds{1}^{\otimes I(\G)}\bigotimes\limits_{a=1}^{n_1(\G)}\Psi_{b_a}\bigotimes\limits_{\a=1}^{n_2(\G)}\cO_{\g^{\gamma_\G}_\a}\otimes\mu_2^{\otimes n_3(\G)}\Biggr\>_\Gamma\!.\!\!
\end{gather}
We introduced the modified observables defined by
\[
\g^{\gamma_\G}_\a=\biggl[\prod_{e\in\gamma_\Gamma(R,\a)}{\rm ev}^\ast_{m_e}\biggr] \g_\a.
\]
The $ \gamma_\Gamma(R, \a)$ is the unique path on special tree $\Gamma$ from the sink vertex $R$ to a vertex $\a$, decorated with $\cO_{\g_a}$. The composition of evaluation maps along the path $ \gamma_\Gamma (R, \a)$ is an evaluation map~(\ref{ev_map_trop}) on the moduli space of tropical curves with marked points, i.e.,
\[
\g^{\gamma_\G}_\a={\rm ev}^\ast_\a \g_\a.
\]
The right-hand side in (\ref{pre_eval}) has trivial decorations on edges and hence can be immediately evaluated into
\begin{gather}\label{simpl_integrand_path_tree}
\Biggl\<\mathds{1}^{\otimes I(\G)} \bigotimes\limits_{a=1}^{n_1(\G)} \Psi_{b_a}\bigotimes\limits_{\a=1}^{n_2(\G)} \cO_{\g^{\gamma_\G}_\a} \otimes \mu_2^{\otimes n_3(\G)} \Biggr\>_\Gamma = \int_{X}\bigwedge_\Gamma {\rm ev}^\ast_\a \g_\a.
\end{gather}
The $\bigwedge_\Gamma$
stands for a particular order in wedge product, determined by the path on a special tree~$\Gamma$. The forms $\g_\a$ represent the Poincar\'e duals to complex cycles. Hence, they are forms of even degree, and the order of product does not affect the result, i.e.,
\[
\int_{X}\bigwedge_\Gamma {\rm ev}^\ast_\a \g_\a= \int_{X}\bigwedge\limits_{\a=1}^n {\rm ev}^\ast_\a \g_\a.
\]
The integrals in (\ref{integral_rep_ampl}) and (\ref{simpl_integrand_path_tree}) combine into the integral over the tropical moduli space (\ref{trop_moduli_marked}) of the tropical curve with discrete data $\G$, i.e.,
\[
\mathcal{A}_\Gamma (\cO_{\g_\a}; \Psi_{b_a})=\int_{(\mathbb{R}^+\times S^1)^{I(\G)}} \int_{X}\bigwedge\limits_{\a=1}^n {\rm ev}^\ast_\a \g_\a = \int_{\mathcal{M}_{0,n} \bigl(X,\beta; \vec{\Gamma}\bigr)} \bigwedge\limits_{\a=1}^n {\rm ev}^\ast_\a \g_\a.
\]
The proof is complete.
\end{proof}

\begin{Remark} The string theory version of Theorem~\ref{theo_htqm_rep_trop_gw} was discussed in \cite{Frenkel:2005ku}. The 2d CFT was constructed, universal for all GW invariants of a toric target $X$. The GW invariant at a~fixed degree was constructed using a particular number of holomortex vertex operators and the vertex operators for observables $\g_\a$. We want to emphasize that the tropicalization of the GW invariants modifies the 2d CFT construction to the HTQM construction, which is much simpler, better understood, and rigorously formalized.
\end{Remark}

\subsection{HTQM for tropical multiplicities}

The recent progress in tropical geometry description for the GW invariants is due to the equivalence of the complex curve counting to the real tropical curve counting with additional multiplicity factors. For a tropical curve $\vec{\G}$ of genus-zero in $\mathbb{R}^2$, the multiplicity is a product of multiplicities at each 3-valent vertex, i.e.,
\begin{gather}\label{eq_Mikhalk_miltipl} 
{\rm mult}\bigl(\vec{\G}\bigr) = \prod_{v \in V_3(\G)} {\rm mult}(v).
\end{gather}
The multiplicity for a vertex $v$ is a vector product of the integer vectors, that is, ${\rm mult}(v) = |\vec{m}_{e_1}\wedge \vec{m}_{e_2}|$ for a pair of edges attached to this vertex. The balancing condition (\ref{vert_bal_con}) ensures that the multiplicity does not depend on the choice of a pair among the three attached edges. The tropical GW invariant for the toric surface $X$ in the Mikhalkin presentation is
\begin{gather}\label{eq_trop_gw_mikhalkin}
TN^X_\beta (C_1,\dots, C_n) = \sum_{\vec{\G}} {\rm mult}\bigl(\vec{\G}\bigr) \cdot \Biggl| \int_{\mathcal{M}^R_{0,n} \bigl(X,\beta; \vec{\Gamma}\bigr)} \bigwedge\limits_{\a=1}^n {\rm ev}^{R\ast}_\a \g^R_\a\Biggr|
\end{gather}
Here $\mathcal{M}^R_{0,n} \bigl(X,\beta; \vec{\Gamma}\bigr)$ is the radial part moduli space (\ref{trop_moduli_marked_components}) for the tropical curves with $n$ marked points of a given combinatorial type $\vec{\G}$. We introduced a radial restriction ${\rm ev}^R$ of the tropical evaluation map (\ref{ev_map_trop}), introduced in Section~\ref{trop_eval_section}, i.e.,
\[
{\rm ev}_\a^R\colon \ \mathcal{M}^R_{0,n}\bigl(X,\beta;\vec{\Gamma}\bigr) \to \mathbb{R}^2\colon \bigl(\vec{r}_c, \tau_1,\dots,\tau_{I(\Gamma)}\bigr) \mapsto \vec{r}_\a=\vec{r}_c + \sum_{e\in \gamma_\G (R, \a)}\pm \vec{m}_e \tau_e.
\]
The $\g_\a^R$ is the radial part of the Poincar\'e dual for the point observable (\ref{point_class_average}), that is, $\g_\a^R=\delta^2(\vec{r} -\vec{r}_\a) {\rm d}r^1{\rm d}r^2$. The moduli space integral in (\ref{eq_trop_gw_mikhalkin}) is the number of tropical curves with $n$ marked points of given combinatorial type $\vec{\G}$, passing through the $n$ points on $\mathbb{R}^2$ at $\vec{r}_1,\dots, \vec{r}_n$.

Our definition of the tropical GW invariants (\ref{form_def_trop_GW_invariants}) is a tropicalization of the Kontsevich--Manin construction (\ref{top_compon_GW}), but it looks different from Mikhalkin's (\ref{eq_trop_gw_mikhalkin}). Using the $U(1)^2$ averaged representatives (\ref{point_class_average}) for $\g_\a$ the moduli space integral in (\ref{form_def_trop_GW_invariants}) factorizes into radial and angular part. The radial part is the moduli space integral in the sum (\ref{eq_trop_gw_mikhalkin}), while the angular part provides the weights. We will denote these weights a Kontsevich--Manin(KM) multiplicities, ${\rm mult}^{\rm KM}\bigl(\vec{\G}\bigr)$. By construction, the KM multiplicies are given by the angular part of the module space integral, i.e.,
\begin{gather}\label{eq_def_KM_multipl}
{\rm mult}^{\rm KM}\bigl(\vec{\G}\bigr) = \Biggl| \int_{\mathcal{M}^\Phi_{0,n} \bigl(X,\beta; \vec{\Gamma}\bigr)} \bigwedge\limits_{\a=1}^n {\rm ev}^{\Phi\ast}_\a \g^\Phi_\a\Biggr|.
\end{gather}
Here $\mathcal{M}^\Phi_{0,n} \bigl(X,\beta; \vec{\Gamma}\bigr)$ is the angular part moduli space (\ref{trop_moduli_marked_components}) for the tropical curves with $n$ marked points of a given combinatorial type $\vec{\G}$. We introduced an angular restriction ${\rm ev}^\Phi$ of the tropical evaluation map (\ref{ev_map_trop}) introduced in Section \ref{trop_eval_section}, i.e.,
\[
{\rm ev}^\Phi_\a\colon \ \mathcal{M}^\Phi_{0,n}\bigl(X,\beta;\vec{\Gamma}\bigr) \to S^1\times S^1\colon (\vec{\phi}_c,\varphi_1,\dots,\varphi_{I(\Gamma)}) \mapsto \vec{\phi}_\a = \vec{\phi}_c + \sum_{e\in \gamma_\G (R, \a)} \pm \vec{m}_e \v_e.
\]
The $\g_\a^\Phi$ is the angular part of the Poincar\'e dual for the point observable (\ref{point_class_average}). Note that the angular parts are independent of the location of the points, i.e.,
\[
\g_\a^\Phi =\g^\Phi = \frac{1}{(2\pi)^2} {\rm d}\phi^1{\rm d}\phi^2.
\]
We use the $A$-model HTQM, defined in Section \ref{TGQ_QM_section} to represent the angular part of the GW invariant. For a directed graph $\vec{\G}$, we introduce the angular amplitudes
\[
\mathcal{A}^\Phi_\Gamma (\cO_\a; v_a)=\Biggl\<(2\pi G_-)^{\otimes I(\G)} \bigotimes\limits_{a=1}^{n_1(\G)}v_a\bigotimes\limits_{\a=1}^{n_2(\G)} \cO_\a \otimes \mu_2^{\otimes n_3(\G)}\Biggr\>_\Gamma.
\]
The key difference to the full amplitude (\ref{A_0_graph_homotopy_short}) is the change of the propagator on internal edges from $KG_-$ to just $G_-$.
\begin{Proposition} For the tropical curve $\vec{\G}$ of degree $\beta$ in toric surface $X$, the KM multiplicity equal to the angular $A$-model HTQM amplitude
\begin{gather}\label{eq_htqm_rep_km_multipl}
{\rm mult}^{\rm KM}\bigl(\vec{\G}\bigr) = \biggl|\mathcal{A}^\Phi_\Gamma ( \cO_{\g^\Phi_1},\dots,\cO_{\g^\Phi_n}; \underbrace {\Psi_{b_1},\dots,\Psi_{b_1}}_{d_1},\dots, \underbrace{\Psi_{b_{B}
},\dots,\Psi_{b_{B}}}_{d_{B}})\biggr|,
\end{gather}
where the number $d_{a}$ of divisor states $\Psi_{b_a}$ is given by the tropical intersection number of the degree class $\beta$ and the corresponding hyperplane, i.e., $d_{a} = \beta \cdot H_{b_a}$.
\end{Proposition}
\begin{proof} The proposition is similar to Theorem \ref{theo_htqm_rep_trop_gw}, so its proof is a modified version of the theorem proof.
\end{proof}

For the toric surface and point observables, the angular $A$-model HTQM amplitude is simple enough, so we can evaluate it explicitly to show that it equals to the Mikhalkin multiplicity~(\ref{eq_Mikhalk_miltipl}).
\begin{Proposition} For the tropical curve $\vec{\G}$, the KM multiplicity is identical to the Mikhalkin multiplicity, i.e.,
\[
{\rm mult}^{\rm KM}\bigl(\vec{\G}\bigr) = \prod_{v \in V_3(\G)} {\rm mult}(v).
\]
\end{Proposition}
\begin{proof} We use the proposition to represent the KM multiplicity as angular $A$-model HTQM amplitude on a special tree $\G$. For a special tree $\G$, we choose a 3-valent vertex $v$, connected to the two 1-valent vertices. There could be some number of 2-valent vertices in between. In the vicinity of $v$, the special tree $\G$ has the schematic form depicted on the left side below:
\[
\begin{tikzpicture}[scale=0.6]
\filldraw[black] (-3,1.5)node[anchor=east]{$\G$};
\draw(-3,1)-- (-4, 1) ;
\draw(-3,1)-- (-3, 2) ;
\draw[pattern=north east lines, pattern color=black] (1.8,1.8) circle (0.45);
\draw [color=black, thick] (0, 0) -- (1.5, 1.5);
\draw [color=black, thick] (-2.1,0)-- (0, 0) ;
\draw [color=black, thick, dashed] (-3,0)-- (-2.1, 0) ;
\draw [color=black, thick, dashed] (0, -3) -- (0, -2.1);
\draw [color=black, thick] (0, -2.1) -- (0, 0);
\filldraw[black] (0,0) circle (2pt) node[anchor=south]{$\mu_2$};
\filldraw[black] (-3,0) circle (2pt) node[anchor=north]{$\Psi_{b_1}$};
\filldraw[black] (-2.1,0) circle (2pt) node[anchor=north]{$\cO^\Phi_{\g}$};
\filldraw[black] (-1.4,0) circle (2pt);
\filldraw[black] (-1.8,0) circle (2pt);
\filldraw[black] (-0.9,0) circle (2pt) node[anchor=north]{$\cO^\Phi_{\g}$};
\filldraw[black] (0,-3) circle (2pt) node[anchor=west]{$\Psi_{b_2}$};
\filldraw[black] (0,-2.1) circle (2pt) node[anchor=west]{$\cO^\Phi_{\g}$};
\filldraw[black] (0,-1.4) circle (2pt);
\filldraw[black] (0,-1.8) circle (2pt);
\filldraw[black] (0,-0.9) circle (2pt) node[anchor=west]{$\cO^\Phi_{\g}$};
\filldraw[black] (0.5,0.5) circle (2pt);
\filldraw[black] (0.8,0.8) circle (2pt);
\filldraw[black] (1.2,1.2) circle (2pt) node[anchor=north]{$\;\cO^\Phi_{\g}$};
\end{tikzpicture}
\begin{tikzpicture}[scale=0.7]
\draw[pattern=north east lines, pattern color=black] (1.3,1.3) circle (0.45);
\draw [color=black, thick] (0, 0) -- (1, 1);
\draw [color=black, thick] (-1,0)-- (0, 0) ;
\draw [color=black, thick, dashed] (-2,0)-- (-1, 0) ;
\draw [color=black, thick, dashed] (0, -2) -- (0, -1);
\draw [color=black, thick] (0, -1) -- (0, 0);
\filldraw[black] (0,0) circle (2pt) node[anchor=south]{$\mu_2$};
\filldraw[black] (-2,0) circle (2pt) node[anchor=north]{$\Psi_{b_1}$};
\filldraw[black] (-1,0) circle (2pt) node[anchor=north]{$\cO^\Phi_{\g}$};
\filldraw[black] (0,-2) circle (2pt) node[anchor=west]{$\Psi_{b_2}$};
\filldraw[black] (0,-1) circle (2pt) node[anchor=west]{$\cO^\Phi_{\g}$};
\end{tikzpicture}
\begin{tikzpicture}[scale=0.7]
\draw[pattern=north east lines, pattern color=black] (1.3,1.3) circle (0.45);
\draw [color=black, thick] (0, 0) -- (1, 1);
\draw [color=black, thick] (-1,0)-- (0, 0) ;
\draw [color=black, thick, dashed] (-2,0)-- (-1, 0) ;
\draw [color=black, thick, dashed] (0, -2) -- (0, 0);
\filldraw[black] (0,0) circle (2pt) node[anchor=south]{$\mu_2$};
\filldraw[black] (-2,0) circle (2pt) node[anchor=north]{$\Psi_{b_1}$};
\filldraw[black] (-1,0) circle (2pt) node[anchor=north]{$\cO^\Phi_{\g}$};
\filldraw[black] (0,-2) circle (2pt) node[anchor=west]{$\Psi_{b_2}$};
\filldraw[black] (0.6,0.6) circle (2pt) node[anchor=north]{$\;\;\cO^\Phi_{\g}$};
\end{tikzpicture}\qquad
\begin{tikzpicture}[scale=0.6]
\filldraw[black] (-2,2.5)node[anchor=east]{$\G'$};
\draw(-2,2)-- (-3, 2) ;
\draw(-2,2)-- (-2, 3) ;
\draw[pattern=north east lines, pattern color=black] (2.3,2.3) circle (0.45);
\draw [color=black, thick] (0, 0) -- (2, 2);
\filldraw[black] (1.2,1.2) circle (2pt) node[anchor=north]{$\;\cO^\Phi_{\g}$};
\filldraw[black] (0,0) circle (2pt) node[anchor=north]{$\Psi_{b_1+b_2}$};
\filldraw[black] (-2.5,-2.5) node[anchor=west]{$$};
\end{tikzpicture}
\]
The filled circle starts with a 3-valent vertex, and each edge is decorated with an arbitrary number of evaluation observables. The non-vanishing amplitudes have at most two evaluation observables on the subtree outside the filled circle. Below, we show that all possible incoming states for the subtrees depicted by the filled circle are proportional to $G_- \cO_{\g^\Phi} \Psi_{b_1+b_2}$. Hence, there is a recursive relation between the amplitudes on special tree $\G$ and a smaller special tree~$\G'$ obtained from $\G$ by removing a vertex $v$, i.e.,
\begin{gather}
 \bigl|\mathcal{A}^\Phi_\Gamma \bigl(\cO_{\g^\Phi_1},\dots,\cO_{\g^\Phi_n};\Psi_{b_1},\dots,\Psi_{b_{n+1}}\bigr)\bigr|\nonumber\\
 \qquad {} = |b_1\wedge b_2|
\bigl|\mathcal{A}^\Phi_{\Gamma'} \bigl(\cO_{\g^\Phi_1},\dots,\cO_{\g^\Phi_{n-1}};\Psi_{b_1+b_2},\Psi_{b_3},\dots,\Psi_{b_{n+1}}\bigr)\bigr|.\label{eq_recursion_angula_ampl}
\end{gather}
Note that the factor $ |b_1\wedge b_2|$ is the vertex multiplicity for the vertex $v$ that we removed. The~matching between the moduli space dimension (\ref{trop_mod_dim}) and the total degree of the form for KM multiplicity integral (\ref{eq_def_KM_multipl}) implies that the $n$ point evaluation observables on toric hypersurface require $n+1$ leaves. At each recursion step, we remove one 3-valent vertex, one evaluation observable, and one tree leaf. Hence, after we eliminate all 3-valent vertices, we end up with the special tree depicted below:
\[
\begin{tikzpicture}[scale=0.7]
\draw[black,dashed](0,0)-- (2, 0);
\draw[black,dashed](2,0)-- (4, 0);
\filldraw[black] (0,0) circle (2pt) node[anchor=north]{$\Psi_{\sum b_k}$} ;
\filldraw[black] (2,0) circle (2pt) node[anchor=south]{$\cO_{\g}^\Phi$} ;
\filldraw[black] (4,0) circle (2pt) node[anchor=north]{$\Psi_{b_{n+1}}$} ;
\end{tikzpicture}
\]
The corresponding amplitude vanishes unless the sum of all vectors $\vec{b}_k$ is zero, otherwise
\[
 \mathcal{A}^\Phi_\Gamma \bigl(\cO_{\g^\Phi};\Psi_{\sum b_k},\Psi_{b_{n+1}}\bigr) = \int_{S^1\times S^1} \gamma^\Phi =1.
\]
Hence, the recursion formula (\ref{eq_recursion_angula_ampl}) allows us to express the amplitude as a product of vertex multiplicities, i.e.,
\[
 \mathcal{A}^\Phi_\Gamma \bigl(\cO_{\g^\Phi_1},\dots,\cO_{\g^\Phi_n};\Psi_{b_1},\dots,\Psi_{b_{n+1}}\bigr) = \prod_{v \in V(\G)} {\rm mult}(v).
\]
To prove the recursion formula (\ref{eq_recursion_angula_ampl}) we first consider a state
\[
2\pi G_- \cO_\g^\Phi |1, \vec{m}\> = 2\pi G_- \bigl| \gamma^\Phi, \vec{m}\bigr\> =\frac{1}{2\pi} \bigl| \vec{m}\wedge {\rm d}\vec{\phi}, \vec{m}\bigr\>.
\]
The state $ \bigl| \vec{m}\wedge {\rm d}\vec{\phi}, \vec{m}\bigr\>$ is a 1-form so a multiplication of it by $\cO_{\g^\Phi}$ equals zero. Hence, the angular amplitude is non-zero only for special trees with at most one 2-valent vertex between~3- and~1-valent ones. Therefore, special trees with non-zero amplitudes are among the five types depicted below:
\[
\begin{tikzpicture}[scale=0.65]
\draw[pattern=north east lines, pattern color=black] (1.3,1.3) circle (0.45);
\draw [color=black, thick, dashed] (-2,0)-- (0, 0) ;
\draw [color=black, thick, dashed] (0, -2) -- (0, 0);
\draw [color=black, thick] (0, 0) -- (1, 1);
\filldraw[black] (0,0) circle (2pt) node[anchor=south]{$\mu_2$};
\filldraw[black] (-2,0) circle (2pt) node[anchor=north]{$\Psi_{b_1}$};
\filldraw[black] (0,-2) circle (2pt) node[anchor=west]{$\Psi_{b_2}$};
\end{tikzpicture}
\begin{tikzpicture}[scale=0.65]
\draw[pattern=north east lines, pattern color=black] (1.3,1.3) circle (0.45);
\draw [color=black, thick, dashed] (-2,0)-- (0, 0) ;
\draw [color=black, thick, dashed] (0, -2) -- (0, 0);
\draw [color=black, thick] (0, 0) -- (1, 1);
\filldraw[black] (0,0) circle (2pt) node[anchor=south]{$\mu_2$};
\filldraw[black] (-2,0) circle (2pt) node[anchor=north]{$\Psi_{b_1}$};
\filldraw[black] (0,-2) circle (2pt) node[anchor=west]{$\Psi_{b_2}$};
\filldraw[black] (0.6,0.6) circle (2pt) node[anchor=north]{$\;\;\cO^\Phi_{\g}$};
\end{tikzpicture}
\begin{tikzpicture}[scale=0.65]
\draw[pattern=north east lines, pattern color=black] (1.3,1.3) circle (0.45);
\draw [color=black, thick] (0, 0) -- (1, 1);
\draw [color=black, thick] (-1,0)-- (0, 0) ;
\draw [color=black, thick, dashed] (-2,0)-- (-1, 0) ;
\draw [color=black, thick, dashed] (0, -2) -- (0, 0);
\filldraw[black] (0,0) circle (2pt) node[anchor=south]{$\mu_2$};
\filldraw[black] (-2,0) circle (2pt) node[anchor=north]{$\Psi_{b_1}$};
\filldraw[black] (-1,0) circle (2pt) node[anchor=north]{$\cO^\Phi_{\g}$};
\filldraw[black] (0,-2) circle (2pt) node[anchor=west]{$\Psi_{b_2}$};
\filldraw[black] (0.6,0.6) circle (2pt) node[anchor=north]{$\;\;\cO^\Phi_{\g}$};
\end{tikzpicture}
\begin{tikzpicture}[scale=0.65]
\draw[pattern=north east lines, pattern color=black] (1.3,1.3) circle (0.45);
\draw [color=black, thick] (0, 0) -- (1, 1);
\draw [color=black, thick] (-1,0)-- (0, 0) ;
\draw [color=black, thick, dashed] (-2,0)-- (-1, 0) ;
\draw [color=black, thick, dashed] (0, -2) -- (0, -1);
\draw [color=black, thick] (0, -1) -- (0, 0);
\filldraw[black] (0,0) circle (2pt) node[anchor=south]{$\mu_2$};
\filldraw[black] (-2,0) circle (2pt) node[anchor=north]{$\Psi_{b_1}$};
\filldraw[black] (-1,0) circle (2pt) node[anchor=north]{$\cO^\Phi_{\g}$};
\filldraw[black] (0,-2) circle (2pt) node[anchor=west]{$\Psi_{b_2}$};
\filldraw[black] (0,-1) circle (2pt) node[anchor=west]{$\cO^\Phi_{\g}$};
\end{tikzpicture}
\begin{tikzpicture}[scale=0.65]
\draw[pattern=north east lines, pattern color=black] (1.3,1.3) circle (0.45);
\draw [color=black, thick] (0, 0) -- (1, 1);
\draw [color=black, thick] (-1,0)-- (0, 0) ;
\draw [color=black, thick, dashed] (-2,0)-- (-1, 0) ;
\draw [color=black, thick, dashed] (0, -2) -- (0, -1);
\draw [color=black, thick] (0, -1) -- (0, 0);
\filldraw[black] (0,0) circle (2pt) node[anchor=south]{$\mu_2$};
\filldraw[black] (-2,0) circle (2pt) node[anchor=north]{$\Psi_{b_1}$};
\filldraw[black] (-1,0) circle (2pt) node[anchor=north]{$\cO^\Phi_{\g}$};
\filldraw[black] (0,-2) circle (2pt) node[anchor=west]{$\Psi_{b_2}$};
\filldraw[black] (0,-1) circle (2pt) node[anchor=west]{$\cO^\Phi_{\g}$};
\filldraw[black] (0.6,0.6) circle (2pt) node[anchor=north]{$\;\;\cO^\Phi_{\g}$};
\end{tikzpicture}
\]
The incoming states for the first and second special trees vanish due to
\[
2\pi G_- \mu_2 \bigl(\Psi_{b_1}, \Psi_{b_2}\bigr) = 2\pi G_- \Psi_{b_1+b_2} = 0.
\]
The incoming state for the third tree
\begin{align*}
2\pi G_- \cO_{\g^\Phi} 2\pi G_- \mu_2 \bigl(2\pi G_- \cO_{\g^\Phi} \Psi_{b_1}, \Psi_{b_2}\bigr) &{}=2\pi G_- \cO_{\g^\Phi} G_- \mu_2 \bigl( \bigl| \vec{b}_1\wedge {\rm d}\vec{\phi}, \vec{b}_1\bigr\>, \bigl| 1, \vec{b}_2\bigr\>\bigr) \\
&{}= 2\pi G_- \cO_{\g^\Phi} G_- \bigl| \vec{b}_1\wedge {\rm d}\vec{\phi}, \vec{b}_1+\vec{b}_2\bigr\>\\
&{}= \bigl(\vec{b}_1\wedge \bigl(\vec{b}_1+\vec{b}_2\bigr)\bigr) 2\pi G_- \cO_{\g^\Phi} \bigl| 1, \vec{b}_1+\vec{b}_2\bigr\> \\
&{}= \bigl(\vec{b}_1\wedge \vec{b}_2\bigr) \cdot 2\pi G_- \cO_{\g^\Phi} \Psi_{b_1+b_2}.
\end{align*}
Note that the state for the special tree with $b_1 \leftrightarrow b_2$ has an extra minus factor, so to make a~recursion relation universal, we need to include the absolute value.
The incoming state for the fourth special tree
\begin{align*}
2\pi G_- \mu_2 \bigl(2\pi G_- \cO_{\g^\Phi} \Psi_{b_1}, 2\pi G_- \cO_{\g^\Phi} \Psi_{b_2}\bigr) &{}= \frac{1}{2\pi} G_- \mu_2 \bigl( \bigl| \vec{b}_1\wedge {\rm d}\vec{\phi}, \vec{b}_1\bigr\>, \bigl| \vec{b}_2\wedge d\vec{\phi}, \vec{b}_2\bigr\>\bigr) \\
&{}= \frac{1}{2\pi} G_- \bigl| \vec{b}_1\wedge {\rm d}\vec{\phi}\wedge \vec{b}_2\wedge {\rm d}\vec{\phi}, \vec{b}_1+\vec{b}_2\bigr\> \\
&{}= \frac{1}{2\pi} \bigl(\vec{b}_1\wedge \vec{b}_2 \bigr) G_- \bigl| {\rm d}\phi^1\wedge {\rm d}\phi^2, \vec{b}_1+\vec{b}_2\bigr\> \\
&{}= \bigl(\vec{b}_1\wedge \vec{b}_2\bigr) \cdot 2\pi G_- \cO_{\g^\Phi} \Psi_{b_1+b_2}.
\end{align*}
The incoming states for the fifth special tree vanishes since the state $G_- \cO_{\g^\Phi} \Psi_{b_1+b_2}$ is already a~1-form and the further multiplication for a 2-form $\gamma^\Phi$ turns the state into a zero state.
\end{proof}

The HTQM representation for the tropical multiplicities (\ref{eq_htqm_rep_km_multipl}) does not include homotopy~$K$, i.e., the moduli integrals are trivial. Trivial moduli space integrals effectively reduce the HTQM to a 1D TQFT. We conjecture that the emerging TQFT is identical to one used in the Mandel and Ruddat construction~\cite{mandel2023tropical}, and we plan to investigate the details of this conjecture further.

\begin{Remark}
Additional evidence for the conjecture is a natural emergence of the Batalin--Vilkovisky (BV) bracket from second-order operator $G_-$. The odd parity of $G_-$ implies super skew-symmetry of the bracket. The 7-term relation (\ref{def_7_term_relation}) for $G_-$ implies the super Jacobi identity for the $G_-$-bracket. In BV formalism, the BV bracket emerges in a similar construction using the BV Laplacian instead of $G_-$.
\end{Remark}

\section[B-model HTQM]{$\boldsymbol{B}$-model HTQM}\label{sec5}
In our construction of the tropical mirror symmetry, we will adopt an approach from \cite{Frenkel:2005ku} to the case of HTQM on special trees.

\subsection{State-observable map for amplitudes}\label{sec_state_oper_map}

 Using the state-operator map (\ref{state_oper_map}) to turn the evaluation operators $\cO_{\g_\a}$ into the corresponding evaluation states $\Psi_{\g_\a}$, we rearrange the amplitude on a special tree $\Gamma$ with 2-valent vertices into the amplitude on a special tree $\Gamma'$ without 2-valent vertices, but with additional leaves. In~particular,
\begin{align*}
\mathcal{A}_\Gamma\bigl(\cO_{\g_\a}; \Psi_{b_a}\bigr)&= \Biggl\< \! (2\pi K G_-)^{\otimes I (\G) } \bigotimes\limits_{a=1}^{n_1(\G)} \Psi_{b_a}\bigotimes\limits_{\a=1}^{n_2(\G)} \cO_{\g_\a} \otimes \mu_2^{\otimes n_3(\G)} \! \Biggr\>_\Gamma \\
&= \Biggl\< \!(2\pi KG_-)^{\otimes I (\G) }\bigotimes\limits_{a=1}^{n_1(\G)} \Psi_{b_a} \bigotimes\limits_{\a=1}^{n_2(\G)} \Psi_{\g_\a} \otimes \mu_2^{\otimes (n_3(\G)+n_2(\G))} \!\Biggr\>_{\Gamma'} \\
& = \mathcal{A}_{\Gamma'} (\Psi_{b_a}, \Psi_{\g_\a}).
\end{align*}
A 2-valent vertex on a special tree $\Gamma$ decorated with the observable $\cO_{\g_\a}$ becomes the 3-valent vertex with a leaf attached to it, decorated by a state $\Psi_{\g_a}$. The new special tree $\Gamma'$ has $n_1(\Gamma')=n_1(\Gamma)+n_2(\Gamma)$ leaves, no 2-valent vertices and $n_3(\Gamma')=n_3(\Gamma)+n_2(\Gamma)$ 3-valent vertices. An~example of such rearrangement is presented at the end of this section.

We can apply the state operator map to the divisor states $\Psi_{b_a}$ to rearrange the 3-valent vertices with leaves attached to them, decorated with $\Psi_{b_a}$ into 2-valent vertices decorated with corresponding divisor operators $\cO_{b_a}$. In particular, we have an equality
\[
\mathcal{A}_\Gamma (\Psi_{b_a}; \cO_{\g_\a})=\mathcal{A}_{\Gamma'}(\Psi_{b_a}, \Psi_{\g_\a})= \mathcal{A}_{\Gamma''}(\Psi_{\g_\a}; \cO_{b_a}).
\]
The special tree $\G''$ has $n_1(\Gamma'') = n_2(\Gamma)$ 1-valent vertices $n_2(\G'') = n_1(\Gamma)$ 2-valent vertices.

For a general, special tree with divisor states on the leaves, the rearrangement might not be possible. An obstruction is the 3-valent vertices with two incoming divisor states. However, the HTQM amplitudes on such special trees are equal to zero.
\begin{Lemma} \label{lemma_two_same_states}
An HTQM amplitude vanishes the special tree when two leaves, with either two evaluation or two divisor states, are connected to the same $3$-valent vertex.
\end{Lemma}
\begin{proof} The graphical representation of the vanishing amplitudes on special trees is presented below:
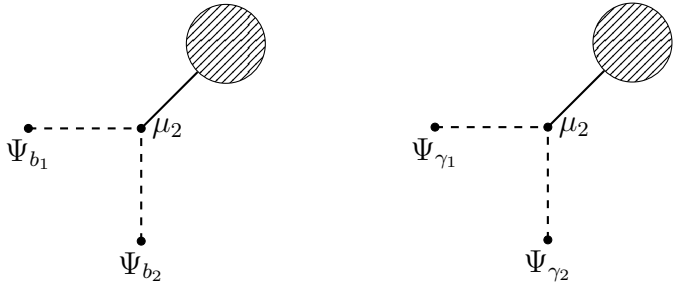
\[
\begin{tikzpicture}[scale=0.7]
 \draw[pattern=north east lines, pattern color=black] (1.5,1.5) circle (0.7);
\draw [color=black, thick] (0, 0) -- (1, 1);
\draw [color=black, thick, dashed] (-2,0)-- (0, 0) ;
\draw [color=black, thick, dashed] (0, -2) -- (0, 0);
\filldraw[black] (0,0) circle (2pt) node[anchor=west]{$\mu_2$};
\filldraw[black] (-2,0) circle (2pt) node[anchor=north]{$\Psi_{b_1}$};
\filldraw[black] (0,-2) circle (2pt) node[anchor=north]{$\Psi_{b_2}$};
\end{tikzpicture}
\qquad
\begin{tikzpicture}[scale=0.7]
 \draw[pattern=north east lines, pattern color=black] (1.5,1.5) circle (0.7);
\draw [color=black, thick, dashed] (-2,0)-- (0, 0) ;
\draw [color=black, thick, dashed] (0, -2) -- (0, 0);
\draw [color=black, thick] (0, 0) -- (1, 1);
\filldraw[black] (0,0) circle (2pt) node[anchor=west]{$\mu_2$};
\filldraw[black] (-2,0) circle (2pt) node[anchor=north]{$\Psi_{\g_1}$};
\filldraw[black] (0,-2) circle (2pt) node[anchor=north]{$\Psi_{\g_2}$};
\end{tikzpicture}
\]
The contributions from 3-valent vertices on the pictures above are equal to
\[
KG_- \mu_2 (\Psi_{b_1}, \Psi_{b_2}) = KG_- \Psi_{b_1+b_2} =0,\qquad KG_- \mu_2 (\Psi_{\g_1}, \Psi_{\g_2}) = KG_-\Psi_{\g_1\wedge \g_2}=0.
\]
Hence, the whole amplitude is zero.
\end{proof}

\begin{Example}The amplitude rearrangement is presented in the pictures below:
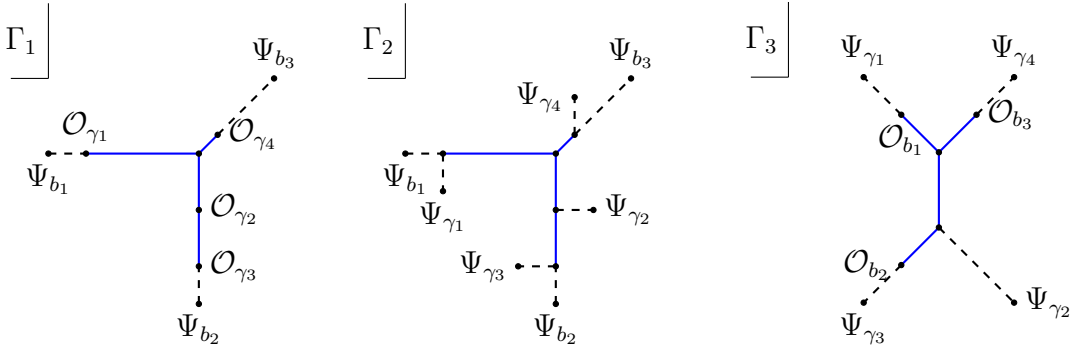
\[
\begin{tikzpicture}[scale=0.5]
\filldraw[black] (-4,3) node[anchor=east]{$\G_1$};
\draw (-5,2)-- (-4, 2) ;
\draw (-4,2)-- (-4, 4) ;
\draw[black, dashed, thick](-4,0)-- (-3, 0) ;
\draw [black, dashed, thick](0, -4) -- (0, -3);
\draw [blue, thick](0, -3) -- (0, 0);
\draw[blue, thick](-3, 0) -- (0, 0);
\draw [black, thick, dashed](0.5, 0.5) -- (2, 2);
\draw [blue, thick](0, 0) -- (0.5, 0.5);
\filldraw[black] (0,0) circle (2pt);
\filldraw[black] (-4,0) circle (2pt) node[anchor=north]{$\Psi_{b_1}$};
\filldraw[black] (0,-1.5) circle (2pt) node[anchor=west]{$\cO_{\g_2}$};
\filldraw[black] (0,-3) circle (2pt) node[anchor=west]{$\cO_{\g_3}$};
\filldraw[black] (2,2) circle (2pt) node[anchor=south]{$\Psi_{b_3}$};
\filldraw[black] (0.5,0.5) circle (2pt) node[anchor=west]{$\cO_{\g_4}$};
\filldraw[black] (0,-4) circle (2pt) node[anchor=north]{$\Psi_{b_2}$};
\filldraw[black] (-3,0) circle (2pt) node[anchor=south]{$\cO_{\g_1}$};
\end{tikzpicture}\qquad
\begin{tikzpicture}[scale=0.5]
\filldraw[black] (-4,3) node[anchor=east]{$\G_2$};
\draw (-5,2)-- (-4, 2) ;
\draw (-4,2)-- (-4, 4) ;
\draw[black,dashed,thick](-4,0)-- (-3, 0) ;
\draw[black,dashed,thick](0, -4) -- (0, -3);
\draw[blue,thick](0, -3) -- (0, 0);
\draw[blue,thick](-3, 0) -- (0, 0);
\draw[black,thick,dashed] (0.5, 0.5) -- (2, 2);
\draw[blue,thick] (0, 0) -- (0.5, 0.5);
\filldraw[black] (0,0) circle (2pt);
\filldraw[black] (-4,0) circle (2pt) node[anchor=north]{$\Psi_{b_1}$};
\filldraw[black] (0,-1.5) circle (2pt);
\filldraw[black] (1,-1.5) circle (2pt) node[anchor=west]{$\Psi_{\g_2}$};
\draw[color=black, thick, dashed] (0,-1.5)--(1,-1.5);
\filldraw[black] (0,-3) circle (2pt);
\filldraw[black] (-1,-3) circle (2pt) node[anchor=east]{$\Psi_{\g_3}$};
\draw[color=black, thick, dashed] (0,-3)--(-1,-3);
\filldraw[black] (2,2) circle (2pt) node[anchor=south]{$\Psi_{b_3}$};
\filldraw[black] (0.5,0.5) circle (2pt);
\filldraw[black] (0.5,1.5) circle (2pt) node[anchor=east]{$\Psi_{\g_4}$};
\draw[color=black, thick, dashed] (0.5,0.5)--(0.5,1.5);
\filldraw[black] (0,-4) circle (2pt) node[anchor=north]{$\Psi_{b_2}$};
\filldraw[black] (-3,0) circle (2pt);
\filldraw[black] (-3,-1) circle (2pt) node[anchor=north]{$\Psi_{\g_1}$};
\draw[color=black, thick, dashed] (-3,0)--(-3,-1);
\end{tikzpicture}\qquad
\begin{tikzpicture}[scale=0.5]
\filldraw[black] (-4,3) node[anchor=east]{$\G_3$};
\draw (-5,2)-- (-4, 2) ;
\draw (-4,2)-- (-4, 4) ;
\draw[blue,thick] (0, -2) -- (-1, -3);
\draw[black,thick,dashed](-1, -3) -- (-2, -4);
\draw[black,thick,dashed](0, -2) -- (2, -4);
\draw[blue,thick](0, -2) -- (0, 0);
\draw[black,dashed,thick](1, 1) -- (2, 2);
\draw[blue,thick](1, 1) -- (0, 0);
\draw[black,dashed, thick](-1, 1) -- (-2, 2);
\draw[blue,thick](0, 0) -- (-1, 1);
\filldraw[black] (0,0) circle (2pt);
\filldraw[black] (0,-2) circle (2pt);
\filldraw[black] (-2,-4) circle (2pt) node[anchor=north]{$\Psi_{\g_3}$};
\filldraw[black] (-1,-3) circle (2pt) node[anchor=east]{$\cO_{b_2}$};
\filldraw[black] (-2,2) circle (2pt) node[anchor=south]{$\Psi_{\g_1}$};
\filldraw[black] (-1,1) circle (2pt) node[anchor=north]{$\cO_{b_1}$};
\filldraw[black] (2,-4) circle (2pt) node[anchor=west]{$\Psi_{\g_2}$};
\filldraw[black] (2,2) circle (2pt) node[anchor=south]{$\Psi_{\g_4}$};
\filldraw[black] (1,1) circle (2pt) node[anchor=west]{$\cO_{b_3}$};
\end{tikzpicture}
\]
The special tree $\G_1$ has 3 leaves, decorated with $\Psi_{b_a}$, four 2-valent vertices, decorated with~$\cO_{\g_\a}$ and the amplitude (\ref{ex_ampl_g1}).
We use the state-operator map to represent the same amplitude and the amplitude (\ref{ex_ampl_g2}) on the special tree $\G_2$ with seven leaves, decorated with states $\Psi_{b_a}$ and $\Psi_{\g_\a}$. The special tree $\G_2$ does not have leaves of the same type connected to the 3-valent vertex. Hence, we can apply the state-operator map to represent the same amplitude as the amplitude~(\ref{ex_ampl_g3}) on a special tree $\G_3$ with four leaves, decorated with $\Psi_{\g_\a}$ and three 2-valent vertices, decorated with the operators $\cO_{b_a}$.
\end{Example}

\subsection[Total A-model amplitudes]{Total $\boldsymbol{A}$-model amplitudes}

In Theorem \ref{theo_htqm_rep_trop_gw}, we showed that the tropical GW invariants can be written as the amplitudes on special trees. Generic special trees may not represent any tropical GW invariant, but we will show below that the amplitudes for such special trees vanish.
\begin{Definition} \label{def_total_ampl_htqm}
The {\it total amplitude} $\< \Psi_{1},\dots,\Psi_{n}\>_Q $ on a special trees $n$ leaves, decorated with the states $\Psi_1,\dots,\Psi_n$ in HTQM $(V,Q, G_{\pm}, g, \mu_2)$ is
\[
\< \Psi_{1},\dots,\Psi_{n}\>_Q = \sum_{\G, \sigma\in S_n} \frac{ \mathcal{A}_\G \bigl(\Psi_{\sigma(1)},\dots,\Psi_{\sigma(n)}\bigr) }{|{\rm Aut}(\G)|},
\]
where $|{\rm Aut}(\G)|$ is the symmetry factor for the tree $\G$. The summation over $\G$ is taken over all distinct special trees $\Gamma$ with $n$ leaves. The summation over $\sigma$ is the summation over the possible assignment of states $\Psi_1,\dots,\Psi_n$ on the leaves of $\G$.
\end{Definition}
\begin{Remark}
The total amplitude in Definition \ref{def_total_ampl_htqm} is the tropical limit of an $n$-point genus 0 string amplitude, i.e., 2d CFT correlation function on a sphere with $n$ marked points, integrated over the corresponding moduli space. For more details about the string amplitudes, see \cite{polchinski2005stringI,polchinski2005stringII}.
\end{Remark}

\begin{Lemma} \label{lemma_trop_degree_selection} The total amplitude $\< \Psi_{\g_1},\dots,\Psi_{\g_n}, \Psi_{b_1},\dots,\Psi_{b_d}\>$ vanishes unless the total number~$d$ of divisor states and the total degree of differential forms $\g_a$ obey the tropical degree selection relation
\[
2(d +n -3)=\sum_{\a=1}^n \deg \g_\a-\dim_{\mathbb{R}} X
\]
and the sum over vectors for the divisor states vanishes
\[
\sum_{a=1}^d \vec{b}_a = 0.
\]
\end{Lemma}
\begin{proof}
The amplitude on a special tree $\G$ is proportional to the $\mu_3^0$-product on three states $|\omega_j, \vec{m}_j\>$, $j=1,2,3$ incoming to the sink vertex. The $A$-model definitions~(\ref{A_model_mu}) and~(\ref{A_model_g}) imply that the
$\mu_3^0$-product vanishes unless the sum of the three integer vectors for the incoming states is zero,
$
\vec{m}_1+\vec{m}_2 + \vec{m}_3= 0$,
and the total degree of three forms $\omega_j$ for the incoming states add up to the dimension of~$X$,~i.e.,
\[
\sum_{j=1}^3 \deg \omega_j = \dim_{\mathbb{R}} X.
\]
The $A$-model multiplication (\ref{A_model_mu}) returns the sum of the integer vectors for the two arguments, while the degree of the resulting form is the sum of degrees for the forms of two arguments. The~propagator $2\pi K G_-$ on internal edges reduces the degree of the form by two while preserving the integer vector. Hence, the total degree of the three arguments for the sink vertex is the total degree of the forms $\g_a$ on the leaves, adjusted by the number on internal edges, i.e.,
\begin{gather}\label{mu_3_selecton_degree}
\sum_{j=1}^3 \deg \omega_j = \sum_{\a=1}^n \deg \g_\a -2 I(\G).
\end{gather}
The sum of the integer vectors equals the sum of the integer vectors of the divisor states~$\Psi_{b_a}$,~i.e.,%
\begin{gather}\label{mu_3_selecton_vectors}
\vec{m}_1+\vec{m}_2 + \vec{m}_3 = \sum_{a=1}^{d} \vec{b}_a .
\end{gather}
The number $I(\G) $ of internal edges for a special tree $\G$ with $n+d$ leaves is
\begin{gather}\label{internal_edges_tree_leaves}
I(\G) = n_1(\G)-3 = n+d-3.
\end{gather}
Expression (\ref{internal_edges_tree_leaves}) and selection conditions (\ref{mu_3_selecton_degree}) and (\ref{mu_3_selecton_vectors}) for non-zero sink vertex multiplication complete the proof of the lemma.
\end{proof}

\begin{Remark}
The tropical degree selection Lemma \ref{lemma_trop_degree_selection} is equivalent to the matching between the total degree of the form and the dimension (\ref{trop_mod_dim}) of the moduli space
\[
\deg \bigwedge {\rm ev}^\ast_\a \g_\a = \sum_{\a=1}^n \deg \g_\a = 2\dim_{\mathbb{C}} \mathcal{M}_{0,n} \bigl(X, \beta; \vec{\G}\bigr) =2 I(\G) +2\dim_{\mathbb{C}} X.
\]
\end{Remark}

\begin{Proposition}\label{prop_total_ampl_trop_gw} The sum over total amplitudes for $n\geq 3$ with an arbitrary number of universal divisor states matches with the weighted sum of tropical GW invariants over the degree of the curve $\beta$, i.e.,
\[
\sum_{d=0}^\infty \frac{1}{d!} \< \Psi_{\g_1},\dots,\Psi_{\g_n}, \underbrace{\Psi_{X},\dots,\Psi_{X}}_{d}\>_Q= \sum_{\beta\in H_2 (X)}q^\beta\<\g_1,\dots,\g_n \>^X_\beta,
\]
where we introduced the universal divisor state
\[
\Psi_X = \sum_{b \in B_X}q_b \Psi_{b}.
\]
The K\"ahler moduli of $X$ are expressed in terms of toric moduli $q_b$ for each compactifying hypersurface $H_b$
\[
q^\beta = \prod_{\vec{b} \in S_\beta } q_{b},
\]
where $S_\beta$ is the tropical curve of degree $\beta$ with all internal length moduli equal to zero $($star with a particular number of rays$)$.
\end{Proposition}
\begin{proof}
Lemma \ref{lemma_trop_degree_selection} tells us that the infinite sum over the number of universal divisors has only one non-zero term with
\[
2(d +n -3) = \sum_{\a=1}^n \deg \g_\a - \dim_{\mathbb{R}} X.
\]
There is a special case $d=0$, when we do not have any divisor states and cannot use Theorem~\ref{theo_htqm_rep_trop_gw} to relate the tropical GW invariant to the $A$-model HTQM amplitudes. The~corresponding tropical GW invariant vanishes unless we have just three observables, then
\[
\<\g_1, \g_2, \g_3\>^X_{\beta=0} = \int_{X} \g_1\wedge \g_2\wedge \g_3.
\]
 According to Lemma \ref{lemma_two_same_states} amplitudes $\< \Psi_{\g_1},\dots,\Psi_{\g_n}\>_Q$ vanish unless the special tree has no internal edges.
The only special tree with no internal edges is a $Y$-shaped tree with $n=3$ leaves. The~total amplitude of such a tree is
\[
\< \Psi_{\g_1},\Psi_{\g_2},\Psi_{\g_3}\>_Q = \mathcal{A}_{Y} (\Psi_{\g_1},\Psi_{\g_2},\Psi_{\g_3}) = \mu^0_3 (\Psi_{\g_1},\Psi_{\g_2},\Psi_{\g_3}) = \int_{X} \g_1\wedge \g_2\wedge \g_3.
\]
Hence, it matches with the tropical GW invariant.

We can write the single amplitude for a fixed $d$ as a sum over total amplitudes with a fixed number of boundary divisors $d_a$ of each type $\Psi_{b_a}$
\begin{gather*}
\frac{1}{d!}\< \Psi_{\g_1},\dots,\Psi_{\g_n}, \underbrace{\Psi_{X},\dots,\Psi_{X}}_{d}\>_Q \\
\qquad{}=\sum_{d_1+\cdots+d_{B} = d} \frac{ q_{b_1}^{d_1}\cdots q_{b_{B}}^{d_{B}}}{d_1!\cdots d_{B}!} \< \Psi_{\g_1},\dots,\Psi_{\g_n}, \underbrace{\Psi_{b_1},\dots,\Psi_{b_1}}_{d_1},\dots, \underbrace{\Psi_{b_B},\dots,\Psi_{b_{B}}}_{d_{B}}\>_Q.
\end{gather*}
Lemma \ref{lemma_trop_degree_selection} tells us that the amplitudes in the sum vanish unless the total sum of all divisor normal vectors equals zero. The sum can be written using the numbers $d_a$ of each type of the normal vectors
\begin{gather}\label{rel_divisor_selection}
\sum^{B}_{a=1} d_a \vec{b}_a = 0.
\end{gather}
The system (\ref{rel_divisor_selection}) is the system of $\dim X$ equations for $B = |B_X|$ variables $d_a$. For a projective toric variety, the vectors $\vec{b}_a \in B_X$ form (overcomplete) basis in $\mathbb{Z}^{\dim X}$, hence the space of possible solutions for the system (\ref{rel_divisor_selection}) has dimension $|B_X|-\dim X$. According to (\ref{toric_betti_numbers}), this dimension matches with the dimension of $H_2(X)$, which is the dimension of the space of possible degrees~$\beta$,~i.e.,{\samepage
\begin{gather*}\begin{split}
&\frac{1}{d!} \< \Psi_{\g_1},\dots,\Psi_{\g_n}, \underbrace{\Psi_{X},\dots,\Psi_{X}}_{d}\>_Q \\
&\qquad{}= \sum_{\beta \in H_2(X)} \frac{ q^\beta}{d_1!\cdots d_{B}!} \< \Psi_{\g_1},\dots,\Psi_{\g_n}, \underbrace{\Psi_{b_1},\dots,\Psi_{b_1}}_{d_1},\dots, \underbrace{\Psi_{b_B},\dots,\Psi_{b_B}}_{d_{B}}\>_Q
\end{split}
\end{gather*}
for $d_{a} = \beta \cdot H_{b_a}$.}

We can recursively construct all special trees: for a given special tree $\G$ with $d$ leaves, we can add a leaf to any of its $2d-3$ edges to produce a special tree with $d+1$ leaves. Moreover, we can construct all possible special trees with $d+1$ leaves by adding an extra leaf to every edge of every special tree with $d$ leaves. Similarly, we can construct all special trees with $d+2$ leaves from special trees with $d$ leaves. However, in addition to adding the two leaves to the edges of every special tree with $d$ leaves, we need to add a $Y$-shaped subtree with two leaves.

We can describe all contributions to the total amplitude $ \< \Psi_{\g_1},\Psi_{\g_2},\Psi_{X},\Psi_{X},\dots,\Psi_X\>_Q$ with two evaluation state insertions starting with special trees with $d$ leaves, decorated with the universal divisor states $\Psi_{X}$ and then add two leaves (on the same or two different edges) or a~$Y$-shaped subtree, decorated with $\Psi_{\g_1}$, $\Psi_{\g_2}$ states. In Lemma \ref{lemma_two_same_states}, we showed that amplitudes on special trees, constructed from adding $\Psi_{\g_1}$, $\Psi_{\g_2}$ on a $Y$-shaped subtree vanish.

We can repeat our analysis for the amplitudes on special trees with $n+d$ leaves to represent them as the amplitudes on special trees with $d$ leaves decorated by universal divisor state $\Psi_X$ and $n$ additional leaves decorated by the evaluation states $\Psi_{\g_1},\dots,\Psi_{\g_n}$ distributed among the edges of a special tree with $d$ leaves. Since we only have evaluation states $\Psi_{\g_1},\dots,\Psi_{\g_n}$ on leaves (no sub-trees), we can use the state-operator map from Section \ref{sec_state_oper_map} to rewrite such amplitudes as the amplitudes with $\cO_{\g_\a}$ operator insertions. Moreover, since we added leaves at every possible location on a special tree, the corresponding amplitudes have operators $\cO_{\g_\a}$ inserted at all possible locations on a special tree, decorated by $\Psi_X$.

The last step of the proof is to trace the symmetry factors. With all symmetry factors included
\begin{gather}
 \frac{ 1}{d_1!\cdots d_{B}!} \< \Psi_{\g_1},\dots,\Psi_{\g_n}, \underbrace{\Psi_{b_1},\dots,\Psi_{b_1}}_{d_1},\dots, \underbrace{\Psi_{b_B},\dots,\Psi_{b_B}}_{d_B}\>_Q \nonumber\\
 \qquad{}=\sum_{\G'_n} \mathcal{A}_{\G_n'} (\cO_{\g_1},\dots,\cO_{\g_n}, \underbrace{\Psi_{b_1},\dots,\Psi_{b_1}}_{d_1},\dots, \underbrace{\Psi_{b_B},\dots,\Psi_{b_B}}_{d_B}).\label{rel_total_ampl_trop_curves}
\end{gather}
The sum is taken over $\G_n'$ the special trees with $n_2(\G_n')=n$ 2-valent vertices, decorated with the~$\cO_{\g_a}$ observables in arbitrary order and a fixed decoration with divisor states $\Psi_{b_a}$ on the leaves.\looseness=-1

According to Theorem \ref{theo_htqm_rep_trop_gw}, each term in the sum on the second row of (\ref{rel_total_ampl_trop_curves}) is the contribution to the tropical GW invariant from the tropical curve associated with the special tree $\G_n'$, hence
\[
\frac{1}{d_1!\cdots d_{B}!} \< \Psi_{\g_1},\dots,\Psi_{\g_n}, \underbrace{\Psi_{b_1},\dots,\Psi_{b_1}}_{d_1},\dots, \underbrace{\Psi_{b_B},\dots,\Psi_{b_B}}_{d_{B}}\>_Q =\<\g_1,\dots,\g_n \>^X_\beta.
\]
To complete the proof, we need to describe the discrete data (directions, integer vector decoration subject to consistency) of a tropical curve $\vec{\G}$ using the amplitude data: special tree $\Gamma$ and its decoration by $\Psi_{b_a}$ states and $\cO_{\g_\a}$ operators.

A special tree $\Gamma$ is a directed tree, so we can identify directions on $\G$ with the directions on tropical curve $\vec{\G}$.
We can equip each edge $e$ of a directed tree $\G$ with an integer vector $\vec{m}_e$ using an integer vector for an (incoming) state $|\omega_e,\vec{m}_e\>$ for the edge $e$. Note that the generic state in HTQM is a linear combination of states $|\omega, \vec{m}\>$ with different vectors $\vec{m}$. Below, we show that the incoming states on all edges of a special tree, decorated with $\Psi_{b_a}$ states and $\cO_{\g_\a}$ operators, are states with a single integer vector. The consistency of such choice of integer vectors follows from the HTQM definitions in Section \ref{TGQ_QM_section}:
\begin{itemize}\itemsep=0pt
\item Each external edge $e$ is decorated with the state $\Psi_{b}=\bigl|1, \vec{b}\bigr\>$ for a primitive normal vector $\vec{b} \in B_X$ hence it is a state with a single integer vector $\vec{m}_e = \vec{b}$, moreover $\vec{m}_e \in B_X$.
\item The propagator $KG_-$ on edge does not change the integer vector of the incoming state.
\item Each 2-valent vertex is decorated with $\cO_{\g_\a}$ operator, with the action
\[
\cO_{\g_\a}|\omega, \vec{m}^{{\rm in}}\> = \bigl|\g_\a \wedge \omega, \vec{m}^{{\rm in}}\bigr\> = \bigl| \omega', \vec{m}^{{\rm out}}\bigr\>.
\]
The outgoing state is a state with a single integer vector. The incoming and outgoing states have $\vec{m}^{{\rm in}} = \vec{m}^{{\rm out}}$, hence satisfy the balance condition (\ref{vert_bal_con}) at each 2-valent vertex.
\item Each 3-valent vertex is decorated with the $\mu_2$ multiplication, such that
\[
\mu_2 \bigl(\bigl|\omega_1, \vec{m}^{{\rm in}}_1\bigr\>,\bigl|\omega_2, \vec{m}^{{\rm in}}_2\bigr\>\bigr)=\bigl|\omega_1\wedge \omega_2, \vec{m}^{{\rm in}}_1+ \vec{m}^{{\rm in}}_2\bigr\> = \bigl|\omega', \vec{m}^{{\rm out}}\bigr\>,
\]
The outgoing state is a state with a single integer vector. The three states have $\vec{m}^{{\rm in}}_1+\vec{m}^{{\rm in}}_2= \vec{m}^{{\rm out}}$ and hence satisfy the balance condition (\ref{vert_bal_con}) at each 3-valent vertex.
\item The sink vertex, decorated with the multiplication $\mu_3^0$, such that
\[
\mu^0_3 \bigl(\bigl|\omega_1, \vec{m}^{{\rm in}}_1\bigr\>,\bigl|\omega_2, \vec{m}^{{\rm in}}_2\bigr\>,\bigl|\omega_3, \vec{m}^{{\rm in}}_3\bigr\>\bigr)\propto \delta\bigl(\vec{m}^{{\rm in}}_1+\vec{m}^{{\rm in}}_2+\vec{m}^{{\rm in}}_3\bigr),
\]
Hence for non-zero amplitude the three incoming states obey $\vec{m}^{{\rm in}}_1+\vec{m}^{{\rm in}}_2+\vec{m}_3^{{\rm in}} = 0$, hence satisfy the balance condition at sink vertex.\hfill \qed
\end{itemize}\renewcommand{\qed}{}
\end{proof}

\subsection[Total B-model amplitudes]{Total $\boldsymbol{B}$-model amplitudes}
In Proposition \ref{prop_total_ampl_trop_gw}, we described evaluation states as extra leaves attached to the special tree decorated with the universal divisor states. We can perform a similar analysis while switching between universal divisor and evaluation states' roles.
\begin{Proposition}\label{prop_total_ampl_b_model} The sum over total $A$-model amplitudes is the $B$-model total amplitude, i.e.,
\[
 \sum_{d=0}^\infty \frac{1}{d!} \< \Psi_{\g_1},\dots,\Psi_{\g_n}, \underbrace{\Psi_{X},\dots,\Psi_{X}}_{d}\>_Q= \bigl\< \Psi^X_{\g_1},\dots,\Psi^X_{\g_n}\bigr\>_{Q^X}.
\]
The $B$-model HTQM $\bigl(V,Q^X,G_{\pm}, g, \mu_2\bigr)$ has a deformed differential
\[
Q^{X}=Q - 2\pi [G_-, \cO_X]
\]
and {\it mirror states}
\[
\Psi_\g^X = \sum_{d=0}^\infty (2\pi KG_- \cO_X)^d \Psi_\g.
\]
\end{Proposition}
\begin{proof} We can recursively construct all special trees: For a given special tree $\G_n$ with $n$ leaves, we can add a leaf to any of its $2n-3$ edges to produce a special tree with $n+1$ leaves. Moreover, we can construct all possible special trees with $n+1$ leaves by adding an extra leaf to every edge of every special tree with $n$ leaves. Similarly, we can construct all special trees with $n+2$ leaves from special trees with $n$ leaves. To get all trees, we need to add two leaves on the edges of every special tree with $n$ leaves and a $Y$-shaped subtree with two leaves.

We can evaluate all contributions to the total amplitude $ \< \Psi_{\g_1},\dots,\Psi_{\g_n}, \Psi_{X},\Psi_{X}\>_Q$ by treating two universal divisor states as two additional leaves or a $Y$-shaped subtree on a special tree with~$n$ leaves, decorated by evaluation states $\Psi_{\g_1},\dots,\Psi_{\g_n}$. In Lemma \ref{lemma_two_same_states}, we showed that amplitudes on special trees, constructed from adding a $Y$-shaped subtree, vanish.

We can repeat our analysis for the amplitudes on special trees with $n+d$ leaves, decorated by $d$ universal divisor states to express them as the amplitudes on special trees with $n$ leaves, decorated by evaluation states
$\Psi_{\g_1},\dots,\Psi_{\g_n}$ and $d$ additional leaves decorated by $\Psi_X$ states distributed among edges on a special tree. The summation over $d$ from zero to infinity with the symmetry factor $1/d!$ is the same as a multivariable sum over numbers of additional leaves $d_e$ on each edge $e$ for the special tree, weighted with the $1/d_e!$ symmetry factor.

There are two types of edges on a special tree: internal and external. The sum over additional leaves for internal edges describes the deformation of the differential $Q$, while the sum for the external edges describes the deformation of evaluation states $\Psi_{\g_1},\dots,\Psi_{\g_n}$. We use the state-operator map to turn the universal divisor states into operators $\cO_X$. For identical operators, the corresponding contributions from an edge will be the same for arbitrary orders of their insertions. For $d_e$ identical operators, there will be $d_e!$ identical contributions which exactly cancel the symmetry factor $1/d_e!$.

For an internal edge, the sum over an arbitrary number of insertions of operators $\cO_X$ has a~graphical representation
\begin{gather*}
\begin{tikzpicture}[scale=0.88]
\draw[ultra thick] (0,0)-- (2, 0) ;
\filldraw[black] (0,0) circle (2pt) ;
\filldraw[black] (2,0) circle (2pt) ;
\node (=) at (3,0) {$=$};
\draw[thick] (4,0)-- (6, 0) ;
\filldraw[black] (4,0) circle (2pt) ;
\filldraw[black] (6,0) circle (2pt) ;
\node (+) at (7,0) {$+$};
\draw[thick] (8,0)-- (10, 0) ;
\filldraw[black] (8,0) circle (2pt) ;
\filldraw[black] (10,0) circle (2pt) ;
\filldraw[black] (9,0) circle (2pt) node[anchor=south]{$\cO_X$};
\node (+) at (11,0) {$+$};
\draw[thick] (12,0)-- (15, 0) ;
\filldraw[black] (12,0) circle (2pt) ;
\filldraw[black] (15,0) circle (2pt) ;
\filldraw[black] (13,0) circle (2pt) node[anchor=south]{$\cO_X$};
\filldraw[black] (14,0) circle (2pt) node[anchor=south]{$\cO_X$};
\node (+) at (16,0) {$+$};
\node (+) at (16.5,0) {$\dots$};
\end{tikzpicture}
\end{gather*}
which evaluates into
\begin{align}
 K^X G_-&{}= K G_- +2\pi K G_- \cO_{X} K G_- +(2\pi)^2 K G_- \cO_{X} K G_- \cO_{X} KG_- +\cdots \nonumber \\
&{}= (K+ K\Phi K+ K \Phi K \Phi K+ \cdots)G_-.\label{prop_def}
\end{align}
From Example \ref{top_ampl_htq_deformation}, we can recognize the sum (\ref{prop_def}) as the deformation $Q \to Q - \Phi$ of the HTQM by an operator $\Phi=2\pi [ G_-, \cO_X]$.

For an external edge, attached to a leaf decorated with the evaluation state $\Psi_\a$, the sum over an arbitrary number of insertions of operators $\cO_X$
\begin{gather*}
\begin{tikzpicture}[scale=0.88]
\draw[ultra thick, dashed] (0,0)-- (2, 0) ;
\filldraw[black] (0,0) circle (2pt) ;
\filldraw[black] (2,0) circle (2pt) ;
\node (=) at (3,0) {$=$};
\draw[thick, dashed] (4,0)-- (6, 0) ;
\filldraw[black] (4,0) circle (2pt) ;
\filldraw[black] (6,0) circle (2pt) node[anchor=south]{$\Psi_\g$};
\node (+) at (7,0) {$+$};
\draw[thick] (8,0)-- (9, 0) ;
\draw[thick, dashed] (9,0)-- (10, 0) ;
\filldraw[black] (8,0) circle (2pt) ;
\filldraw[black] (10,0) circle (2pt) node[anchor=south]{$\Psi_\g$};
\filldraw[black] (9,0) circle (2pt) node[anchor=south]{$\cO_X$};
\node (+) at (11,0) {$+$};
\draw[thick] (12,0)-- (14, 0) ;
\draw[thick, dashed] (14,0)-- (15, 0) ;
\filldraw[black] (12,0) circle (2pt) ;
\filldraw[black] (15,0) circle (2pt) node[anchor=south]{$\Psi_\g$};
\filldraw[black] (13,0) circle (2pt) node[anchor=south]{$\cO_X$};
\filldraw[black] (14,0) circle (2pt) node[anchor=south]{$\cO_X$};
\node (+) at (16,0) {$+$};
\node (+) at (16.5,0) {$\dots$};
\end{tikzpicture}
\end{gather*}
evaluates into
\begin{align}
\Psi^X_\g &{}=\Psi_\g + 2\pi KG_-\cO_{X}\Psi_\g + (2\pi)^2 KG_-\cO_{X} KG_-\cO_{X} \Psi_\g +\cdots\nonumber\\
&{}= \sum_{d=0}^\infty (2\pi KG_-\cO_X)^d \Psi_\g.\label{B_type_QM_obs_deform}
\end{align}
This concludes the proof.
\end{proof}

\begin{Remark} The differential forms on $X$ of complex dimension $\dim_{\mathbb{C}}X=N$ have degree less or equal to $2N$. Each term in the expansion (\ref{prop_def}) decreases the degree of the form by 2, so the terms with $N$ or more divisor operator insertions will act trivially on any state. Hence, we conclude that $K^X$ is a polynomial in $\cO_X$ of degree $N-1$.
\end{Remark}

\subsection{Mirror states}
\begin{Definition}
The {\it mirror state} $\Psi_\g^X$ for toric space $X$ is the deformation (\ref{B_type_QM_obs_deform}) of an $A$-model evaluation state $\Psi_\g$ by the universal divisor operators $\cO_{X}$, i.e.,
\begin{gather}\label{form_mirr_state_def}
\Psi^X_\g =\sum_{d=0}^\infty (2\pi K G_- \cO_{X})^d \Psi_\g .
\end{gather}
\end{Definition}
Though the mirror state $\Psi_\g^X$ was defined as a series in toric moduli $q_b$, it is a polynomial in~$q_b$ due to the proposition below. We evaluated these polynomials for observables on arbitrary complex toric surfaces in \cite{Losev:2023uxa}.
\begin{Proposition} For an $A$-model state $\Psi^X_\g$ with $\g\in\Omega_{{\rm trop}}^{k,k}(X)$, the sum \eqref{form_mirr_state_def} contains at most $k+1$ terms.
\end{Proposition}
\begin{proof}
The action of $G_-K$ lowers the degree of the form by $(1,1)$, while the action of the $\cO_X$ preserves the degree. Hence, $G_-K$ can be applied at most $k$ times.
\end{proof}

\begin{Proposition} The mirror state $\Psi_\g^X$ is $Q^X$- and $G_-$-closed if the $A$-model state $\Psi_\g$ is $Q$- and $G_-$-closed.
\end{Proposition}
\begin{proof}
 The property $G_-\Psi^X_\g =0$ immediately follows from the relation $G_-KG_- = - G_-^2 K = 0$. The other property
\[
Q_X \Psi^X_\g =0
\]
requires careful usage of the QM properties from Section \ref{TGQ_QM_section}. In particular, we can evaluate
\begin{gather}\label{Q_X_mirror_state}
Q^X \Psi_\g^X = Q \Psi^X_\g - 2\pi [G_-, \cO_X] \Psi_\g^X = 2\pi K G_- \cO_{X} Q\Psi^X_\g = 2\pi K G_- \cO_{X} Q^X \Psi^X_\g
\end{gather}
and since $2\pi KG_-\cO_{X}\neq1$, the equality (\ref{Q_X_mirror_state}) completes the proof of the proposition.
\end{proof}

\subsection{Dual variables}
 It is convenient to introduce mirror angular variables $Y_j \in S^1 $ dual to the integer vector components $m^j \in \mathbb{Z}$. We introduce a Fourier transform of a state
\begin{gather}\label{A_B_htqm_states_transform}
\Psi= \sum_{\vec{m}\in \mathbb{Z}^N} {\rm e}^{{\rm i} \< \vec{m}, \vec{Y} \>} c_{\vec{m}}|\omega, \vec{m}\> \in V_B = \Omega^\ast_{{\rm trop}}(X )\otimes C^{\infty} \bigl(\mathbb{T}^N\bigr).
\end{gather}
The Fourier transform (\ref{A_B_htqm_states_transform}) is the tropical analog of a T-duality, in the context of the mirror for toric varieties see \cite{Frenkel:2005ku}.

We describe the differential forms on $X$ using Grassmann variables $\psi_R^j$ and $\psi_\Phi^j$. The~$Q$ and~$G_\pm$ on the states (\ref{A_B_htqm_states_transform}) become
\begin{gather}\label{diff_op_repres}
Q\Psi = \psi_R^k \frac{\p}{\p r^k} \Psi,\qquad G_-\Psi = - {\rm i} \frac{\p}{\p Y_k} \frac{\p}{\p \psi_{\Phi}^k}\Psi, \qquad G_+\Psi = -{\rm i} \frac{\p}{\p Y_k} \frac{\p}{\p \psi_{R}^k}\Psi.
 \end{gather}
The multiplication $\mu_2$, on forms becomes multiplication of functions on superspace with coordinates $r$, $Y$, $\psi_R$, $\psi_\Phi$, i.e.,
\[
\mu_2 (\Psi_1 ,\Psi_2) =\Psi_1\cdot \Psi_2,
\]
The pairing $g$ is the integration over superspace, i.e.,
\[
g(\Psi_1, \Psi_2) =\int {\rm d}\mu\, \Psi_1\Psi_2.
\]
The Berezin integration measure for $\dim_{\mathbb{C}}X=N$ is
\[
{\rm d}\mu = {\rm d}^N r {\rm d}^N Y {\rm d}^N\psi_{\Phi} {\rm d}^N \psi_{R}.
\]
The integrating region (for Grassmann-even variables) is the $N$-dimensional torus \smash{$\bigl(S^1\bigr)^N$} for $Y$-variables and Euclidean space $\mathbb{R}^N$ for $r$-variables.

The differential operator representation (\ref{diff_op_repres}) of the HTQM data $(V, Q, G_{\pm}, \mu_2, g)$ allows for an easy check of HTQM definitions from Section \ref{sec_htqm_spec_tees}.
In particular, the 7-term relation (\ref{def_7_term_relation}) for $G_-$ is a property of the second-order differential operator in representation (\ref{diff_op_repres}).

The divisor state $|1, \vec{b}\>$ becomes an exponential function of $Y$, i.e.,
\[
\Psi_b = {\rm e}^{{\rm i}\<\vec{b},\vec{Y}\>}={\rm e}^{{\rm i} b^kY_k}.
\]
For the tropical form
\[
\g = \g_{i_1\dots i_kj_1 \dots j_l}(r) {\rm d}\phi^{i_1}\wedge\cdots\wedge {\rm d}\phi^{i_k}\wedge {\rm d}r^{j_1}\wedge\cdots\wedge {\rm d}r^{j_l},
\]
the corresponding evaluation state is
\[
\Psi_\g = \g_{i_1\dots i_kj_1\dots j_l}(r) \psi_{\Phi}^{i_1}\cdots \psi_{\Phi}^{i_k} \cdot\psi_R^{j_1}\cdots\psi_R^{j_l}.
\]
The deformation of the $A$-model by the universal divisor operator $\cO_X$ is
\[
Q^X=Q-2\pi [G_-, \cO_X]=\psi_R^j \frac{\p}{\p r^j}+ 2\pi {\rm i} \sum_{\vec{b} \in B_X} q_{b} \frac{\p \Psi_b}{ \p Y_j }\frac{\p}{\p \psi_{\Phi}^j} .
\]
The same differential can be written as
\[
Q^X=Q + 2\pi {\rm i} \frac{\p W_X}{ \p Y_j }\frac{\p}{\p\psi_{\Phi}^j},
\]
for the {\it mirror superpotential}
\begin{gather}\label{eq_mirr_superp_explicit}
W_X = \sum_{\vec{b}\in B_X} q_{b} \Psi_{b}=\sum_{\vec{b} \in B_X} q_{b} {\rm e}^{{\rm i}\<\vec{b} ,\vec{Y}\>}.
\end{gather}
We can remove some number of toric moduli $q_{b}$ by a redefinition of $Y_j$ to obtain a more familiar \big(at least for the $\mathbb{P}^N$ case\big) form of the superpotential with fewer parameters $q_{b}$.

Our expression for the tropical mirror superpotential (\ref{eq_mirr_superp_explicit}) for toric space $X$ matches with the mirror superpotentials in complex geometry, derived by Givental \cite{givental1995quantum}, Hori and Vafa \cite{Hori:2000kt}, Frenkel and Losev \cite{Frenkel:2005ku} using different methods. For $X = \mathbb{P}^2$, the same mirror superpotential was derived by Gross \cite{gross2010mirror} using the tropical curve counting with Mikhalkin's vertex multiplicities.

\subsection{Tropical mirror relation}
\begin{Theorem} The $n\geq 3$ point tropical GW invariants for toric variety $X$ and cycles $\g_1,\dots,\g_n$ equal to the total amplitude in the $B$-model HTQM for the corresponding mirror states and the superpotential, defined by the compactifying divisors $B_X$ of $X$, i.e.,
\begin{gather*}
 \sum_{\beta\in H_2 (X)}q^\beta\<\g_1,\dots,\g_n \>^X_\beta = \<\Psi^X_{\g_1},\dots,\Psi^X_{\g_n}\>_{Q^X},
\\
\Psi^X_\g =\sum_{d=0}^\infty (2\pi K G_- \cO_{X})^d \Psi_\g,\qquad Q_X=Q + 2\pi {\rm i} \frac{\p W_X}{ \p Y_j }\frac{\p}{\p \psi_{\Phi}^j} ,\qquad W_X = \sum_{\vec{b} \in B_X} q_{b} {\rm e}^{{\rm i}\<\vec{b} ,\vec{Y}\>}.
\end{gather*}
\end{Theorem}
\begin{proof}
The proof of the theorem follows from Propositions \ref{prop_total_ampl_trop_gw} and \ref{prop_total_ampl_b_model}.
\end{proof}

The $B$-model HTQM with superpotential is the quantum mechanical version of the Landau--Ginzburg theory in 2 dimensions.
\begin{Remark} The $B$-model HTQM is a version of supersymmetric quantum mechanics with~4 supercharges \cite{Cooper:1994eh}, where the anti-holomorphic superpotential equals zero and $Y$ and $r$ are not complex conjugates.
\end{Remark}

\section[Tropical mirror for P\^{}1]{Tropical mirror for $\boldsymbol{\mathbb{P}^1}$}
The second homology $H_2\bigl(\mathbb{P}^1\bigr)$ is one-dimensional, so the degree of a curve is a single positive integer number $d$. There is only one non-trivial cycle on $\mathbb{P}^1$, a point cycle. The genus 0 GW invariants for $\mathbb{P}^1$ with $3$ or more cycles are non-zero for degree one curves only. Moreover, for any number $n\geq 2$,
$
N^{\mathbb{P}^1}_1 (p_1,\dots,p_n) = 1$.
The tropical correspondence theorem \cite{mikhalkin2005enumerative} holds for $\mathbb{P}^1$, hence we can evaluate the tropical GW invariants
\begin{gather}\label{trop_complex_gw_invarians_p_1}
TN^{\mathbb{P}^1}_1 (p_1,\dots,p_n) =N^{\mathbb{P}^1}_1 (p_1,\dots,p_n) = 1.
\end{gather}
We will use the $U(1)$-averaged representative to the Poincar\'e dual to the tropical point cycle at radial position $r_\a$ is
\[
\g_\a = \frac{1}{2\pi} \delta (r-r_\a){\rm d}\phi\wedge {\rm d}r.
\]

\subsection{Tropical Gromov--Witten invariants}
Our analysis for the tropical GW invariant for two points on $\mathbb{P}^1$ from Section \ref{moduli_space_integr_section} immediately generalizes to the case of $n$ points.
The discrete data of tropical curves of degree 1 with $n$ marked points is the order of marked points on a line, labeled by a permutation $\sigma \in S_n$. The tropical curve $\vec{\G}_{(12\dots n)}$ is presented below:
 
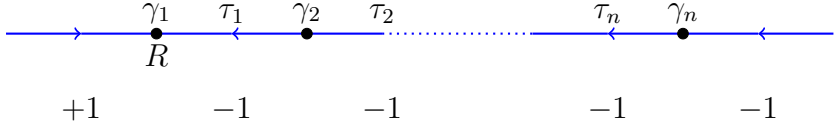
\[
 \begin{tikzpicture}[scale=1]
\draw[blue, thick, ->] (0, 0) -- (1, 0);
\draw[ blue, thick](1, 0)-- (3,0);
\draw[blue, thick, ->](4, 0)-- (3,0);
\draw[blue, thick](4, 0)-- (5,0);
\draw[blue, thick,dotted](5, 0)-- (7,0);
\draw [blue, thick](8, 0)-- (7,0);
\draw[blue, thick,->](9, 0)-- (8,0);
\draw [blue, thick, ->](11, 0)-- (10,0);
\draw [blue, thick](10, 0)-- (9,0);
\filldraw[black] (2,0) circle (2pt) node[anchor=south]{$\g_1$} node[anchor=north]{$R$};
\filldraw[black] (4,0) circle (2pt) node[anchor=south]{$\g_2$};
\filldraw[black] (9,0) circle (2pt) node[anchor=south]{$\g_n$};
\node at (3,0) [anchor=south]{$\tau_1$};
\node at (5,0) [anchor=south]{$\tau_2$};
\node at (8,0) [anchor=south]{$\tau_{n}$};
\node at (1,-1) {$+1$};
\node at (3,-1) {$-1$};
\node at (5,-1) {$-1$};
\node at (8,-1) {$-1$};
\node at (10,-1) {$-1$};
\end{tikzpicture}
\]
The tropical GW invariant is a sum over permutations, i.e.,
 \begin{gather}\label{trop_gw_n_point_p1}
\< \gamma_1,\dots, \gamma_n\>_{0,1}^{\mathbb{P}^1} = \sum_{\sigma\in S_n} \< \gamma_1,\dots, \g_n\>_{\vec{\G}_\sigma}= \sum_{\sigma \in S_n} \int_{(S^1)^{n}} \int_{\mathbb{R} \times (\mathbb{R}^{+})^{n-1} } \bigwedge_{\a=1}^n {\rm ev}_\a^\ast \g_\a.
 \end{gather}
Each contribution is evaluated into
\[
 \< \gamma_1,\dots, \g_n\>_{\vec{\G}_\sigma} = \Theta \bigl(r_{\sigma(1)}, r_{\sigma(2)},\dots,r_{\sigma(n)}\bigr),
\]
where we used the multivariable $\Theta$-function
\[
\Theta (r_{1}, r_{2},\dots,r_{n}) =
\begin{cases}
1& \hbox{when}\ r_1<r_2<\dots<r_{n-1}<r_n, \\
0& \hbox{otherwise}.
\end{cases}
\]
The sum over permutations in (\ref{trop_gw_n_point_p1}) matches with the tropical correspondence theorem prediction~(\ref{trop_complex_gw_invarians_p_1}).

\subsection[3-point invariant via A-model HTQM]{3-point invariant via $\boldsymbol{A}$-model HTQM}\label{sec_3_point_A_model_p_1}
For $n=3$, we have $3! = 6$ distinct tropical curves and the same number of special trees in HTQM. For the permutation $(123)$, the decorated special tree $\G_{(123)}$ is presented in the left picture below. In the right picture, we performed the state-operator map for the evaluation operators $\cO_{\g_\a}$:
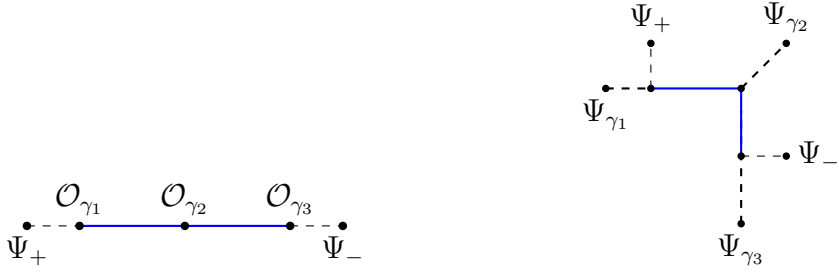
\[
\begin{tikzpicture}[scale=0.7]
\draw [blue,thick](1, 0)--(3,0)--(5,0);
\draw[black,dashed](0,0)-- (1, 0);
\draw[black,dashed](5,0)-- (6, 0);
\filldraw[black] (0,0) circle (2pt) node[anchor=north]{$\Psi_+$} ;
\filldraw[black] (1,0) circle (2pt) node[anchor=south]{$\cO_{\g_1}$} ;
\filldraw[black] (3,0) circle (2pt) node[anchor=south]{$\cO_{\g_2}$} ;
\filldraw[black] (5,0) circle (2pt) node[anchor=south]{$\cO_{\g_3}$} ;
\filldraw[black] (6,0) circle (2pt) node[anchor=north]{$\Psi_-$} ;
\end{tikzpicture}
\qquad
\begin{tikzpicture}[scale=0.6]
\draw[black,thick,dashed] (-3,0)-- (-2, 0) ;
\draw[blue,thick] (-2,0)-- (0, 0) ;
\draw[black,thick,dashed] (0, -3) -- (0, -.5);
\draw[blue,thick] (0, -1.5) -- (0, 0);
\draw[black,thick,dashed] (0, 0) -- (1, 1);
\filldraw[black] (0,0) circle (2pt);
\filldraw[black] (-3,0) circle (2pt) node[anchor=north]{$\Psi_{\g_1}$};
\filldraw[black] (0,-1.5) circle (2pt);
\filldraw[black] (1,-1.5) circle (2pt) node[anchor=west]{$\Psi_{-}$};
\draw[color=black, dashed] (0,-1.5)--(1,-1.5);
\filldraw[black] (1,1) circle (2pt) node[anchor=south]{$\Psi_{\g_2}$};
\filldraw[black] (0,-3) circle (2pt) node[anchor=north]{$\Psi_{\g_3}$};
\filldraw[black] (-2,0) circle (2pt);
\filldraw[black] (-2,1) circle (2pt) node[anchor=south]{$\Psi_{+}$};
\draw[color=black, dashed ] (-2,0)--(-2,1);
\end{tikzpicture}
\]
The $A$-model amplitude for the special tree $\G_{(123)}$ is
\begin{align*}
 \< \gamma_1, \gamma_2, \gamma_3\>_{\G_{(123)}} &{} = g(\Psi_+,\cO_{\g_1} 2\pi KG_- \cO_{\g_2} 2\pi KG_-\cO_{\g_3} \Psi_-) \\
&{}=g(\Psi_+, \cO_{\g_1} 2\pi KG_- | \gamma_2 \Theta (r_3-r) , -1\> ) \\
&{}=g(|1,1\>, | \gamma_1 \Theta (r_3-r_2) \Theta (r_2-r), -1\> ) \\
&{}=\int_{{S}^1} {\rm d}\phi \int_{\mathbb{R}} {\rm d}r \, \frac{1}{2\pi}\delta(r-r_1) \Theta (r_3-r_2) \Theta (r_2-r) = \Theta (r_1, r_2, r_3).
\end{align*}

\subsection{3-point invariant via total amplitude}
We turn three evaluation observables $\cO_\g$ into states $\Psi_\g$, so we have five states in total. Hence, we need a total amplitude on a special tree $\G_5$ with five leaves. There is a single special tree $\G_5$ with 5 leaves, with symmetry factor $|{\rm Aut}(\G_5)|=8$. The total amplitude on $\G_5$ evaluates into
\[
\< \Psi_{\g_1}, \Psi_{\g_2}, \Psi_{\g_3}, \Psi_{+}, \Psi_{-}\>_{Q} = \sum_{\sigma \in S_5} \frac{1}{|{\rm Aut}(\G_5)|} \mathcal{A}_\G \bigl(\Psi_{\g_{\sigma(1)}}, \Psi_{\g_{\sigma(2)}}, \Psi_{\g_{\sigma(3)}}, \Psi_{\sigma(+)}, \Psi_{\sigma(-)}\bigr).
\]
Among $|S_5| =120$ amplitudes there are $|S_5|/|{\rm Aut}(\G_5)|$=15 distinct ones, but only 6 are non-zero. Below, we present all 15 distinct amplitudes: 6 non-zero contributions for the first special tree and three groups of 3 amplitudes, which vanish in the remaining special trees:
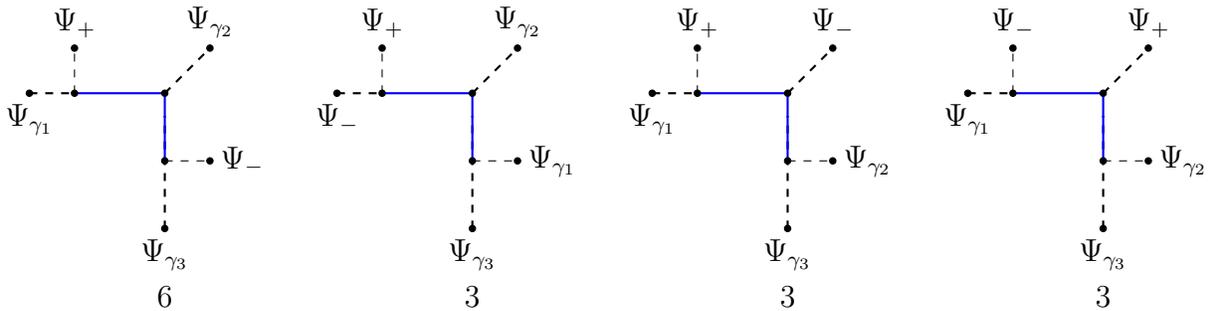
\begin{gather*}
\begin{tikzpicture}[scale=0.54]
\draw[black, thick, dashed] (-3,0)-- (-2, 0) ;
\draw[blue, thick] (-2,0)-- (0, 0) ;
\draw[black, thick, dashed] (0, -3) -- (0, -.5);
\draw[blue, thick] (0, -1.5) -- (0, 0);
\draw[black, thick, dashed] (0, 0) -- (1, 1);
\filldraw[black] (0,0) circle (2pt);
\filldraw[black] (-3,0) circle (2pt) node[anchor=north]{$\Psi_{\g_1}$};
\filldraw[black] (0,-1.5) circle (2pt);
\filldraw[black] (1,-1.5) circle (2pt) node[anchor=west]{$\Psi_{-}$};
\draw[color=black, dashed] (0,-1.5)--(1,-1.5);
\filldraw[black] (1,1) circle (2pt) node[anchor=south]{$\Psi_{\g_2}$};
\filldraw[black] (0,-3) circle (2pt) node[anchor=north]{$\Psi_{\g_3}$};
\filldraw[black] (-2,0) circle (2pt);
\filldraw[black] (-2,1) circle (2pt) node[anchor=south]{$\Psi_{+}$};
\draw[color=black, dashed ] (-2,0)--(-2,1);
\node (=) at (0,-4.5) {$6$};
\end{tikzpicture}\quad
\begin{tikzpicture}[scale=0.54]
\draw[black, thick, dashed] (-3,0)-- (-2, 0) ;
\draw[blue, thick] (-2,0)-- (0, 0) ;
\draw[black, thick, dashed] (0, -3) -- (0, -.5);
\draw[blue, thick] (0, -1.5) -- (0, 0);
\draw[black, thick, dashed] (0, 0) -- (1, 1);
\filldraw[black] (0,0) circle (2pt);
\filldraw[black] (-3,0) circle (2pt) node[anchor=north]{$\Psi_{-}$};
\filldraw[black] (0,-1.5) circle (2pt);
\filldraw[black] (1,-1.5) circle (2pt) node[anchor=west]{$\Psi_{\g_1}$};
\draw[color=black, dashed] (0,-1.5)--(1,-1.5);
\filldraw[black] (1,1) circle (2pt) node[anchor=south]{$\Psi_{\g_2}$};
\filldraw[black] (0,-3) circle (2pt) node[anchor=north]{$\Psi_{\g_3}$};
\filldraw[black] (-2,0) circle (2pt);
\filldraw[black] (-2,1) circle (2pt) node[anchor=south]{$\Psi_{+}$};
\draw[color=black, dashed ] (-2,0)--(-2,1);
\node (=) at (0,-4.5) {$3$};
\end{tikzpicture}\quad
\begin{tikzpicture}[scale=0.54]
\draw [black, thick, dashed] (-3,0)-- (-2, 0) ;
\draw [blue, thick] (-2,0)-- (0, 0) ;
\draw [black, thick, dashed] (0, -3) -- (0, -.5);
\draw [blue, thick] (0, -1.5) -- (0, 0);
\draw [black, thick, dashed] (0, 0) -- (1, 1);
\filldraw[black] (0,0) circle (2pt);
\filldraw[black] (-3,0) circle (2pt) node[anchor=north]{$\Psi_{\g_1}$};
\filldraw[black] (0,-1.5) circle (2pt);
\filldraw[black] (1,-1.5) circle (2pt) node[anchor=west]{$\Psi_{\g_2}$};
\draw[color=black, dashed] (0,-1.5)--(1,-1.5);
\filldraw[black] (1,1) circle (2pt) node[anchor=south]{$\Psi_{-}$};
\filldraw[black] (0,-3) circle (2pt) node[anchor=north]{$\Psi_{\g_3}$};
\filldraw[black] (-2,0) circle (2pt);
\filldraw[black] (-2,1) circle (2pt) node[anchor=south]{$\Psi_{+}$};
\draw[color=black, dashed ] (-2,0)--(-2,1);
\node (=) at (0,-4.5) {$3$};
\end{tikzpicture}\quad
\begin{tikzpicture}[scale=0.54]
\draw [black, thick, dashed] (-3,0)-- (-2, 0) ;
\draw [blue, thick] (-2,0)-- (0, 0) ;
\draw [black, thick, dashed] (0, -3) -- (0, -.5);
\draw [blue, thick] (0, -1.5) -- (0, 0);
\draw [black, thick, dashed] (0, 0) -- (1, 1);
\filldraw[black] (0,0) circle (2pt);
\filldraw[black] (-3,0) circle (2pt) node[anchor=north]{$\Psi_{\g_1}$};
\filldraw[black] (0,-1.5) circle (2pt);
\filldraw[black] (1,-1.5) circle (2pt) node[anchor=west]{$\Psi_{\g_2}$};
\draw[color=black, dashed] (0,-1.5)--(1,-1.5);
\filldraw[black] (1,1) circle (2pt) node[anchor=south]{$\Psi_{+}$};
\filldraw[black] (0,-3) circle (2pt) node[anchor=north]{$\Psi_{\g_3}$};
\filldraw[black] (-2,0) circle (2pt);
\filldraw[black] (-2,1) circle (2pt) node[anchor=south]{$\Psi_{-}$};
\draw[color=black, dashed ] (-2,0)--(-2,1);
\node (=) at (0,-4.5) {$3$};
\end{tikzpicture}
\end{gather*}
The correlation function is given by the six non-zero contributions, labeled by the permutations of evaluation observables $\g_1$, $\g_2$, $\g_3$
\begin{align*}
\< \Psi_{\g_1}, \Psi_{\g_2}, \Psi_{\g_3}, q_+\Psi_{+}, q_-\Psi_{-}\>_{Q} &= \sum_{\sigma \in S_3} \mathcal{A}_\G (\Psi_{\g_{\sigma(1)}}, \Psi_{\g_{\sigma(2)}}, \Psi_{\g_{\sigma(3)}}, q_+\Psi_{+}, q_-\Psi_{-}) \\
&=q_+q_- \sum_{\sigma\in S_3} \Theta (r_{\sigma(1)}, r_{\sigma(2)}, r_{\sigma(3)}) = q_+q_-.
\end{align*}
Each of the six amplitudes has the following form:
\begin{gather*}
\mathcal{A}_\G (\Psi_{\g_1}, \Psi_{\g_2}, \Psi_{\g_3}, \Psi_{+}, \Psi_{-}) = \mu_3^0 (2\pi KG_- \mu_2 (\Psi_{\g_1}, \Psi_{+}), 2\pi KG_- (\Psi_{\g_3}, \Psi_{-}), \Psi_{\g_2}) \\
\qquad{}=\frac{1}{2\pi} \int {\rm d}Y \int {\rm d}\psi_R {\rm d}\psi_\Phi \int {\rm d}r \, {\rm e}^{{\rm i}Y} \Theta (r-r_1) \cdot {\rm e}^{-{\rm i}Y} \Theta (r_3-r) \cdot \delta (r-r_2) \psi_\Phi \psi_R \\
\qquad{}=q \int {\rm d}r\, \Theta (r-r_1) \Theta (r_3-r) \delta (r-r_2) = \Theta (r_1, r_2, r_3).
\end{gather*}
We used the $KG_-$ action on the product of divisor state $\Psi_\pm$ and an evaluation state $\Psi_\g$
\begin{align*}
2\pi KG_- \mu_2 (\Psi_{\g_1}, \Psi_{\pm}) &{}=2\pi \int^\infty_0 {\rm d}t\, {\rm e}^{-tH} G_+G_- (\Psi_{\g_1}, \Psi_{\pm}) \\
&{}=2\pi \int^\infty_0 {\rm d}t\, {\rm e}^{-tH} G_+G_- \biggl(\frac{1}{2\pi}\delta (r-r_1) \psi_\Phi \psi_R \cdot {\rm e}^{\pm {\rm i} Y} \biggr) \\
&{}={\rm e}^{\pm {\rm i}Y} \int^\infty_0 {\rm d}t\, \delta (r-r_1 \mp t) = {\rm e}^{\pm {\rm i}Y} \Theta (\pm(r-r_1)).
\end{align*}

\subsection{Mirror map}
In toric description of $\mathbb{P}^1$, we have $B_{\mathbb{P}^1}=\{ 1, -1\}$. The corresponding divisor operators are $\cO_{\pm}={\rm e}^{\pm {\rm i}Y}$. The deformation of differential
\[
Q^{\mathbb{P}^1} = Q - 2\pi q_+ [G_-, \cO_+] - 2\pi q_- [G_-, \cO_-] =\psi_R\frac{\p}{\p r }- 2\pi \bigl(q_+ {\rm e}^{{\rm i}Y} - q_-{\rm e}^{-{\rm i}Y}\bigr) \frac{\p}{\p \psi_\Phi}.
\]
Hence, the mirror superpotential is
\begin{gather}\label{w_p_1_mirror}
W_{\mathbb{P}^1} = q_+ {\rm e}^{{\rm i}Y} +q_-{\rm e}^{-{\rm i}Y}.
\end{gather}
 We can shift $Y$ and express the K\"ahler module $q = q_+q_-$ in terms of two toric moduli $q_+$, $q_-$ to rewrite the superpotential to the more familiar form
\[
W_{\mathbb{P}^1} = {\rm e}^{{\rm i}Y} +q {\rm e}^{-{\rm i}Y}.
\]
The mirror state for the point cycle at $r=r_0$
\begin{align}
\Psi^{\mathbb{P}^1}_{\g}&{}=\Psi_{\g} + 2\pi KG_- q_+\cO_+ \Psi_{\g} + 2\pi KG_- q_-\cO_-\Psi_{\g} \nonumber\\
&{}=\Psi_{\g} + 2\pi q_+ \int^\infty_0{\rm d}t\, {\rm e}^{-tH} G_+G_- \bigl({\rm e}^{{\rm i}Y}\Psi_{\g}\bigr) +2\pi q_- \int^\infty_0{\rm d}t\, {\rm e}^{-tH} G_+G_- \bigl({\rm e}^{-{\rm i}Y} \Psi_{\g}\bigr) \nonumber\\
&{}= \frac{1}{2\pi}\delta (r-r_0)\psi_\Phi \psi_R + q_+ {\rm e}^{{\rm i}Y} \Theta (r-r_0) + q_- {\rm e}^{-{\rm i}Y} \Theta(r_0-r).\label{form_mirr_state_point_p1}
\end{align}

\subsection[3-point invariant via B-model]{3-point invariant via $\boldsymbol{B}$-model}
 The $B$-model total amplitude on a special tree, decorated with three mirror states $\Psi^{\mathbb{P}^1}_{\g_\a}$, is given by a single $Y$-shaped tree.
 The amplitude evaluates into
\begin{gather*}
\bigl\<\Psi^{\mathbb{P}^1}_{\g_1} , \Psi^{\mathbb{P}^1}_{\g_2}, \Psi^{\mathbb{P}^1}_{\g_3}\bigr\>_{Q^{\mathbb{P}^1}} = \mu_{3}^0 \bigl(\Psi^{\mathbb{P}^1}_{\g_1}, \Psi^{\mathbb{P}^1}_{\g_2}, \Psi^{\mathbb{P}^1}_{\g_3}\bigr)
=\int_{S^1} {\rm d}Y \int {\rm d}\psi_R {\rm d}\psi_\Phi \int^{+\infty}_{-\infty} {\rm d}r\, \Psi^{\mathbb{P}^1}_{\g_1} \cdot \Psi^{\mathbb{P}^1}_{\g_2}\cdot \Psi^{\mathbb{P}^1}_{\g_3} \\
\qquad{}=q_+q_- \int^{+\infty}_{-\infty} {\rm d}r \, \delta (r-r_1) ( \Theta (r-r_2) \Theta (r_3-r) + \Theta (r_2-r) \Theta (r-r_3) ) +(\text{c.p.}) \\
\qquad{}=q_+q_- \Theta (r_2, r_1, r_3) + q_+q_-\Theta (r_3, r_1, r_2) +(\text{c.p.}) \\
\qquad{}=q_+q_- \sum_{\sigma\in S_3} \Theta \bigl(r_{\sigma(1)}, r_{\sigma(2)}, r_{\sigma(3)}\bigr) = q_+q_-.
\end{gather*}
Alternatively, we can express the same amplitude in $A$-model notations. Each mirror state $\Psi^X_{\g}$ is a sum of 3 terms (\ref{form_mirr_state_point_p1}), hence a single $B$-model amplitude is a sum of 27 $A$-model amplitudes. We arranged the 27 amplitudes in several groups in the picture below. The number below the special tree indicates the number of diagrams of the same type \big(different choice of vectors $\vec{b}_a$ and a permutation of $\g_1$, $\g_2$, $\g_3$\big):
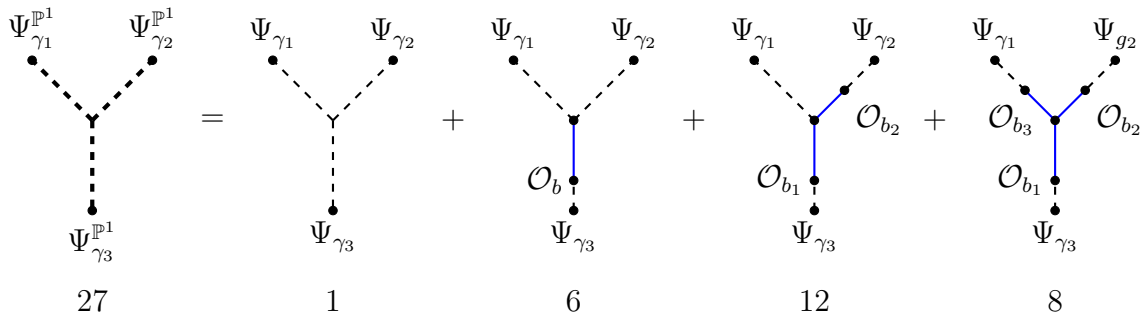
\begin{gather*}
\begin{tikzpicture}[scale=0.78]
\draw[ultra thick, dashed] (0,0)-- (0, -1.5) ;
\draw[ultra thick, dashed ] (0,0)-- (1, 1) ;
\draw[ultra thick, dashed] (0,0)-- (-1, 1) ;
\filldraw[black] (0,-1.5) circle (2pt) node[anchor=north]{$\Psi_{\g_3}^{\mathbb{P}^1}$} ;
\filldraw[black] (1,1) circle (2pt) node[anchor=south]{$\Psi_{\g_2}^{\mathbb{P}^1}$} ;
\filldraw[black] (-1,1) circle (2pt) node[anchor=south]{$\Psi_{\g_1}^{\mathbb{P}^1}$} ;
\node (=) at (2,0) {$=$};
\node (=) at (0,-3) {$27$};
\draw[ thick, dashed] (4,0)-- (4, -1.5) ;
\draw[thick, dashed ] (4,0)-- (5, 1) ;
\draw[thick, dashed] (4,0)-- (3, 1) ;
\filldraw[black] (4,-1.5) circle (2pt) node[anchor=north]{$\Psi_{\g_3}$} ;
\filldraw[black] (5,1) circle (2pt) node[anchor=south]{$\Psi_{\g_2}$} ;
\filldraw[black] (3,1) circle (2pt) node[anchor=south]{$\Psi_{\g_1}$} ;
\node at (6,0) {$+$};
\node at (4,-3) {$1$};
\draw[ thick, dashed] (8,-1.5)-- (8, -1) ;
\draw[ thick, blue] (8,-1)-- (8, 0) ;
\draw[thick, dashed ] (8,0)-- (9, 1) ;
\draw[thick, dashed] (8,0)-- (7, 1) ;
\filldraw[black] (8,0) circle (2pt) ;
\filldraw[black] (8,-1) circle (2pt) node[anchor=east]{$\cO_b$} ;
\filldraw[black] (8,-1.5) circle (2pt) node[anchor=north]{$\Psi_{\g_3}$} ;
\filldraw[black] (9,1) circle (2pt) node[anchor=south]{$\Psi_{\g_2}$} ;
\filldraw[black] (7,1) circle (2pt) node[anchor=south]{$\Psi_{\g_1}$} ;
\node at (10,0) {$+$};
\node at (8,-3) {$6$};
\draw[ thick, dashed] (12,-1.5)-- (12, -1) ;
\draw[ thick, blue] (12,-1)-- (12, 0) ;
\draw[thick, blue ] (12,0)-- (12.5, 0.5) ;
\draw[thick, dashed ] (13,1)-- (12.5, 0.5) ;
\draw[thick, dashed] (12,0)-- (11, 1) ;
\filldraw[black] (12,0) circle (2pt) ;
\filldraw[black] (12,-1) circle (2pt) node[anchor=east]{$\cO_{b_1}$} ;
\filldraw[black] (12.5,0.5) circle (2pt);
\node at (12.5,0) [anchor=west]{$\cO_{b_2}$};
\filldraw[black] (12,-1.5) circle (2pt) node[anchor=north]{$\Psi_{\g_3}$} ;
\filldraw[black] (13,1) circle (2pt) node[anchor=south]{$\Psi_{\g_2}$} ;
\filldraw[black] (11,1) circle (2pt) node[anchor=south]{$\Psi_{\g_1}$} ;
\node at (14,0) {$+$};
\node at (12,-3) {$12$};
\draw[thick, dashed] (16,-1.5)-- (16, -1) ;
\draw[thick, blue] (16,-1)-- (16, 0) ;
\draw[thick, blue ] (16,0)-- (16.5, 0.5) ;
\draw[thick, dashed ] (17,1)-- (16.5, 0.5) ;
\draw[thick, blue] (16,0)-- (15.5, 0.5) ;
\draw[thick, dashed] (15.5, 0.5) -- (15,1) ;
\filldraw[black] (16,0) circle (2pt) ;
\filldraw[black] (16,-1) circle (2pt) node[anchor=east]{$\cO_{b_1}$} ;
\filldraw[black] (16.5,0.5) circle (2pt) ;
\node at (16.5,0) [anchor=west]{$\cO_{b_2}$};
\filldraw[black] (15.5,0.5) circle (2pt);
\node at (14.7,0) [anchor=west]{$\cO_{b_3}$};
\filldraw[black] (16,-1.5) circle (2pt) node[anchor=north]{$\Psi_{\g_3}$} ;
\filldraw[black] (17,1) circle (2pt) node[anchor=south]{$\Psi_{g_2}$} ;
\filldraw[black] (15,1) circle (2pt) node[anchor=south]{$\Psi_{\g_1}$} ;
\node at (16,-3) {$8$};
\end{tikzpicture}
\end{gather*}
The first special tree does not have any divisor operators, so it is a degree 0 $A$-model contribution, which vanishes by Lemma \ref{lemma_trop_degree_selection}, since $\deg \g_1+ \deg \g_2 + \deg \g_3 = 6 \neq 2 = \dim_{\mathbb{R} } \mathbb{P}^1$. The special trees from the second and fourth groups have an odd number of divisor operators. Hence, the sum of the corresponding normal vectors is always non-zero, and according to Lemma \ref{lemma_trop_degree_selection}, the corresponding amplitudes vanish. Six of the third group of 12 special trees are non-vanishing and have $\vec{b}_1+\vec{b}_2 = 0$. The state-operator map can rearrange the non-vanishing amplitudes into a sum over $6= 3!$ permutations from Section \ref{sec_3_point_A_model_p_1}.

\subsection[4-point invariant via B-model]{4-point invariant via $\boldsymbol{B}$-model}
The $B$-model description of the tropical GW invariant for 4 point cycles on $\mathbb{P}^1$ is the total amplitude on a special tree with four leaves. There is only one such tree with the symmetry factor $|{\operatorname{Aut}}\,\G_4|=8$. Hence, the sum over $4! = 24$ permutations of four states becomes the sum of $24/8 =3$ terms
 \begin{gather}\label{total_ampl_4_pt_p1}
\bigl\< \Psi^{\mathbb{P}^1}_{1},\Psi^{\mathbb{P}^1}_2,\Psi^{\mathbb{P}^1}_{3}, \Psi^{\mathbb{P}^1}_4\bigr\>_{Q^{\mathbb{P}^1}}=\mathcal{A}_{\G_s} \bigl(\Psi^{\mathbb{P}^1}_{\g_\a}\bigr)
+\mathcal{A}_{\G_t} \bigl(\Psi^{\mathbb{P}^1}_{\g_\a}\bigr)+\mathcal{A}_{\G_u} \bigl(\Psi^{\mathbb{P}^1}_{\g_\a}\bigr).
\end{gather}
The three special trees $\G_s$, $\G_t$ and $\G_u$, decorated with states $\Psi_{\g_\a}^{\mathbb{P}^1}$ are presented below:
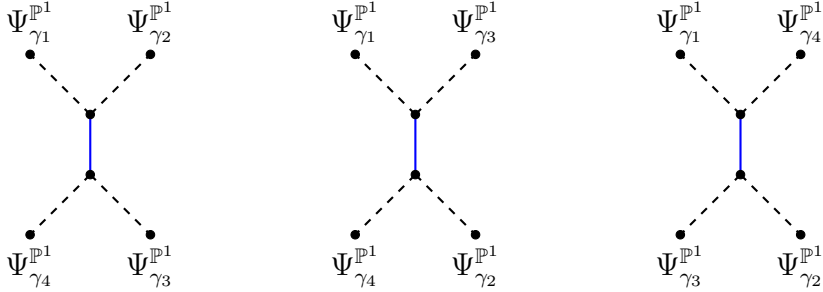
\[
 \begin{tikzpicture}[scale=0.8]
\draw[color=black,dashed,thick] (0, -1) -- (-1, -2);
\draw[color=black,dashed,thick] (0, -1) -- (1, -2);
\draw[color=blue,thick](0, -1) -- (0, 0);
\draw[color=black,dashed,thick](0, 0) -- (1, 1);
\draw[color=black,dashed,thick](0, 0) -- (-1, 1);
\filldraw[black] (0,0) circle (2pt);
\filldraw[black] (0,-1) circle (2pt);
\filldraw[black] (-1,-2) circle (2pt) node[anchor=north]{$\Psi^{\mathbb{P}^1}_{\g_4}$};
\filldraw[black] (-1,1) circle (2pt) node[anchor=south]{$\Psi^{\mathbb{P}^1}_{\g_1}$};
\filldraw[black] (1,-2) circle (2pt) node[anchor=north]{$\Psi^{\mathbb{P}^1}_{\g_3}$};
\filldraw[black] (1,1) circle (2pt) node[anchor=south]{$\Psi^{\mathbb{P}^1}_{\g_2}$};
\end{tikzpicture}
\qquad
 \begin{tikzpicture}[scale=0.8]
\draw[color=black, dashed, thick] (0, -1) -- (-1, -2);
\draw[color=black, dashed, thick] (0, -1) -- (1, -2);
\draw[color=blue, thick] (0, -1) -- (0, 0);
\draw[color=black, dashed, thick] (0, 0) -- (1, 1);
\draw[color=black, dashed, thick] (0, 0) -- (-1, 1);
\filldraw[black] (0,0) circle (2pt);
\filldraw[black] (0,-1) circle (2pt);
\filldraw[black] (-1,-2) circle (2pt) node[anchor=north]{$\Psi^{\mathbb{P}^1}_{\g_4}$};
\filldraw[black] (-1,1) circle (2pt) node[anchor=south]{$\Psi^{\mathbb{P}^1}_{\g_1}$};
\filldraw[black] (1,-2) circle (2pt) node[anchor=north]{$\Psi^{\mathbb{P}^1}_{\g_2}$};
\filldraw[black] (1,1) circle (2pt) node[anchor=south]{$\Psi^{\mathbb{P}^1}_{\g_3}$};
\end{tikzpicture}
\qquad
 \begin{tikzpicture}[scale=0.8]
\draw[color=black,dashed,thick] (0, -1) -- (-1, -2);
\draw[color=black,dashed,thick] (0, -1) -- (1, -2);
\draw[color=blue,thick](0, -1) -- (0, 0);
\draw[color=black,dashed,thick] (0, 0) -- (1, 1);
\draw[color=black,dashed,thick] (0, 0) -- (-1, 1);
\filldraw[black] (0,0) circle (2pt);
\filldraw[black] (0,-1) circle (2pt);
\filldraw[black] (-1,-2) circle (2pt) node[anchor=north]{$\Psi^{\mathbb{P}^1}_{\g_3}$};
\filldraw[black] (-1,1) circle (2pt) node[anchor=south]{$\Psi^{\mathbb{P}^1}_{\g_1}$};
\filldraw[black] (1,-2) circle (2pt) node[anchor=north]{$\Psi^{\mathbb{P}^1}_{\g_2}$};
\filldraw[black] (1,1) circle (2pt) node[anchor=south]{$\Psi^{\mathbb{P}^1}_{\g_4}$};
\end{tikzpicture}
\]
The $\G_s$-amplitude evaluates into
\begin{align}
\mathcal{A}_{\G_s} \bigl(\Psi^{\mathbb{P}^1}_{\g_\a}\bigr) &= g \bigl(\mu_2\bigl(\Psi^{\mathbb{P}^1}_{\g_4},\Psi^{\mathbb{P}^1}_{\g_3}\bigr), 2\pi KG_-\mu_2\bigl(\Psi^{\mathbb{P}^1}_{\g_1},\Psi^{\mathbb{P}^1}_{\g_2}\bigr)\bigr)\nonumber\\
 & = q_+q _-\Theta (\min(r_3, r_4)-\max(r_1, r_2)) + q_+q_-\Theta (\min(r_1, r_2)-\max(r_3, r_4) ).\label{ampl_g_s_b_model}
\end{align}
We used the following relation:
\[
2\pi K G_- \mu_2 \bigl(\Psi^{\mathbb{P}^1}_{\g_\a}, \Psi^{\mathbb{P}^1}_{\a_\b}\bigr) = q_+ {\rm e}^{{\rm i}Y} \Theta (r-\max(r_\a,r_\b) ) + q_- {\rm e}^{-{\rm i}Y} \Theta (\min(r_\a, r_\b)-r).
\]
The amplitude (\ref{ampl_g_s_b_model}) describes the eight out of 24 possible permutations for positions of four points $r_1$, $r_2$, $r_3$, $r_4$ on the real line. Indeed, the first $\Theta$-function describes the four permutations when the pair of points~$r_1$, $r_2$ lies to the left of the other pair of points $r_3$, $r_4$ on the real line. The second $\Theta$-function describes the situation when the pair $r_1$, $r_2$ lies to the right of the pair~$r_3$,~$r_4$. The sum of three amplitudes in (\ref{total_ampl_4_pt_p1}) describes all 24 permutations, so the correlation function simplifies to the familiar $q_+q_-$ expression (\ref{trop_complex_gw_invarians_p_1}).

\subsection{Hints for localization}
We can add $Q^{\mathbb{P}^1}$-exact term (which is also $G_-$-closed) to turn the mirror state (\ref{form_mirr_state_point_p1}) into a~function of $Y$, i.e.,
\begin{gather}\label{cp_1_dress_+}
\Psi^{\mathbb{P}^1}_\g +Q^{\mathbb{P}^1}\biggl( -\frac{1}{2\pi} \Theta (r_0-r)\psi_{\Phi} \biggr)=q_+ {\rm e}^{{\rm i}Y}.
\end{gather}
Alternatively, we can use different $Q_{\mathbb{CP}^1}$-exact term (which is also $G_-$-closed) to derive
\begin{gather}\label{cp_1_dress_-}
\Psi^{\mathbb{P}^1}_\g +Q^{\mathbb{P}^1} \biggl( \frac{1}{2\pi} \Theta (r-r) \psi_{\Phi} \biggr)=q_- {\rm e}^{-{\rm i}Y}.
\end{gather}
We can express (\ref{cp_1_dress_+}) and (\ref{cp_1_dress_-}) as the equality of cohomology classes
\[
\bigl[\Psi^{\mathbb{P}^1}_\g\bigr]=\bigl[q_+ {\rm e}^{{\rm i}Y} \bigr]=l[q_- {\rm e}^{-{\rm i}Y} \bigr] \in H^\ast \bigl(Q^{\mathbb{P}^1}\bigr).
\]
Since we used $G_-$-closed exact terms, all three forms represent the same class in $Q^{\mathbb{P}^1}+zG_-$ cohomology. Hence, we conjecture that they correspond to K. Saito's good section in the theory of primitive form \cite{saito1983period} for exponential mirror superpotential (\ref{w_p_1_mirror}).

\section{Conclusion}\label{sec7}
Our paper showed that the tropical GW invariants at genus zero could be written as the amplitudes in higher topological quantum mechanics on special trees. The higher topological quantum mechanics on special trees admit an analog of the state-operator correspondence for divisor and evaluation states. We used the state-operator map to summate over amplitudes in $A$-model HTQM and showed that the result is a total amplitude in $B$-model HTQM. For a tropical Gromov--Witten theory on a toric variety $X$, we showed that the $B$-model HTQM is a deformation of the $A$-model by an exponential mirror superpotential, written in terms of compactification polyhedron data $B_X$ of $X$. For tropical observables $\g$, we found the mirror-dual states $\Psi_\g^X$ in the $B$-model and formulated the mirror relation.

We showed that the mirror states for a $\mathbb{P}^1$ toric variety can be written as holomorphic functions in $Q^{\mathbb{P}^1} +z G_-$ cohomology and conjectured that such functions define K.~Saito's good section for the mirror superpotential. In our work \cite{Losev:2023bhj}, we showed that a similar relation holds for a smooth toric variety and indeed defines K.~Saito's good section for the exponential mirror superpotential.

\subsection*{Acknowledgments}

We are grateful to Pavel Mnev and Yasha Neiman for many discussions on the topics presented in this paper. We thank the referees for their valuable suggestions for improving our manuscript. V.L.'s work was supported by the Quantum Gravity Unit of the Okinawa Institute of Science and Technology Graduate University (OIST).

\pdfbookmark[1]{References}{ref}
\LastPageEnding

\end{document}